\documentclass[final,1p,times]{elsarticle}

\usepackage{amssymb}
\usepackage{mathrsfs}
\usepackage{epsfig}
\usepackage{amssymb}
\usepackage{amsfonts}
\usepackage{latexsym}
\usepackage{amsthm}

\biboptions{numbers,sort&compress}

\usepackage{bm}
\usepackage{graphicx}
\usepackage{amsmath}
\usepackage{hyperref}
\usepackage{slashed}

\journal{Computer Physics Communications}

\begin{document}

\begin{frontmatter}



\title{The {\tt iEBE-VISHNU} code package for relativistic heavy-ion collisions}


\author[a,b]{Chun Shen\corref{cor1}}
\ead{chunshen@physics.mcgill.ca}

\author[a]{Zhi Qiu}
\author[a,c]{Huichao Song}
\author[d]{Jonah Bernhard}
\author[d]{Steffen Bass}
\author[a]{Ulrich Heinz}
\ead{heinz@mps.ohio-state.edu}

\cortext[cor1]{Corresponding author}

\address[a]{Department of Physics, The Ohio State University, Columbus, Ohio 43210-1117, USA}
\address[b]{Department of Physics, McGill University, 3600 University Street, Montreal, Quebec, H3A 2T8, Canada}
\address[c]{Department of Physics and State Key Laboratory of Nuclear Physics and Technology, Peking University, Beijing 100871, China}
\address[d]{Department of Physics, Duke University, Durham, North Carolina 27708, USA}

\begin{abstract}
The {\tt iEBE-VISHNU} code package performs event-by-event simulations for relativistic heavy-ion collisions using a hybrid approach based on (2+1)-dimensional viscous hydrodynamics coupled to a hadronic cascade model. We present the detailed model implementation, accompanied by some numerical code tests for the package. {\tt iEBE-VISHNU} forms the core of a general theoretical framework for model-data comparisons through large scale Monte-Carlo simulations. A numerical interface between the hydrodynamically evolving medium and thermal photon radiation is also discussed. This interface is more generally designed for calculations of all kinds of rare probes that are coupled to the temperature and flow velocity evolution of the bulk medium, such as jet energy loss and heavy quark diffusion. 
\end{abstract}

\begin{keyword}
Relativistic heavy-ion collision, relativistic viscous hydrodynamics, quark-gluon plasma


\end{keyword}

\end{frontmatter}

\newpage 


% Computer program descriptions should contain the following
% PROGRAM SUMMARY.
\begin{center}
{\bf PROGRAM SUMMARY}
\end{center}

\begin{small}
\noindent
{\em Manuscript Title: }The {\tt iEBE-VISHNU} code package for relativistic heavy-ion collisions \\
{\em Authors:} Chun Shen, Zhi Qiu, Huichao Song, Jonah Bernhard, Steffen Bass, Ulrich Heinz\\
{\em Program Title:} iEBE-VISHNU \\
{\em Journal Reference:}                                      \\
  %Leave blank, supplied by Elsevier.
{\em Catalogue identifier:}                                   \\
  %Leave blank, supplied by Elsevier.
{\em Licensing provisions:} none                                   \\
  %enter "none" if CPC non-profit use license is sufficient.
{\em Programming language:} Fortran, C++, python, bash, SQLite  \\
{\em Computer:} Laptop, desktop, cluster                                              \\
  %Computer(s) for which program has been designed.
{\em Operating system:} Tested on GNU/Linux Ubuntu 12.04 x64, Red Hat Linux 6, Mac OS X 10.8+              \\
  %Operating system(s) for which program has been designed.
{\em RAM: } 2G bytes                                              \\
  %RAM in bytes required to execute program with typical data.
{\em Number of processors used:} 1                             \\
  %If more than one processor.
{\em Keywords:} Relativistic viscous hydrodynamics, quark-gluon plasma, Monte-Carlo simulation \\
  % Please give some freely chosen keywords that we can use in a
  % cumulative keyword index.
{\em Classification:} 17 Nuclear Physics                                         \\
  %Classify using CPC Program Library Subject Index, see (
  % http://cpc.cs.qub.ac.uk/subjectIndex/SUBJECT_index.html)
  %e.g. 4.4 Feynman diagrams, 5 Computer Algebra.
{\em External routines/libraries:} GNU Scientific Library (GSL), HDF5, Numpy, UrQMD {\it v3.3}       \\
  % Fill in if necessary, otherwise leave out.
%{\em Subprograms used:}                                       \\
  %Fill in if necessary, otherwise leave out.
{\em Nature of problem:}\\
  %Describe the nature of the problem here.
  Relativistic heavy-ion collisions are tiny in size ($V \sim 10^{-42}$ m$^3$) and live in a flash ($\sim 5 \times 10^{-23}$ s). It is impossible to use external probes to study the properties of the quark-gluon plasma (QGP), a novel state of matter created during the collisions. Experiments can only measure the momentum information of stable hadrons, who are the remnants of the collisions. In order to extract the thermal and transport properties of the QGP, one needs to rely on Monte-Carlo event-by-event model simulations, which reverse-engineer the experimental measurements to the early time dynamics of the relativistic heavy-ion collisions. 
   \\
{\em Solution method:}\\
  %Describe the method solution here.
  Relativistic heavy-ion collisions contain multiple stages of evolution. The physics that governs each stage is implemented into individual code component.  A general driver script glues all the modular packages as a whole to perform large-scale Monte-Carlo simulations. The final results are stored into SQLite database, which supports standard querying for massive data analysis. By tuning transport coefficients of the QGP as free parameters, e.g. the specific shear viscosity $\eta/s$, we can constrain various transport properties of the QGP through model-data comparisons. 
   \\
%{\em Restrictions:} none\\
  %Describe any restrictions on the complexity of the problem here.
%{\em Unusual features:} none \\
  %Describe any unusual features of the program/problem here.
%{\em Additional comments:}\\
  %Provide any additional comments here.
{\em Running time:}\\
  %Give an indication of the typical running time here.
The following running time is tested on a laptop computer with a 2.4 GHz Intel Core i5 CPU, 4GB memory. All the C++ and Fortran codes are compiled with the GNU Compiler Collection (GCC) 4.9.2 and {\tt -O3} optimization. 
\begin{table}[h!]
\centering
\begin{tabular}{c|c|c|c}
\hline \hline
 & p+p & 0-5\% p+Pb & 0-5\% Pb+Pb  \\ \hline
initial condition generator {\tt superMC}  & 20$s$ & 20$s$ & 50$s$ \\
(100 events, 400$\times$400 grid) & & &  \\ \hline
(2+1)-d hydrodynamics {\tt VISHNew} & 120$s$ & 200$s$ & 690$s$ \\
 (1 event, 400$\times$400 grid) & & & \\ \hline
Cooper-Frye freeze-out {\tt iSS}& 4$s$ & 15$s$ & 350$s$ \\ 
 (500 events, $\vert y \vert < 4$) & & & \\ \hline
hadron cascade {\tt UrQMD} & 0.03$s$ & 0.18$s$ & 150$s$ \\ 
 (1 event, $\vert y \vert < 4$) & & & \\ \hline \hline
\end{tabular}
\caption{Summary of typical running time (in second) of individual component in the package. Different types of collisions are simulated at $\sqrt{s_\mathrm{NN}} = 5.02$ TeV.}
\label{runningtimetable}
\end{table}

%It takes about 60 minutes to simulate a typical 0-5\% central Au+Au collision at $\sqrt{s_\mathrm{NN}} = 200$ GeV. The (2+1)-d viscous hydrodynamics ({\tt VISH2+1}) takes $\sim 10$ minutes with lattice spacing $dx = dy = 0.1$ fm and $d\tau = 0.02$ fm on a $26 \times 26$ fm$^2$ grid in the transverse plane. The Cooper-Fyre freeze-out ({\tt iS}) takes about 45 minutes to compute the 319 hadron spectra and their anisotropic flow coefficients $v_n (n = 1 - 9)$. In the hybrid approach, the hadronic cascade code ({\tt UrQMD}) takes about 30 mins to simulate 500 oversampled events of particles collisions. 
%\begin{thebibliography}{0}
%\bibitem{1}Reference 1         % This list should only contain those items referenced in the                 
                                              % Program Summary section.   
                                              % Type references in text as [1], [2], etc.
                               % This list is different from the bibliography at the end of 
                               % the Long Write-Up.
%\end{thebibliography}

\end{small}

\newpage

\section{Introduction}
\label{sec1}
%%%%%%%%%%%%%%%%%%%%%%%%%%%%%%%%%%%%%%%%%%%%%

The Relativistic Heavy-Ion Collider (RHIC) at Brookhaven National Laboratory and the Large Hadron Collider (LHC) at CERN provide unique experimental access to a new state of matter at extremely high densities and temperatures: the Quark Gluon Plasma (QGP), in which quarks and gluons are no longer confined inside individual nucleons. Studying the thermodynamic and transport properties of the QGP will help us understand emergent phenomena in hot and dense many-body systems governed by the strong interaction. However, these ``little bangs'' are almost point-like in size ($V$ $\sim10^{-42}$ m$^3$) and disappear almost instantaneously ($\sim 5 \times 10^{-23}$s). This makes it impossible to use external probes to measure the properties of the QGP. In order to extract the dynamical evolution of relativistic heavy-ion collisions, one has to rely on realistic theoretical model simulations, which back trace the final experimental observables to the early stage of the collisions.  

%
%=======================================
\begin{figure}[h!]
  \centering
  \includegraphics[width=0.85\linewidth]{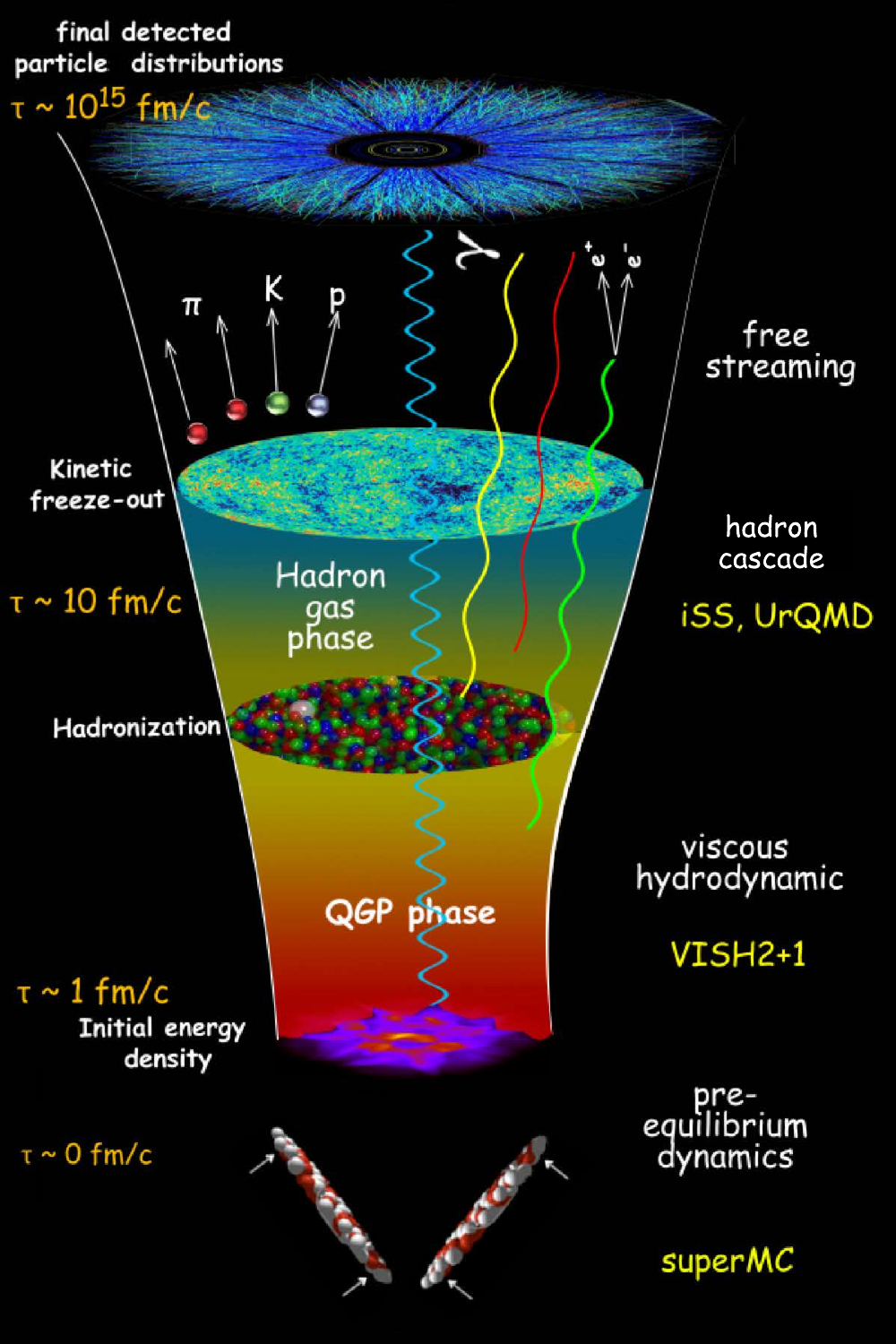}
  \caption{Illustration of the evolution of the fireball created in relativistic heavy-ion collisions, together with the theoretical model used in each stage. }
  \label{Intro.codes}
\end{figure}
%=======================================
%
The evolution of a relativistic heavy-ion collision contains multiple stages which are governed by different underlying physics. Right after the initial overlap of the colliding nuclei, the system is dominated by gluons characterized by an over populated phase-space distribution \cite{Kovchegov:1537858}. The number of gluons is of order $\sim \frac{1}{g^2}$ with $ g < 1$ and these gluons carry each a very small fraction of the longitudinal momentum of the incoming nucleus (small-$x$ gluons). During the first 1 fm/$c$, due to the large occupation number of gluons at leading order in strong coupling $g$, these saturated small-$x$ gluons will evolve according to the classical Yang-Mills equation of motion. It is believed that the next-to-leading order quantum corrections to the classical field evolution drive the system rapidly towards local isotropy in momentum space \cite{Gelis:2013rba,Dusling:2010rm} and somewhat later to local thermal equilibrium.  After $0.3-0.5$ fm/$c$, the system achieves approximately local momentum isotropy; local thermal equilibrium is reached after a few fm/$c$. The quarks and gluons that are produced after the collision form a strongly coupled plasma (QGP). The dynamics of the QGP can be described by macroscopic viscous hydrodynamics where the viscous corrections account for the remaining deviation from local isotropy and thermal equilibrium \cite{Song:2007ux,Dusling:2007gi,Luzum:2008cw,Bozek:2009dw,Holopainen:2010gz,Schenke:2010rr,Pang:2012he,Karpenko:2013wva,DelZanna:2013eua}. As the system expands and cools, it will smoothly cross over from the QGP phase to a hadron gas phase according to the equation of state (EOS) determined from Lattice QCD calculations \cite{Chojnacki:2007jc, Song:2008si, Huovinen:2009yb, Borsanyi:2013bia, Bazavov:2014pvz}. At hadronization, the quark-gluon fluid will convert into hadrons due to confinement. In the hadronic phase, the hadron cascade model can provide us with a detailed microscopic description of the evolution \cite{Bass:1998ca,Bleicher:1999xi}. 

As the fireball continues to expand and cool, the collision rates between the hadronic resonances decrease. First, the inelastic collisions between particles cease and the system reaches chemical freeze-out almost directly after hadronization \cite{Hirano:2002ds}. After this point, only resonance decays and baryon-antibaryon annihilation can change the particle yields \cite{Song:2013qma}. Regeneration of baryon-antibaryon pairs is a rare process that can be neglected. As the system evolves further, the density of the fireball becomes so low that the mean free time of the particles becomes much larger than the Hubble time (i.e. the time over which the inter particle spacing doubles.) \cite{Heinz:2013wva}. The particles reach kinetic freeze out and subsequently free-stream to the detectors. In Fig. \ref{Intro.codes}, we schematically summarize the theoretical models and the corresponding codes that we will use to simulate the different stages of heavy-ion collisions. We will explain them in detail in the following Sections.

In relativistic heavy-ion collisions, rare electromagnetic observables like photons and dileptons only interact with the medium through the electromagnetic interaction, which is much weaker than the strong interaction. For this reason, their mean free path is much longer than the system size, and hence they suffer negligible final state interactions after they are produced during the fireball evolution. This advantage over strongly interacting probes makes them the cleanest penetrating probe for the heavy-ion collisions. Hadrons can only break free at the final kinetic freeze-out surface. Their measured momentum distribution carries indirect time integrated evolution information about the fireball. On the other hand, a large fraction of the thermal photons are produced early inside the fireball. Their momentum distribution preserves the dynamical information of the medium directly at their production points. Electromagnetic probes can thus provide us with constraints on the early dynamics of the fireball that are complementary to those obtained from the much more abundant hadronic observables. In Sec. \ref{sec9}, we will discuss the interface which coupled the event-by-event viscous hydrodynamic evolution with thermal photon radiations. 

The entire integrated package is open source\footnote{%
   Except for the {\tt UrQMD} component, the {\tt iEBE-VISHNU} 
   package is made available under the GNU general public license v3.0.} 
and can be freely downloaded from \url{https://u.osu.edu/vishnu}. Other viscous relativistic hydrodynamic codes for application to relativistic heavy-ion collisions have been developed, and results obtained with them have been reported in the literature. These include the (2+1)-d pure viscous hydrodynamic code {\tt v-USPhydro} \cite{Noronha-Hostler:2013gga}; the (2+1)-dimensional hybrid code {\tt SONIC} \cite{Habich:2014jna}, publicly available at \url{https://sites.google.com/site/revihy/home}, which interfaces a strongly-coupled pre-equilibrium phase based on the AdS/CFT correspondence with the (2+1)-d viscous hydrodynamic code {\tt VH2+1} and a hadronic cascade; the (3+1)-dimensional hybrid code {\tt IPGlasma+MUSIC+UrQMD} \cite{Ryu:2015vwa} which couples a weakly-coupled pre-equilibrium stage based on classical Yang-Mills-evolution of fluctuating gluon fields \cite{Schenke:2012wb} to (3+1)-dimensional viscous hydrodynamics \cite{Schenke:2010rr, Gale:2012rq} and the {\tt UrQMD} hadron cascade; and a number of pure viscous hydrodynamic codes in 3+1 dimensions: the Frankfurt-Kiev code \cite{Karpenko:2013wva}, the Jyv\"askyl\"a-Frankfurt-Debrecen code \cite{Niemi:2012ry,Molnar:2014zha}, the Krakow code \cite{Bozek:2011ua, Bozek:2011if}, the {\tt ECHO-QGP} code \cite{DelZanna:2013eua}, the {\tt CLVisc} code \cite{Pang:2014ipa}, and the Nagoya code \cite{Akamatsu:2013wyk}.

\section{General Framework}

Every relativistic heavy-ion collision is a multi-stage system. In our hybrid package, there is a specific code simulating each stage of the evolution. A python shell script links all the individual programs together to perform large-scale event-by-event simulations of relativistic heavy-ion collisions. The major components include the initial condition generator ({\tt superMC}), a (2+1)-d viscous hydrodynamic simulator ({\tt VISHNew}), a particle sampler ({\tt iSS}), and a hadron cascade simulator ({\tt UrQMD}). In the next section, we will discuss in some detail the physics implemented in these codes. 

To perform event-by-event simulations on multiple computing cores, for example using $N$ cores on a cluster, we divide the total number of events, $N_\mathrm{ev}$, into $N$ jobs with $M = N_\mathrm{ev}/N$ events in each jobs. Then we submit these $N$ jobs in parallel. The $M$ events within each job run sequentially. 
 
\subsection{Work flow for a single sequential simulation}

%=======================================
\begin{figure}[h!]
  \centering
  \includegraphics[width=0.9\linewidth]{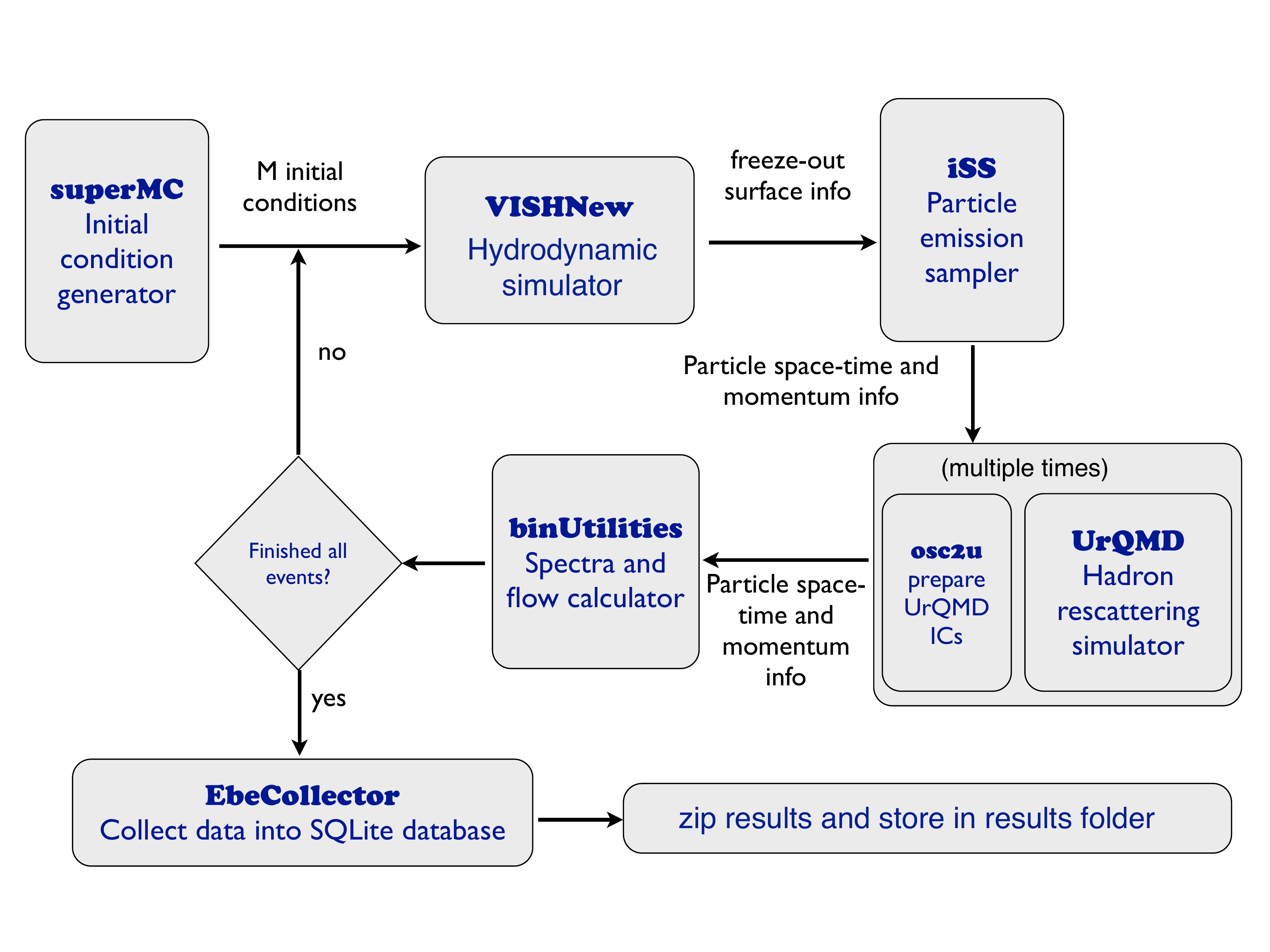}
  \caption{The work flow for a single job with $M$ events.}
  \label{singleJobworkflow}
\end{figure}
%=======================================

For each job, the work flow is summarized in Fig. \ref{singleJobworkflow}.  The job is started by generating $M$ fluctuating initial conditions with the Monte-Carlo generator {\tt superMC}\footnote{{\tt superMC} is based on the code package {\tt mckt} \cite{ALbacete:2010ad, Albacete:2012xq}}. Then each initial entropy density profile is evolved with the viscous hydrodynamic model, {\tt VISHNew} \cite{Song:2007ux, Song:2008si}. At the end of the hydrodynamic simulation, a switching hypersurface is identified and fluid cells on this switching hyper-surface are converted into individual particles using the particle sampler, {\tt iSS}. These particles are fed into {\tt UrQMD} \cite{Bass:1998ca}, a hadronic rescattering cascade which follows the particles microscopically until they stop interacting and (if unstable) decay.%
\footnote{%
    To accumulate statistics, the {\tt UrQMD} casacade is optionally 
    run multiple times (with different sampled particles from {\tt iSS}) 
    for each hydrodynamic simulation.}  
The combination of the hydrodynamic evolution algorithm for the QGP stage with a microscopic hadronic cascade forms a hybrid algorithm with the name {\tt VISHNU} ({\tt VISH2+1} 'n' {\tt UrQMD}) \cite{Song:2010aq}. In the end, we collect the final particle information (momenta and positions of their last interactions or decays) from all the $M$ events using {\tt binUtilities}, store the final analyzed results in the  {\tt SQLite} database using {\tt EbeCollector}, and zip everything. 

\subsection{Large scale event-by-event simulations}

\begin{figure}[h!]
  \centering
  \includegraphics[width=0.7\linewidth]{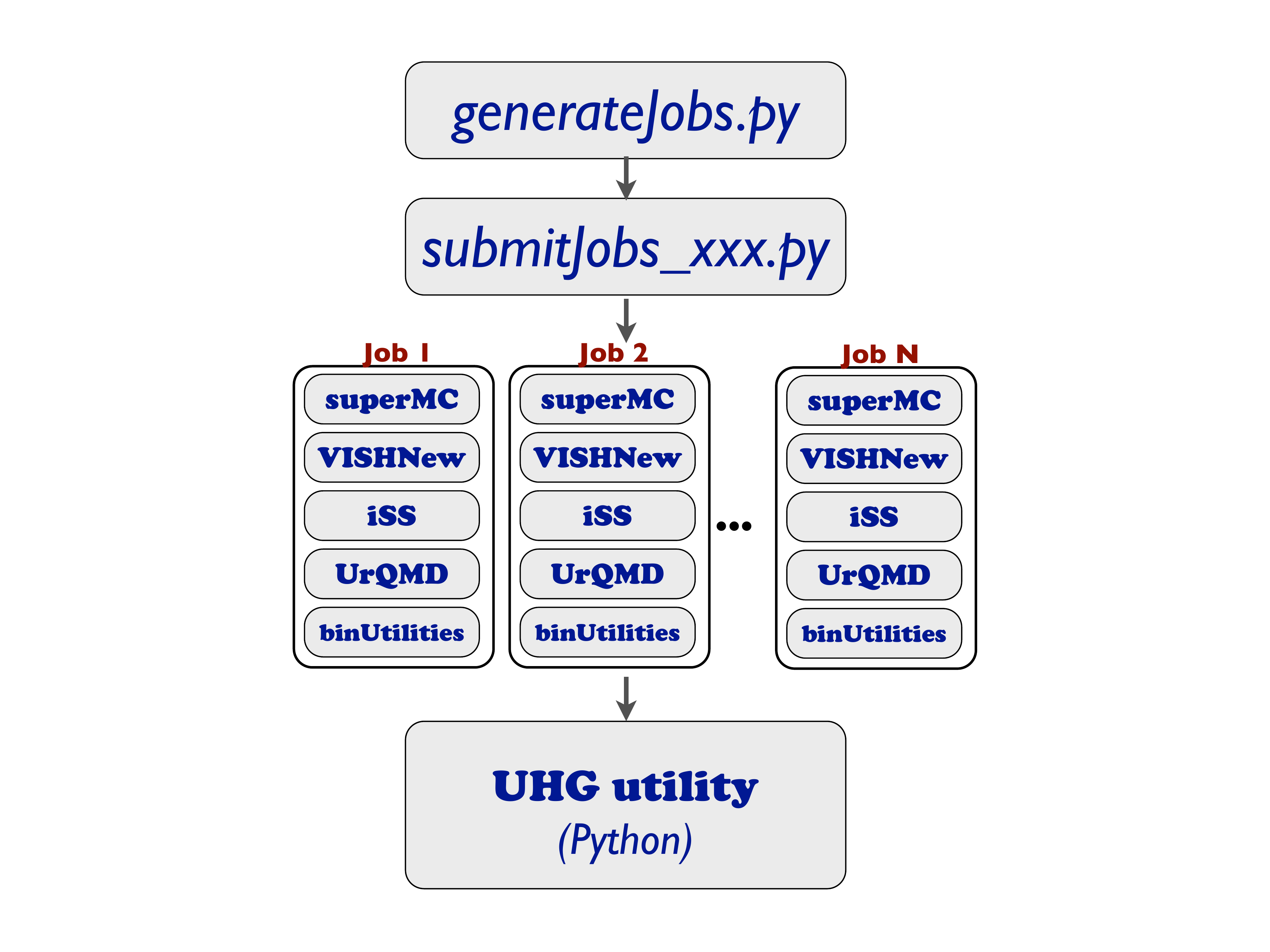}
  \caption{Work flow for multiply jobs in the large scale of event-by-event simulations.}
  \label{multiJobworkflow}
\end{figure}

For large-scale event-by-event simulations, two additional python scripts are used to generate and submit multiple jobs as illustrated in Fig. \ref{multiJobworkflow}. Users specify the number of jobs and the number of events within each job through {\tt generateJobs.py} which sets up the entire simulation and then use the script {\tt submitJobs\_local.py} or {\tt submitJobs\_qsub.py} to submit all the jobs to a local cluster or to a qsub system on the Ohio Supercomputer Center (OSC), respectively. Easy adjustments of those latter python scripts can adapt the package to other supercomputing facilities or the Open Science Grid.  

%=======================================
%=======================================

After all $N$ jobs are finished, the database files from each job will be combined into one for future physics analysis of the output by users. A user friendly python package tool, {\tt UHG utility}, is provided for querying the database and computing experimental observables and performing various statistical analyses. 

%%%%%%%%%%%%%%%%%%%%%%%%%%%%%%%%%%%%%%%%%%%%%
\section{Initial condition generator {\tt SuperMC}}
\label{sec3}
%%%%%%%%%%%%%%%%%%%%%%%%%%%%%%%%%%%%%%%%%%%%%

{\tt SuperMC} generates fluctuating initial conditions using the Monte-Carlo Glauber (MC-Glauber) \cite{Broniowski:2007nz, Alver:2008aq, Loizides:2014vua} or Monte-Carlo Kharzeev-Levin-Nardi (MCKLN) \cite{Kharzeev:2001yq,Kharzeev:2004if} models. These models can be run in several distinct modes as selected by the user.   

\subsection{Collision geometry}

In relativistic heavy-ion collisions, the colliding nuclei are accelerated almost to the speed of light. Due to time dilation, nucleons' intrinsic orbital motion is frozen during the interaction period. Thus we can use a Monte Carlo procedure to sample the position of every nucleon inside the projectile and target nuclei according to their Woods-Saxon distribution.  

We take into account the finite size for each individual nucleon. The density distribution of strongly interacting matter for each nucleon is given by
\begin{equation}
\rho_n(\vec{\bf r}) = \left\{ \begin{array}{cl} \frac{\theta(r_\perp-r_n)}{\pi r_n^2}\frac{\theta(L - \vert z \vert)}{L}, & \mbox{cylindrical nucleon,} \\ \frac{1}{(2\pi B)^{3/2}} e^{-r^2/(2B)}, & \mbox{gaussian nucleon.} \end{array}\right. 
\label{superMC.eq1}
\end{equation}
The approximation of a homogeneous cylindrical nucleon density distribution has been popular in the past since it leads to a very simple collision criterium. In this approximation, the transverse radius $r_n = \frac{1}{2} \sqrt{\frac{\sigma^\mathrm{inel}_\mathrm{NN}}{\pi}}$, where the factor of 2 accounts for the quantum mechanical nature of the nucleon-nucleon scattering process. Along the $z$ direction, $L = 2 r_n$. A more realistic modeling takes a gaussian density distribution for the nucleon with an energy-dependent width $B = B(\sqrt{s_\mathrm{NN}}) = \frac{\sigma^\mathrm{in}_\mathrm{NN}(\sqrt{s_\mathrm{NN}})}{8\pi}$\cite{Heinz:2011mh}.%
\footnote{%
      In this expression we changed the denominator from the value 14.30 suggested 
      in \cite{Heinz:2011mh} to $8\pi$ because this yields a better description of the
      measured multiplicity distributions in p+Pb collisions at the LHC, discussed in 
      Sec.~\ref{chap2.multiplicityFluct}. This choice agrees with the naive ansatz
      $\sigma_\mathrm{NN}^\mathrm{inel} = 2 \pi\,(2R)^2$, expressing the inelastic 
      cross section as the area of a disk with radius $2R$ where $R=\sqrt{B}$ is the 
      rms radius of a nucleon.} 
The corresponding nucleon thickness functions in the transverse plane are
\begin{equation}
T_n(r_\perp) = \left\{ \begin{array}{cl} \frac{\theta(r_\perp-r_n)}{\pi r_n^2}, & \mbox{cylindrical nucleon,} \\ \frac{1}{(2\pi B)} e^{-r_\perp^2/(2B)}, & \mbox{gaussian nucleon.} \end{array}\right. .
\label{superMC.eq2}
\end{equation}
With the finite size of each nucleon, in order to reproduce the correct Woods-Saxon distribution for the density of the entire nucleus, we need to sample the nucleon positions according to a {\it modified} Woods-Saxon distribution such that, when folded with the nucleon density distribution \cite{Filip:2007tj, Filip:2009zz, Hirano:2012kj}, it reproduces the correct experimentally measured Woods-Saxon distribution: 
\begin{equation}
\rho^\mathrm{exp.}_\mathrm{WS}(\vec{\bf r}) = \int d^3 {\bf r'} \tilde{\rho}_\mathrm{WS}(\vec{\bf r'}) \rho_n(\vec{\bf r} - \vec{\bf r'}).
\label{superMC.eq3}
\end{equation}
\begin{equation}
\tilde{\rho}_\mathrm{WS}(\vec{\bf r}) =  \frac{\rho_0}{1 + \exp\left(\frac{r - R_A \Omega(\theta)}{\xi} \right)},
\label{superMC.eq4}
\end{equation}
where $\rho_0$ is the nucleon number density in infinite nuclear matter, $R_A$ is the rms charge radius of nucleus $A$, and $\xi$ is the surface width parameter. For a deformed nucleus with non-zero quadrupole and hexadecupole ground state deformation \cite{Filip:2007tj, Filip:2009zz, Broniowski:2007nz, Hirano:2012kj,Loizides:2014vua} $R(\theta) = R_A \Omega(\theta) = R_A( 1 + \beta_2 Y^2_0(\theta) + \beta_4 Y^4_0(\theta))$, where $Y^2_0(\theta)$ and $Y^4_0(\theta)$ are the spherical harmonics, describes the angular dependence of the nuclear radius. In Table \ref{superMC.WStable}, we list the parameters used in {\tt superMC} for some typical colliding nuclei. 
\begin{table}[h]
\centering
\begin{tabular}{cc|ccccc}
\hline \hline
Element & Atomic Mass & $\rho_0$ (fm$^{-3}$) & $R_A$ (fm) & $\xi$ (fm) & $\beta_2$ & $\beta_4$ \\ \hline
Cu & 63 & 0.1686 & 4.28 & 0.50 & 0.162 & 0.006 \\
Au & 197& 0.1695 & 6.42 & 0.45 & -0.130 & -0.030 \\
Pb & 208 & 0.1610 & 6.67 & 0.44 & 0 & 0 \\
U & 238 & 0.1660 & 6.86 & 0.44& 0.280 & 0.093 \\ \hline \hline
\end{tabular}
\caption{Parameters for the modified Woods-Saxon density distribution, $\tilde{\rho}_\mathrm{WS}$, for some heavy nuclei that have been used in relativistic heavy-ion collisions.}
\label{superMC.WStable}
\end{table}

Note that our parameterization of the nuclear density distribution does not account for the existence of a neutron skin in large nuclei. Inclusion of a neutron skin is left for a future improvement of the {\tt superMC} code.

\subsection{The MC-Glauber approach}

The density of the inelastic cross section $\sigma^\mathrm{inel}_\mathrm{NN}$ at impact parameter $\vec{\bf b}$ is
\begin{equation}
P(\vec{\bf b}) = \left\{ \begin{array}{cl} \theta(2r_n - b), & \mbox{cylindrical nucleon,} \\ 1 - \exp(-\sigma_\mathrm{gg} T_\mathrm{nn}(b)), & \mbox{gaussian nucleon.} \end{array}\right. 
\label{superMC.eq5}
\end{equation}
where $\sigma_\mathrm{gg}$ is the inelastic gluon-gluon cross-section \cite{Heinz:2011mh} and $T_\mathrm{nn}$ is the nucleon-nucleon overlap function,
\begin{equation}
T_\mathrm{nn}(b) = \int d^2 \vec{\bf r}_\perp T_n(\vec{\bf r}_\perp) T_n(\vec{\bf b} - \vec{\bf r}_\perp) = \frac{e^{-b^2/(4B)}}{4\pi B}.
\label{superMC.eq6}
\end{equation}
For unpolarized nucleons the nucleon density is spherically symmetric, so $T_\mathrm{nn}(b)$ has no directional dependence. A binary collision involving nucleon pair $(i,j)$ will deposit a certain amount of energy in the medium around the collision point $\vec{\bf R}_{ij,\perp} = \frac{1}{2}(\vec{\bf r}_{i\perp} + \vec{\bf r}_{j\perp})$. After thermalization, this energy density is associated with a corresponding amount of entropy density computable from the equation of state. For cylindrical nucleons, we choose a disk-like profile for the deposited energy or entropy density in the transverse plane. For Gaussian nucleons, the deposited energy density is modeled by a gaussian distribution. Thus, the entropy or energy density generated by all the binary collision pairs in the transverse plane is proportional to,
\begin{equation}
BC(\vec{\bf r}_\perp) = \left\{ \begin{array}{cl} \sum_{(i, j) \in \mathrm{pairs}} \frac{\theta(r_n - \vert \vec{\bf r}_\perp - \vec{\bf R}_{ij,\perp} \vert)}{\pi r_n^2}, & \mbox{cylindrical nucleons,} \\ \sum_{(i, j) \in \mathrm{pairs}} \frac{1}{2\pi B}e^{-\vert \vec{\bf r}_\perp - \vec{\bf R}_{ij,\perp} \vert^2/(2B)}, & \mbox{gaussian nucleons.} \end{array}\right.
\label{superMC.eq7}
\end{equation}
The parameters $r_n$ and $B$ are chosen to be the same as in the definition of the shape of the nucleon, Eq. (\ref{superMC.eq1}).

Every nucleon that participates in an inelastic collision is ``wounded'' and will ``bleed'' energy density into the medium. In {\tt superMC}, two distinct ways to distribute the energy deposited by the wounded nucleons are implemented.  

The first option is to deposit the energy azimuthally symmetrically around the center of the wounded nucleon. The total energy or entropy density contributed by all wounded nucleons is then proportional to
\begin{equation}
WN(\vec{\bf r}_\perp) = \left\{ \begin{array}{cl} \sum_{i \in \mathrm{wounded}} \frac{\theta(r_n - \vert \vec{\bf r}_\perp - \vec{\bf r}_{i\perp} \vert)}{\pi r_n^2}, & \mbox{cylindrical nucleons,} \\ \sum_{i \in \mathrm{wounded}} \frac{1}{2\pi B}e^{-\vert \vec{\bf r}_\perp - \vec{\bf r}_{i\perp} \vert^2/(2B)}, & \mbox{gaussian nucleons.} \end{array}\right.
\label{superMC.eq8}
\end{equation}
where the index $i$ runs over all wounded nucleons in both nuclei A and B. 

In the second approach, the energy bled from each wounded nucleon is distributed evenly over its binary collision partners and deposited azimuthally symmetrically around their correspond binary collision points. In this case, the total energy or entropy density contributed by all wounded nucleons is proportional to
\begin{equation}
WN(\vec{\bf r}_\perp) = \left\{ \begin{array}{cl} \sum_{i \in \mathrm{wounded}} \sum_{j=1}^{N_{b,i}} \frac{1}{N_{b,i}}\frac{\theta(r_n - \vert \vec{\bf r}_\perp - \vec{\bf R}_{ij,\perp} \vert)}{\pi r_n^2}, & \mbox{cylindrical nucleon,} \\ \sum_{i \in \mathrm{wounded}} \sum_{j=1}^{N_{b,i}} \frac{1}{N_{b,i}} \frac{1}{2\pi B}e^{-\vert \vec{\bf r}_\perp - \vec{\bf R}_{ij,\perp} \vert^2/(2B)}, & \mbox{gaussian nucleon.} \end{array}\right.
\label{superMC.eq9}
\end{equation}
where $N_{b,i}$ is the number of binary collision partners associated with wounded nucleon $i$. This way of distributing the energy density is motivated by the idea that the inelastic collisions between nucleons that generate wounded nucleons or binary collision events are fundamentally the same.

The second approach distributes the entropy or energy density of wounded nucleons over a more compact transverse area, which in the end result increases the initial eccentricity of the fireball created in the collision at large impact parameters. In central collisions, the difference in eccentricity between the two energy deposition schemes is negligible.  

In the MC-Glauber model, the total energy density produced in the transverse plane after thermalization is taken to be a mixture of the wounded nucleon and binary collision density profiles \cite{Kharzeev:2000ph, Back:2001xy}: 
\begin{equation}
\left\{ \begin{array}{c} s_0(\vec{ \bf r}_\perp) \\ e_0(\vec{ \bf r}_\perp) \end{array} \right\} = \frac{1}{\tau_0} \left\{ \begin{array}{c} \kappa_s \\ \kappa_e \end{array} \right\} \left(\frac{1-\alpha}{2} WN(\vec{\bf r}_\perp) + \alpha BC(\vec{\bf r}_\perp) \right),
\label{superMC.eq10}
\end{equation}
Here $\alpha$ is the binary mixing parameter and $\kappa$ is an overall normalization factor $\kappa$ is tuned to reproduce to measured final charged multiplicity in the most central collisions, while $\alpha$ is adjusted to reproduce its observed dependence on collision centrality. Due to viscous heating during the hydrodynamic expansion, the normalization $\kappa$ depends on the specific shear viscosity $\eta/s$. In Table~\ref{superMC.MCGlbnormalizationfactor}, we list the values of $\kappa_s$ for several values of $\eta/s$ at RHIC and LHC energies. 
\begin{table}[h]
\centering
\begin{tabular}{c|c|c|c}
\hline \hline
 & Au+Au @ 200 A GeV & Pb+Pb @ 2.76 A TeV & Pb+Pb @ 5.5 A TeV  \\ \hline
$\eta/s = 0.08$ & 17.900 & 34.591 & 40.132 \\
$\eta/s = 0.12$ & 16.694 & 32.759 & 38.161 \\ 
$\eta/s = 0.16$ & 15.492 & 30.908 & 36.148 \\ 
$\eta/s = 0.20$ & 14.290 & 29.040 & 34.116 \\ \hline \hline
\end{tabular}
\caption{The normalization factor $\kappa_s$ for the different values of $\eta/s$ at the RHIC and LHC energies for MC-Glauber model. }
\label{superMC.MCGlbnormalizationfactor}
\end{table}

\subsection{The MCKLN approach}

The MCKLN model \cite{Kharzeev:2001yq,Kharzeev:2004if} is based on a $k_T$-factorization ansatz \cite{Kharzeev:2001yq,Kharzeev:2004if} in which the produced gluon density distribution can be calculated as
\begin{eqnarray}
\frac{dN_g}{dy d^2 p_\perp d^2 x_\perp} &=& \frac{2\pi^3 N_c}{N_c^2 - 1} \int^{p_\perp}_0 d^2 k_\perp \frac{\alpha_s(\mathrm{max}\{((\vec{\bf p}_\perp + \vec{\bf k}_\perp)/2)^2, ((\vec{\bf p}_\perp - \vec{\bf k}_\perp)/2)^2\})}{p_\perp^2} \notag \\
&\times& \phi_A\left(x_1, \left(\frac{\vec{\bf p}_\perp + \vec{\bf k}_\perp}{2}\right)^2; \vec{\bf x}_\perp + \vec{\bf b}/2\right) \notag \\
&\times& \phi_B\left(x_2, \left(\frac{\vec{\bf p}_\perp - \vec{\bf k}_\perp}{2}\right)^2; \vec{\bf x}_\perp - \vec{\bf b}/2\right),
\end{eqnarray}
where $\alpha_s$ is the strong coupling constant and $\phi_A$ and $\phi_B$ are the unintegrated gluon distribution functions of the two colliding nucleus. $\vec{\bf p}_\perp = \frac{\vec{\bf p}_{1\perp} + \vec{\bf p}_{2\perp}}{2}$ and $\vec{\bf k}_\perp = \vec{\bf p}_{1\perp} - \vec{\bf p}_{2\perp}$, where $\vec{\bf p}_{1(2)\perp}$ are the transverse momenta of the fusing gluons from the two nuclei and $x_{1(2)} = \frac{p_\perp}{\sqrt{s_\mathrm{NN}}} e^{\pm y}$ are their corresponding light-cone momentum fractions. The unintegrated gluon distribution function is parameterized as,
\begin{equation}
\phi (x, k^2; \vec{\bf x}_\perp)  = \kappa \frac{N_c^2 - 1}{2N_c} \frac{Q_s^2(x, \vec{\bf x}_\perp)}{2\pi^3 \alpha_s(Q_s^2)} \left\{\begin{array}{cl} \frac{1}{Q_s^2 + \Lambda^2}, & k \le Q_s \\ \frac{1}{k^2 + \Lambda^2}, & k > Q_s \end{array}\right.,
\end{equation}
where $\Lambda = \Lambda_\mathrm{QCD} = 0.2$\,GeV, and $\kappa = 1.8$ is a phenomenological parameter adjusted \cite{Hirano:2004en} to fit the measured charged multiplicity at mid rapidity in the most central Au+Au collisions at $\sqrt{s_\mathrm{NN}} = 200$\,GeV at RHIC. The saturation scale is given by the implicit relation,
\begin{equation}
Q_s^2 (x, \vec{\bf x}_\perp) = \frac{4\pi^2 N_c}{N_c^2 - 1} \alpha_s(Q_s^2) xG(x, Q_s^2) T_A(\vec{\bf x}_\perp)
\label{superMC.eq13}
\end{equation}
The running coupling strength is parameterized as,
\begin{equation}
\alpha_s(k^2) = \left\{\begin{array}{cl} \frac{4\pi}{\beta_0 \ln\left((k^2+\Lambda^2)/\Lambda^2_\mathrm{QCD}\right) }, & \alpha_s \le 0.5 \\ 0.5, & \alpha_s \ge 0.5 \end{array}\right.,
\end{equation}
with $\beta_0 = 11 - \frac{2}{3}N_f$. Kharzeev, Levin, and Nardi \cite{Kharzeev:2004if} use the parameterization
$x G(x, k^2) = K \ln (\frac{k^2 + \Lambda^2}{\Lambda_\mathrm{QCD}^2} x^{-\lambda} (1 - x)^4$ with $\lambda = 0.2$ and $K = 0.7$ adjusted such that the average $Q_s^2$ in the transverse plane of a central 200 A GeV Au + Au collision, $\langle Q_s^2 (x = 0.01)\rangle \simeq 2.0$ GeV$^2$.\cite{Hirano:2004en} Inserting this into Eq. (\ref{superMC.eq13}) and dropping the $(1-x)^4$ factor since $x$ is small in the kinematic region of interest leads to
\begin{equation}
Q_s^2 (x, \vec{\bf x}_\perp) = 2 \mbox{GeV}^2 \left(\frac{T(\vec{ \bf x}_\perp)}{T_0}\right) \left( \frac{x_0}{x}\right)^\lambda,
\end{equation}
where $T_0 = 1.53$ fm$^{-2}$ and $x_0 = 0.01$ \cite{Hirano:2009ah}.

The initial entropy density in the transverse is assumed to be proportional to the $p_T$-integrated produced gluon density,
\begin{equation}
    s(\vec{\bf {x}}_\perp) = \frac{\kappa}{\tau_0} \int d^2 p_\perp \frac{dN_g}{dy d^2 p_\perp d^2 x_\perp}. 
\end{equation}
Table~\ref{superMC.MCKLNnormalizationfactor} lists the values of the normalization factor $\kappa$ for different $\eta/s$, which are fixed to reproduce the top 0-5\% final charged hadron multiplicity at the mid-rapidity.

\begin{table}[h]
\centering
\begin{tabular}{c|c|c|c}
\hline \hline
 & Au+Au @ 200 A GeV & Pb+Pb @ 2.76 A TeV & Pb+Pb @ 5.5 A TeV  \\ \hline
$\eta/s = 0.08$ & 5.692 & 6.998 & 7.628 \\
$\eta/s = 0.12$ & 5.309 & 6.625 & 7.250 \\ 
$\eta/s = 0.16$ & 4.923 & 6.255 & 6.871 \\ 
$\eta/s = 0.20$ & 4.541 & 5.878 & 6.486 \\ \hline \hline
\end{tabular}
\caption{The normalization factor $\kappa$ for the different values of $\eta/s$ at the RHIC and LHC energies for MC-KLN model. }
\label{superMC.MCKLNnormalizationfactor}
\end{table}

\subsection{Collision-by-collision multiplicity fluctuations}
\label{chap2.multiplicityFluct}

The entropy (or energy) density dumped into the medium from each binary collision and wounded nucleon can fluctuate. These fluctuations lead to the measured multiplicity fluctuation in pp collisions. We denote such fluctuation as collision-by-collision multiplicity fluctuations. 

In 2012, the CMS collaboration measured flow observables in 0-0.2\% ultra-central Pb + Pb collisions at the LHC \cite{CMS:2013bza}. For these extremely high multiplicity and extremely rare heavy-ion collision events,  the event selection is strongly biased towards upward fluctuations in the particles production of the system. Thus, we would expect collision-by-collision multiplicity fluctuations to become important for the event selection in such ultra-central collisions. 

In {\tt superMC}, we implement collision-by-collision multiplicity fluctuations in the MC-Glauber model based on the phenomenological KNO scaling observed in pp collisions \cite{Khachatryan:2010nk,Ansorge:1988kn}. In the MC-Glauber model, we regard each binary collision and each wounded nucleon as an independent source of energy with stochastic norm. This can be expressed through the following modification of Eq. (\ref{superMC.eq7}) and (\ref{superMC.eq8}):
\begin{equation}
BC(\vec{\bf r}_\perp) = \left\{ \begin{array}{cl} \sum_{(i, j) \in \mathrm{pairs}} \gamma_{i,j} \frac{\theta(r_n - \vert \vec{\bf r}_\perp - \vec{\bf R}_{ij,\perp} \vert)}{\pi r_n^2}, & \mbox{cylindrical nucleons,} \\ \sum_{(i, j) \in \mathrm{pairs}}  \gamma_{i,j}  \frac{1}{2\pi B}e^{-\vert \vec{\bf r}_\perp - \vec{\bf R}_{ij,\perp} \vert^2/(2B)}, & \mbox{gaussian nucleon.} \end{array}\right. 
\label{superMC.eq16}
\end{equation}
and
\begin{equation}
WN(\vec{\bf r}_\perp) = \left\{ \begin{array}{cl} \sum_{i \in \mathrm{wounded}} \gamma_i \frac{\theta(r_n - \vert \vec{\bf r}_\perp - \vec{\bf r}_{i\perp} \vert)}{\pi r_n^2}, & \mbox{cylindrical nucleon,} \\ \sum_{i \in \mathrm{wounded}} \gamma_i \frac{1}{2\pi B}e^{-\vert \vec{\bf r}_\perp - \vec{\bf r}_{i\perp} \vert^2/(2B)}, & \mbox{gaussian nucleon.} \end{array}\right. 
\label{superMC.eq17}
\end{equation}
where the multiplicity scaling factors $\gamma_{i,j}$ and $\gamma_i$ are continuous random variables with unit mean values. 
In practice, we use the Gamma distribution as the probability distribution for $\gamma_{i,j}$ and $\gamma_i$. The Gamma distribution for a random variable $X$ is defined as
\begin{equation}
\mathrm{Gamma}(X) = \frac{1}{\Gamma(k) \theta^k} x^{k-1} e^{-x/\theta},
\label{superMC.eq18}
\end{equation}
where $k$ and $\theta$ are the so-called shape and scale parameters of the Gamma distribution, respectively. 
The Gamma distribution is positive semi-definite and has the following properties:

(1) If $X_i = \mathrm{Gamma}(k_i, \theta)$, then $\sum_i X_i = \mathrm{Gamma}(\sum_i k_i, \theta)$. 

(2) If $X = \mathrm{Gamma}(k, \theta)$, then $c X = \mathrm{Gamma}(k, c\theta)$ for any $c > 0$.

By using these two properties of the Gamma distribution, we can assign two different sets of $(k, \theta)$ parameters for $\gamma_i$ and $\gamma_{i,j}$ in Eqs. (\ref{superMC.eq16}) and Eqs. (\ref{superMC.eq17}), respectively, to ensure that the final total entropy or energy density, which is a weighted sum of all the collisions in the event according to Eq. (\ref{superMC.eq10}) also fluctuates according a Gamma distribution with a desired shape and scale. For $WN(\vec{\bf r}_\perp)$, we write,
\begin{equation}
\gamma_i = \mathrm{Gamma} \left( k_{WN}, \theta_{WN} \right)
\end{equation}
and for $BC(\vec{\bf r}_\perp)$
\begin{equation}
\gamma_{i,j} = \mathrm{Gamma} \left(  k_{BC}, \theta_{BC} \right).
\end{equation}

Based on Eqs. (\ref{superMC.eq10}), (\ref{superMC.eq16}) and (\ref{superMC.eq17}), we then have
\begin{equation}
s = \kappa \left(\mathrm{Gamma}\left(\sum_{i=1}^{N_\mathrm{part}} k_{WN,i}, \frac{1 - \alpha}{2} \theta_{WN} \right)  + \mathrm{Gamma}\left( \sum_{i=1}^{N_\mathrm{coll}} k_{BC,i}, \alpha\theta_{BC} \right) \right).
\label{superMC.eq21}
\end{equation}
By requiring $\frac{1 - \alpha}{2} \theta_{WN} = \alpha\theta_{BC} = \theta$, Eq. (\ref{superMC.eq21}) can be further simplified  to,
\begin{equation}
\left\{ \begin{array}{c} s \\ e \end{array} \right\} = \left\{ \begin{array}{c} \kappa_s \\ \kappa_e \end{array} \right\} \left(\mathrm{Gamma}\left(N_\mathrm{part} k_{WN} + N_\mathrm{coll} k_{BC},   \theta \right)   \right).
\label{superMC.eq22}
\end{equation}
By further writing $k_{WN} = \frac{1 - \alpha}{2} k$ and $k_{BC} = \alpha k$, we finally obtain,
\begin{equation}
\left\{ \begin{array}{c} s \\ e \end{array} \right\} = \frac{1}{\tau_0} \left\{ \begin{array}{c} \kappa_s \\ \kappa_e \end{array} \right\} \left(\mathrm{Gamma}\left( \left( \frac{1 - \alpha}{2} N_\mathrm{part} + \alpha N_\mathrm{coll}\right) k,   \theta \right)   \right).
\label{superMC.eq23}
\end{equation}
Choosing the upper variant gives for the mean entropy density $\langle s \rangle = \frac{\kappa_s}{\tau_0} \left( \frac{1 - \alpha}{2} N_\mathrm{part} + \alpha N_\mathrm{coll}\right) k \theta$, and similarly for the mean energy density if the lower variant is chosen. Setting $k \theta = 1$ ensures that with the perviously adjusted normalizations $\kappa_s$ or $\kappa_e$ the event-averaged total entropy continues to reproduce the value from the conventional MC-Glauber model (and thus the observed final charged multiplicity).

\begin{figure}[h!]
  \centering
  \begin{tabular}{cc}
  \includegraphics[width=0.6\linewidth]{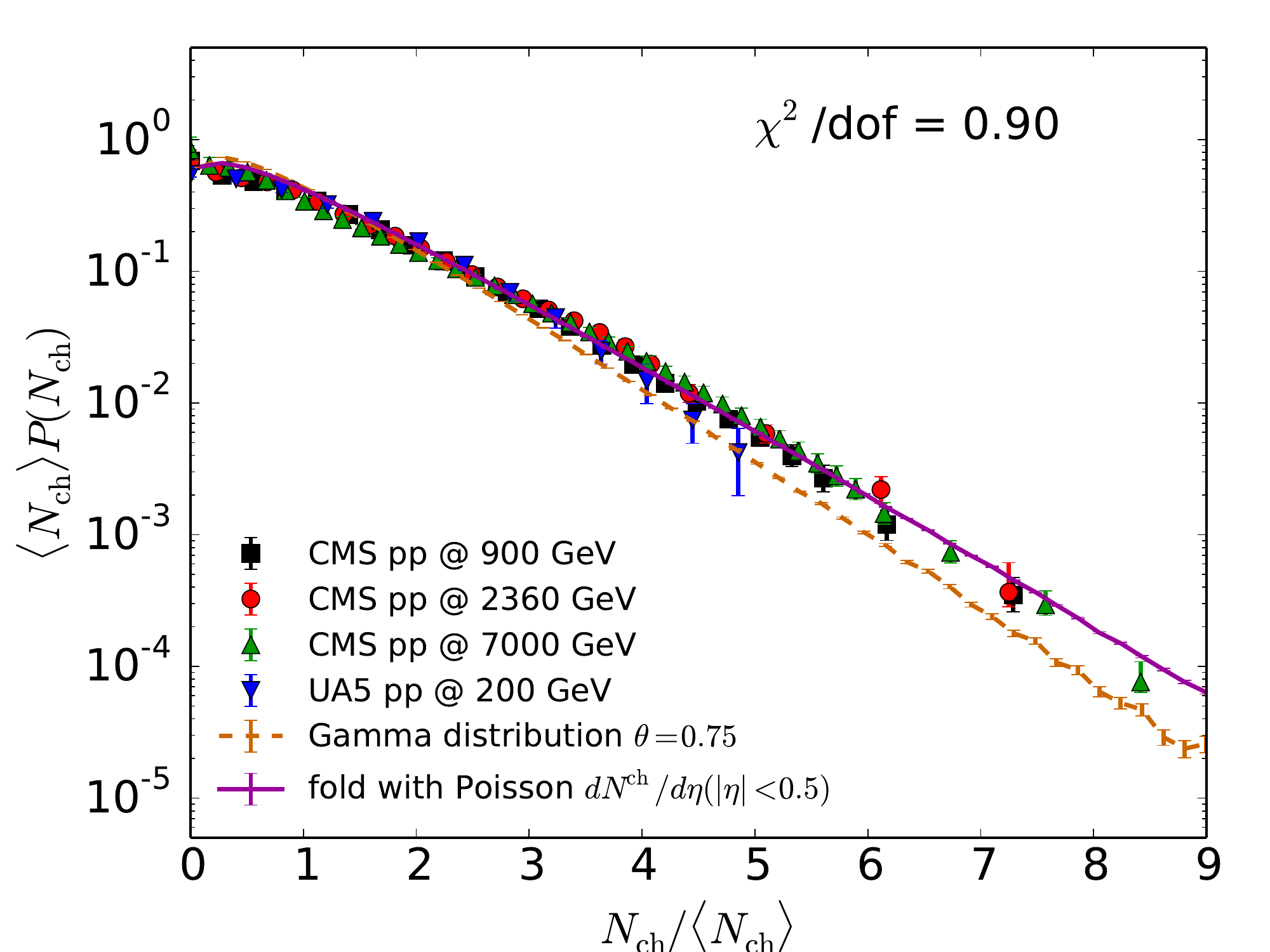}
  \end{tabular}
  \caption{Normalized charged hadron multiplicity distribution for minimum bias 
         pp collisions at $\sqrt{s} = 5.02$ TeV. The dashed line shows the result from a 
         Gamma distribution with $\theta=0.75$, the solid line the distribution obtained
         by folding this Gamma distribution with a Poisson distribution whose mean was
         adjusted to the measured mean charged hadron multiplicity at this collision
         energy. The solid line is compared with experimental data
         \cite{Khachatryan:2010nk,Ansorge:1988kn} showing the KNO 
         scaling of the pp multiplicity distribution in $\vert \eta \vert < 0.5$ for 
         $\sqrt{s}=200$, 900, 2360, and 7000 $A$ GeV.}
\label{ppMultfit}
\end{figure}

The actual value of $\theta$ with ($k = 1/\theta$) in Eq.~(\ref{superMC.eq23}) can be fit to the multiplicity distribution measured in pp collisions (in which $N_\mathrm{part}=2$
and $N_\mathrm{coll}=1$), after folding the initial-state fluctuations with an additional Poisson distribution describing the multiplicity fluctuations generated by the hadronization process.%
\footnote{%
     A Gamma distribution folded with a Poisson distribution results in 
     a negative binomial distribution.}  
At LHC energies, the multiplicity fluctuations in pp collision have been measured at $\sqrt{s} =$ 0.9, 2.36, and 7 TeV \cite{Khachatryan:2010nk}. Additionally, the UA5 Collaboration measured pp multiplicity distributions at $\sqrt{s} = 200$ GeV \cite{Ansorge:1988kn}. According to the KNO scaling hypothesis, $\langle N_\mathrm{ch} \rangle P(N_\mathrm{ch})$ should be a universal (energy independent) function of the normalized multiplicity $N_\mathrm{ch}/\langle N_\mathrm{ch} \rangle$ as shown in Fig. \ref{ppMultfit}. Because the mean $dN^\mathrm{ch}/d\eta$ in minimum bias pp collisions depends on $\sqrt{s}$, the variance of the Poisson distribution differs from one collision energy to another. Thus, the $\theta$ parameter in the Gamma distribution also depends on $\sqrt{s}$. In Table~\ref{superMC.Gammadistribution.theta}, we list the appropriate choice of the $\theta$ parameter at several collision energies. Our minimum-$\chi^2$-fit at $\sqrt{s} = 5.02$ TeV is shown in Fig.~\ref{ppMultfit}.

%=======================================
%=======================================

\begin{table}[h!]
\centering
\begin{tabular}{c|c|c|c}
\hline \hline
collision energy & $dN^\mathrm{ch}/d\eta \vert_{\vert \eta \vert < 0.5}$ & $\theta$  &  $\chi^2/\mathrm{d.o.f}$ \\ \hline
pp @ 200 GeV & 2.47 & 0.61  & 2.02 \\
pp @ 2760 GeV & 4.54 & 0.73 & 0.96 \\ 
pp @ 5020 GeV & 5.28 & 0.75  & 0.90\\ \hline \hline
\end{tabular}
\caption{The choice of the $\theta$ parameter in the Gamma-distribution at several collision energies.}
\label{superMC.Gammadistribution.theta}
\end{table}

%=======================================
\begin{figure}[h!]
  \centering
  \begin{tabular}{cc}
  \includegraphics[width=0.47\linewidth]{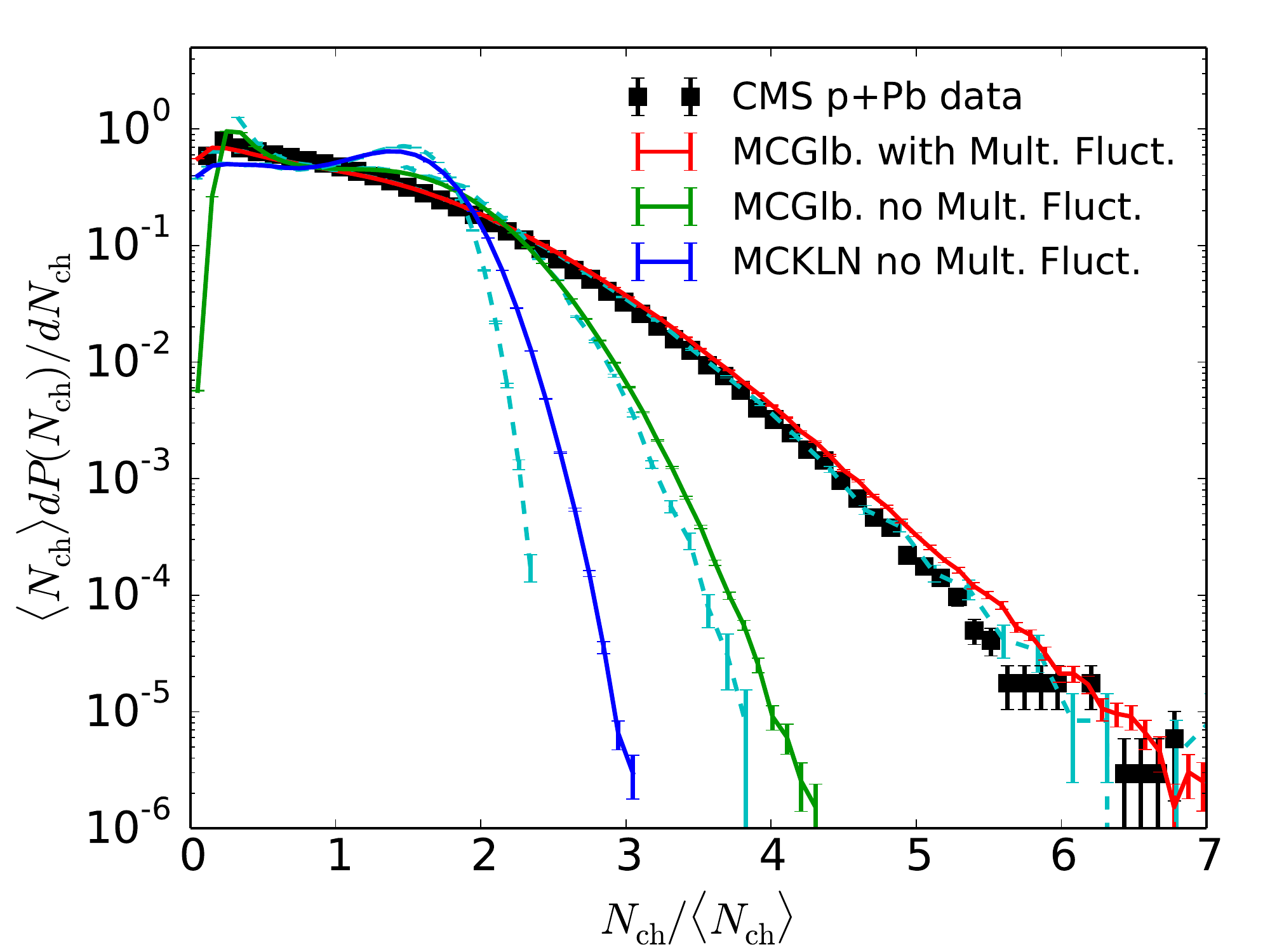} &
  \includegraphics[width=0.47\linewidth]{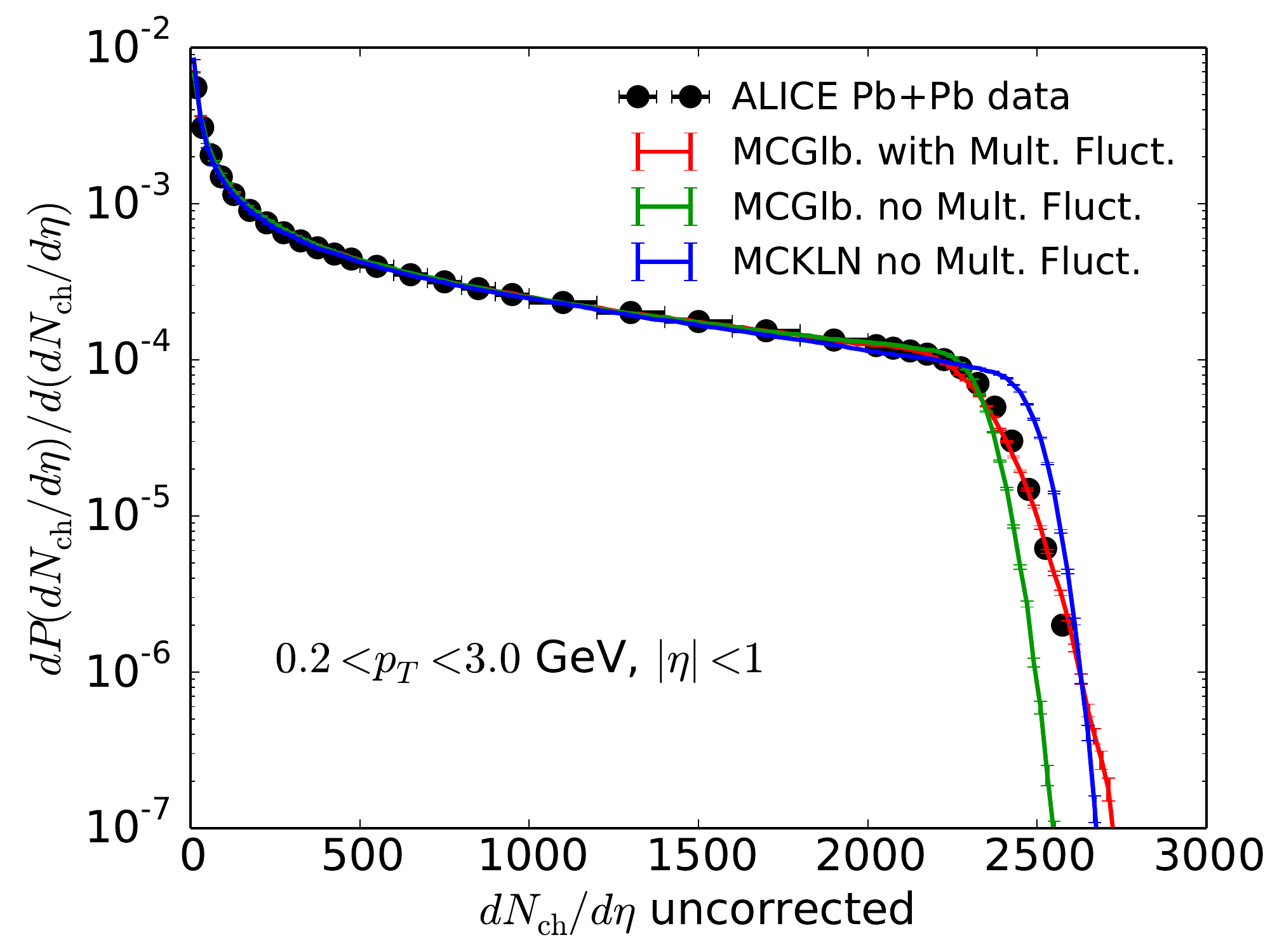}
  \end{tabular}
  \caption{{\it Left panel:} Normalized multiplicity distributions for the MC-Glauber and the MCKLN models in p+Pb collisions at $\sqrt{s_\mathrm{NN}} = 5.02$\,TeV compared with the CMS measurements \cite{Chatrchyan:2013nka}. The normalized distribution of the initial entropy density, $dS/dy$ are shown as the dashed cyan lines. {\it Right panel:} Comparisons of charged hadron multiplicity distribution in Pb+Pb collisions at $\sqrt{s_\mathrm{NN}} = 2.76$\,TeV with ALICE measurement \cite{Abelev:2014mda}. }
  \label{pPbMult_pred}
\end{figure}
%=======================================

Once the parameters of the Gamma distribution are fixed by the phenomenological KNO scaling, we use this model to make a parameter-free postdiction for the multiplicity distributions in p+Pb collisions at $\sqrt{s_\mathrm{NN}} = 5.02$\,TeV. In the MC-Glauber model, collision-by-collision multiplicity fluctuations significantly increase the probability for upward fluctuations in the multiplicity for p+Pb collisions. The MC-KLN model (which does not account for pp multiplicity fluctuations) produces the narrowest distribution for the initial total entropy at mid-rapidity. In order to compare with the p+Pb multiplicity distribution measured by CMS we first convert the initial total entropy to final charged hadron multiplicity, assuming they are proportional to each other. Choosing the same kinematic cuts as used in the CMS measurement, $p_T > 0.4$\,GeV and $\vert \eta \vert < 2.4$ \cite{Chatrchyan:2013nka}, we map $dN^\mathrm{ch}/d\eta \simeq \frac{4.8 \times 0.75}{8.9}\, dS/dy\vert_{y=0}$. The mean charged hadron multiplicity in minimum bias p+Pb collisions obtained by this mapping lies within the measured value $50 \pm 2$ \cite{Chatrchyan:2013nka}. We then fold the distribution of $dN^\mathrm{ch}/d\eta \vert_{\{p_T > 0.4\,\mathrm{GeV},\,\,\vert \eta \vert < 2.4\}}$ values calculated from the initial entropy density distribution with a Poisson distribution of multiplicity fluctuations produced at hadronization (here taken to be controlled by the mean multiplicity at kinetic freeze-out). By oversampling 20 Poisson distributions from each event, we obtained 20 million samples for each set of initial conditions. Their normalized distributions are compared with the CMS measurement in left panel of Fig.~\ref{pPbMult_pred}. The MC-Glauber model with collision-by-collision multiplicity fluctuations can reproduce the shape of the measured p+Pb multiplicity distributions \cite{Bozek:2013uha}. By comparing with the distribution of initial total entropy values we find that the broadening of the multiplicity distribution due to final state Poisson fluctuations becomes less important as $dN^\mathrm{ch}/d\eta$ increases from pp collisions to pPb collisions. This is because the normalized variance of a Poisson distribution with mean $\lambda$ decreases as $1/\sqrt{\lambda}$. In the right panel of Fig.~\ref{pPbMult_pred}, we further compute the charged hadron multiplicity distribution in Pb+Pb collisions at $\sqrt{s_\mathrm{NN}} = 2.76$\,TeV. The MC-Glauber model with collision-by-collision fluctuations can reproduce the ALICE measurements \cite{Abelev:2014mda} very well.

%=======================================
\begin{figure}[h!]
  \centering
  \begin{tabular}{cc}
  \includegraphics[width=0.48\linewidth]{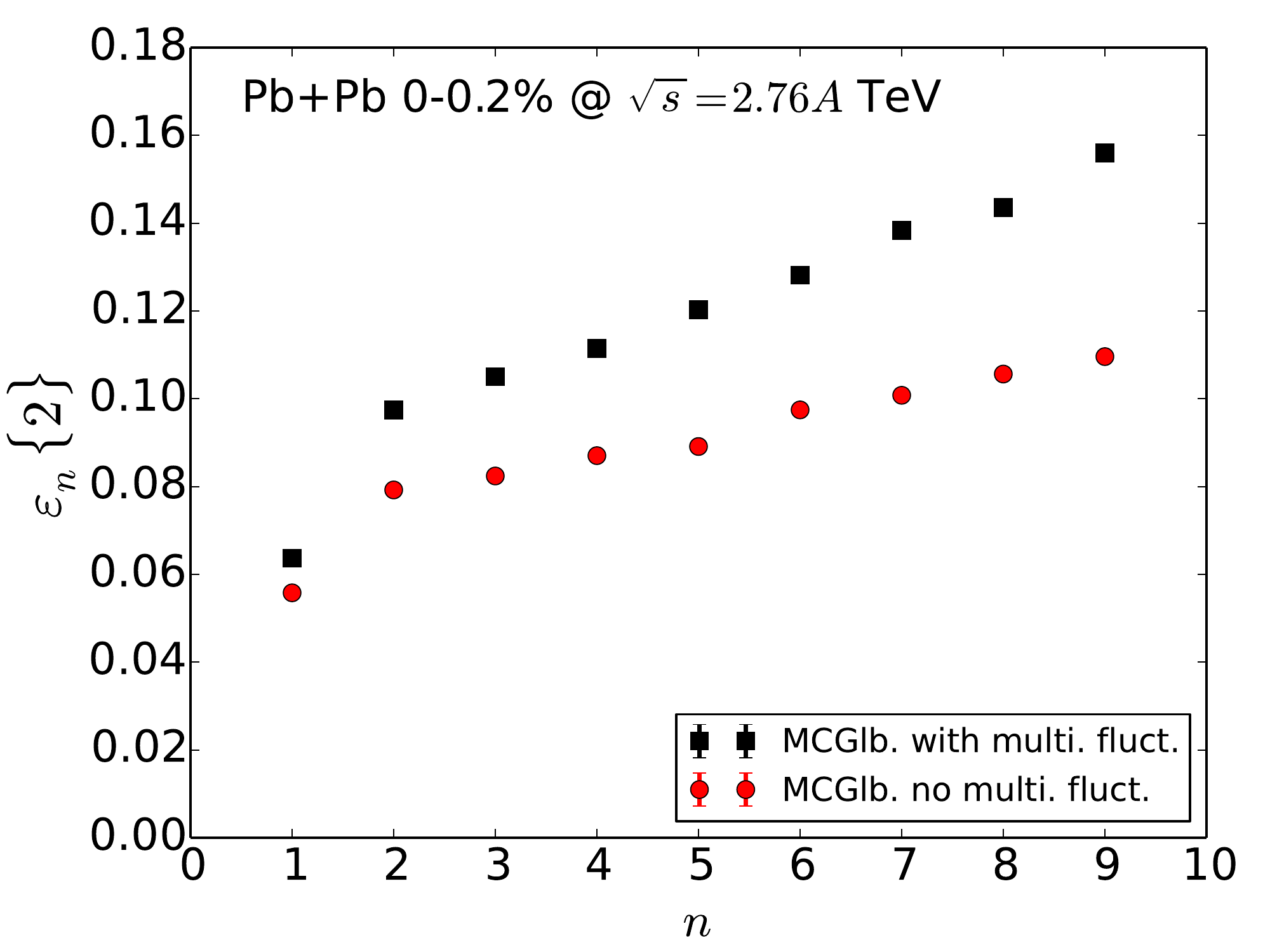} & 
  \includegraphics[width=0.48\linewidth]{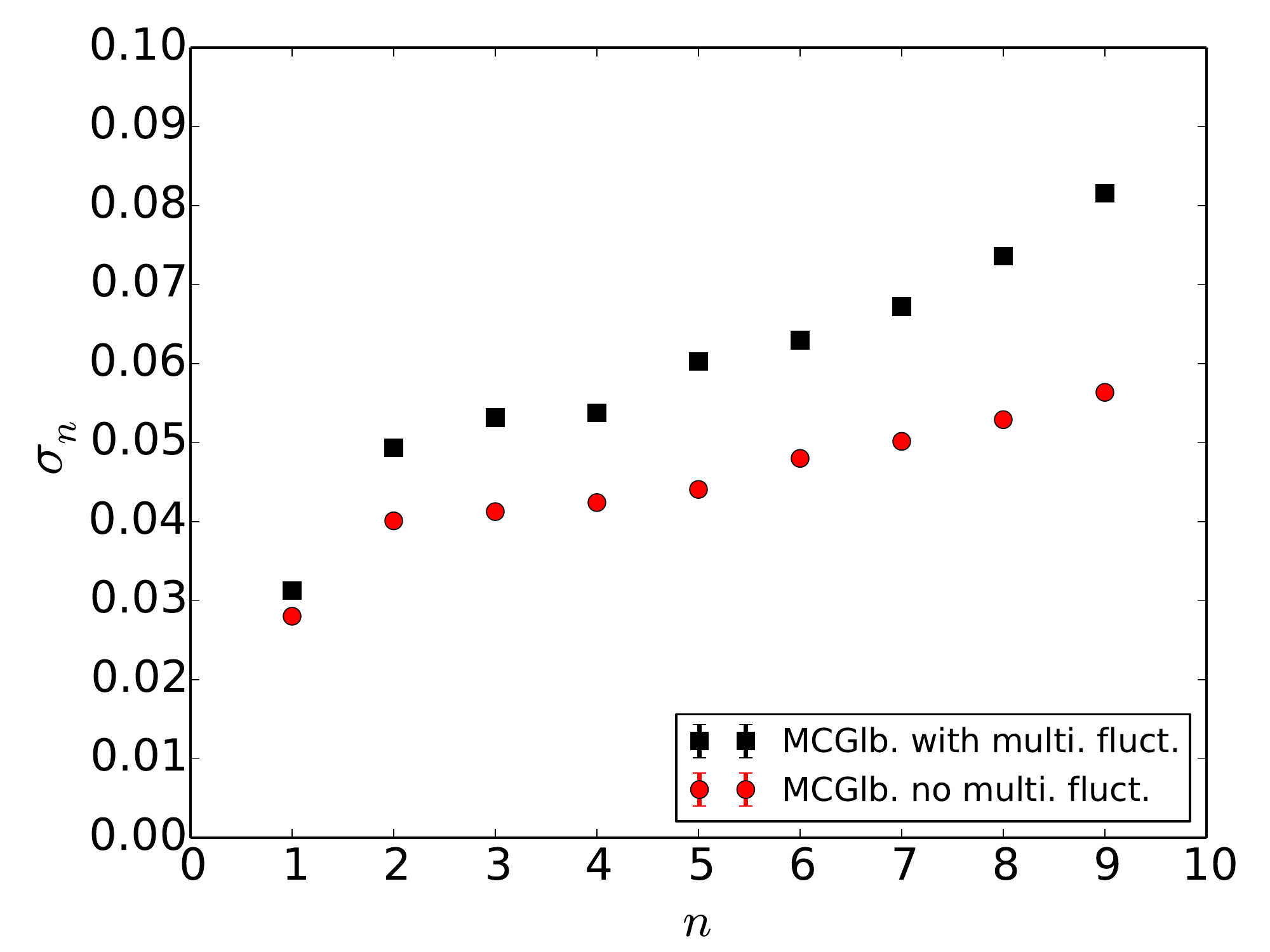} 
  \end{tabular}
  \caption{Left Panel: Root mean square of the $n$-th order initial spatial eccentricity as a function of the harmonic order $n$. Right Panel: The variance of $\varepsilon_n$ as a function of $n$. A repulsive hard core with minimum inter-nucleon distance $r_\mathrm{min} = 0.9$ fm is imposed when sampling nucleon spatial configuration inside the nucleus.}
  \label{eccn_vs_n}
\end{figure}
%=======================================

In Fig. \ref{eccn_vs_n}, we show a comparison of the initial spatial eccentricity $\varepsilon_n\{2\}$ as a function of the harmonic order $n$ for 0-0.2\% ultra-central Pb + Pb collisions at LHC energy. $\varepsilon_n\{2\} = \sqrt{\langle \varepsilon_n^2 \rangle}$ is the rms of $\varepsilon_n$, defined in terms of the fluctuating initial energy density profile $e (r_\perp, \phi)$ as 
\begin{equation}
\varepsilon_1 e^{i\Phi_1} = -\frac{\int d^2 {\bf r}_\perp r_\perp^3 e(r_\perp, \phi) e^{i\,\phi}}{\int d^2 {\bf r}_\perp r_\perp^3 e(r_\perp, \phi)}
\label{superMC.eccdef}
\end{equation}
and
\begin{equation}
\varepsilon_n e^{i n \Phi_n} = -\frac{\int d^2 {\bf r}_\perp r_\perp^n e(r_\perp, \phi) e^{i\,n\,\phi}}{\int d^2 {\bf r}_\perp r_\perp^n e(r_\perp, \phi)}, \quad\quad\quad\mathrm{for }\quad n \ge 2.
\label{superMC.eccdef}
\end{equation}
We can clearly see that the collision-by-collision multiplicity fluctuations increase the eccentricity coefficients for all harmonic orders by 20-40\%. The increase is larger for higher order $n$. The multiplicity fluctuations also increase the variance of $\varepsilon_n$ similar amount. The existence of such fluctuations therefore changes the mean values and their variances of the initial fluctuation spectrum of the MC-Glauber model dramatically.  

%=======================================
\begin{figure*}[h!]
  \centering
  \begin{tabular}{cc}
  \includegraphics[width=0.45\linewidth]{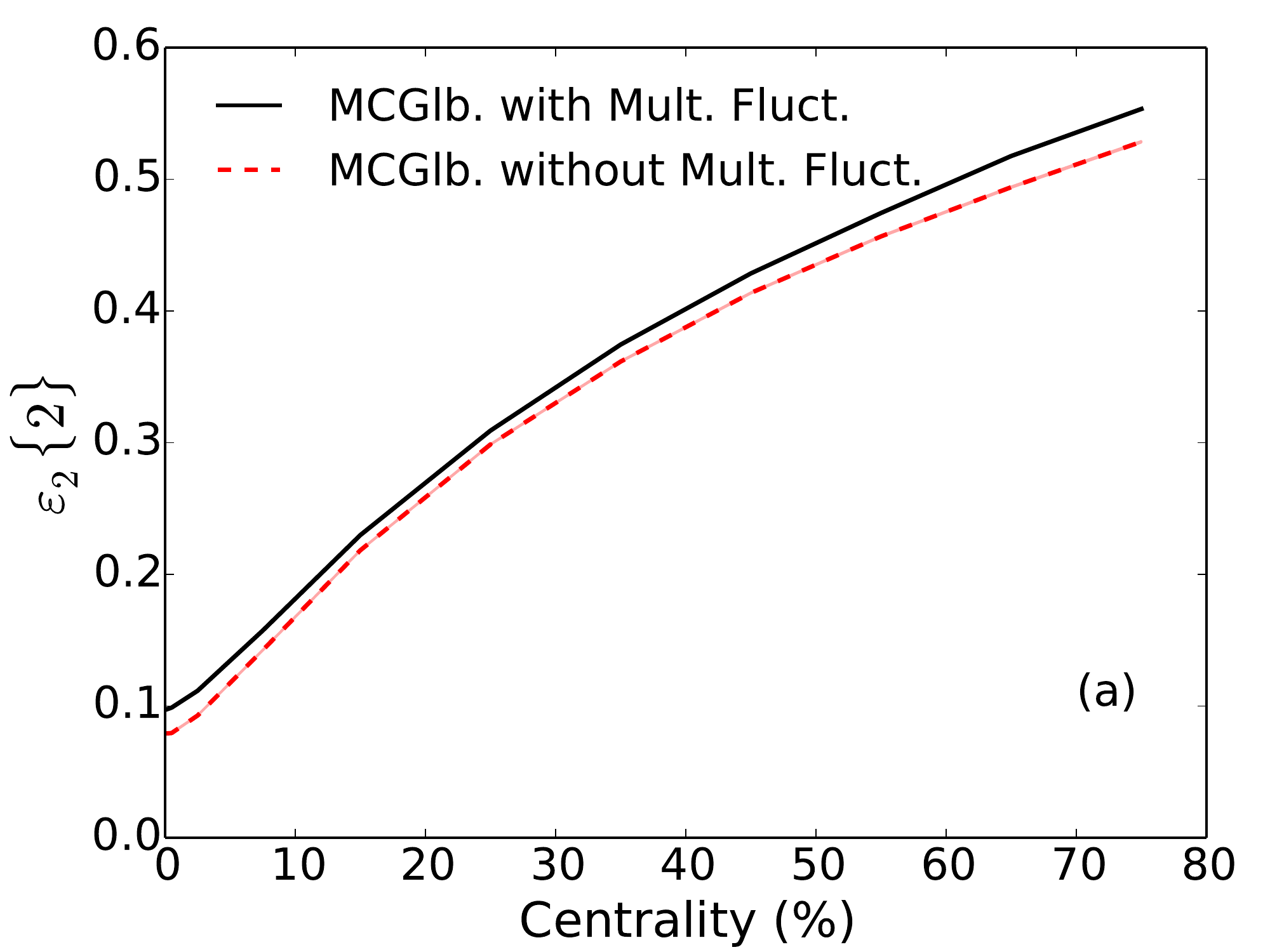} &
  \includegraphics[width=0.45\linewidth]{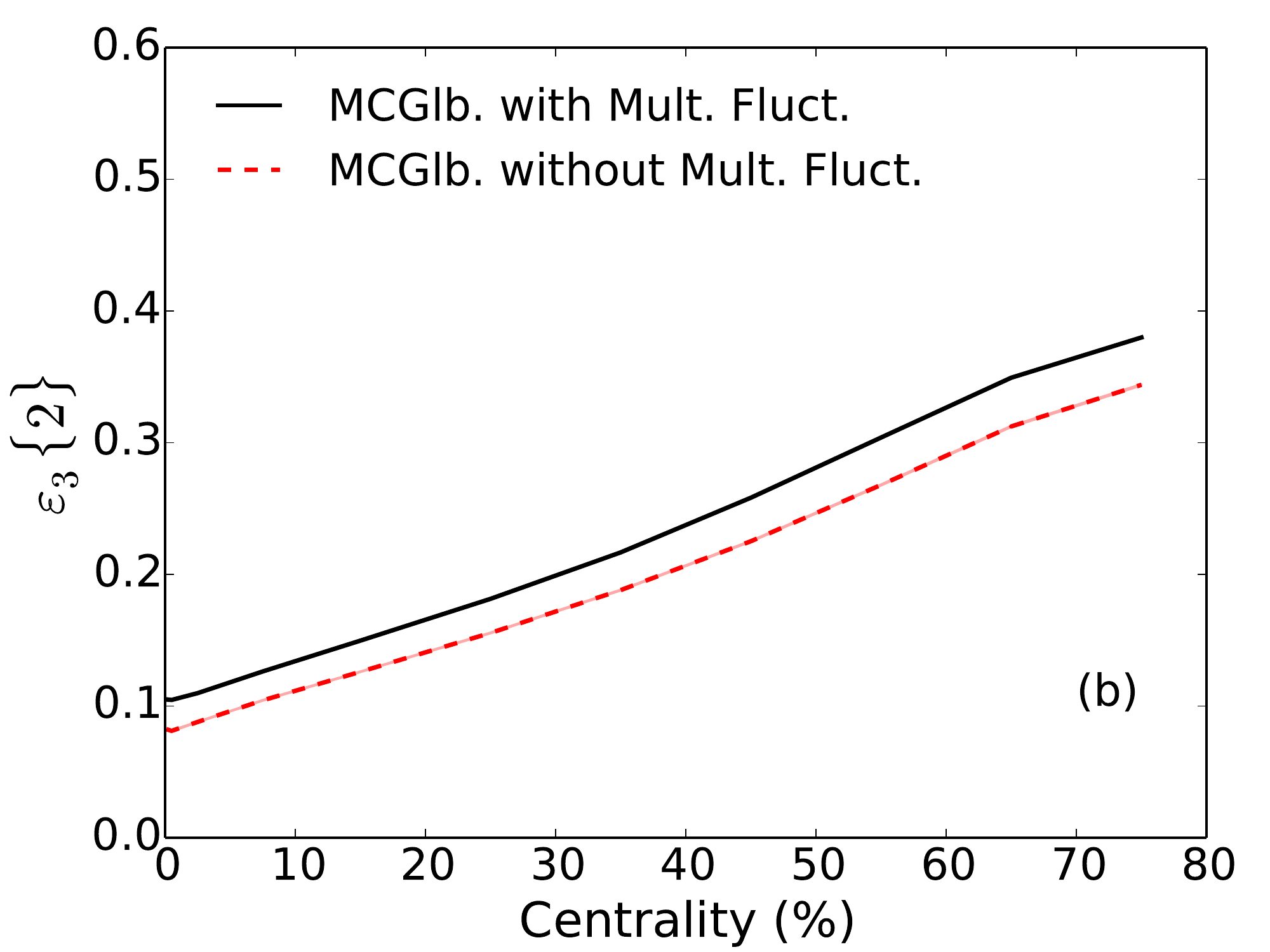} \\
  \includegraphics[width=0.45\linewidth]{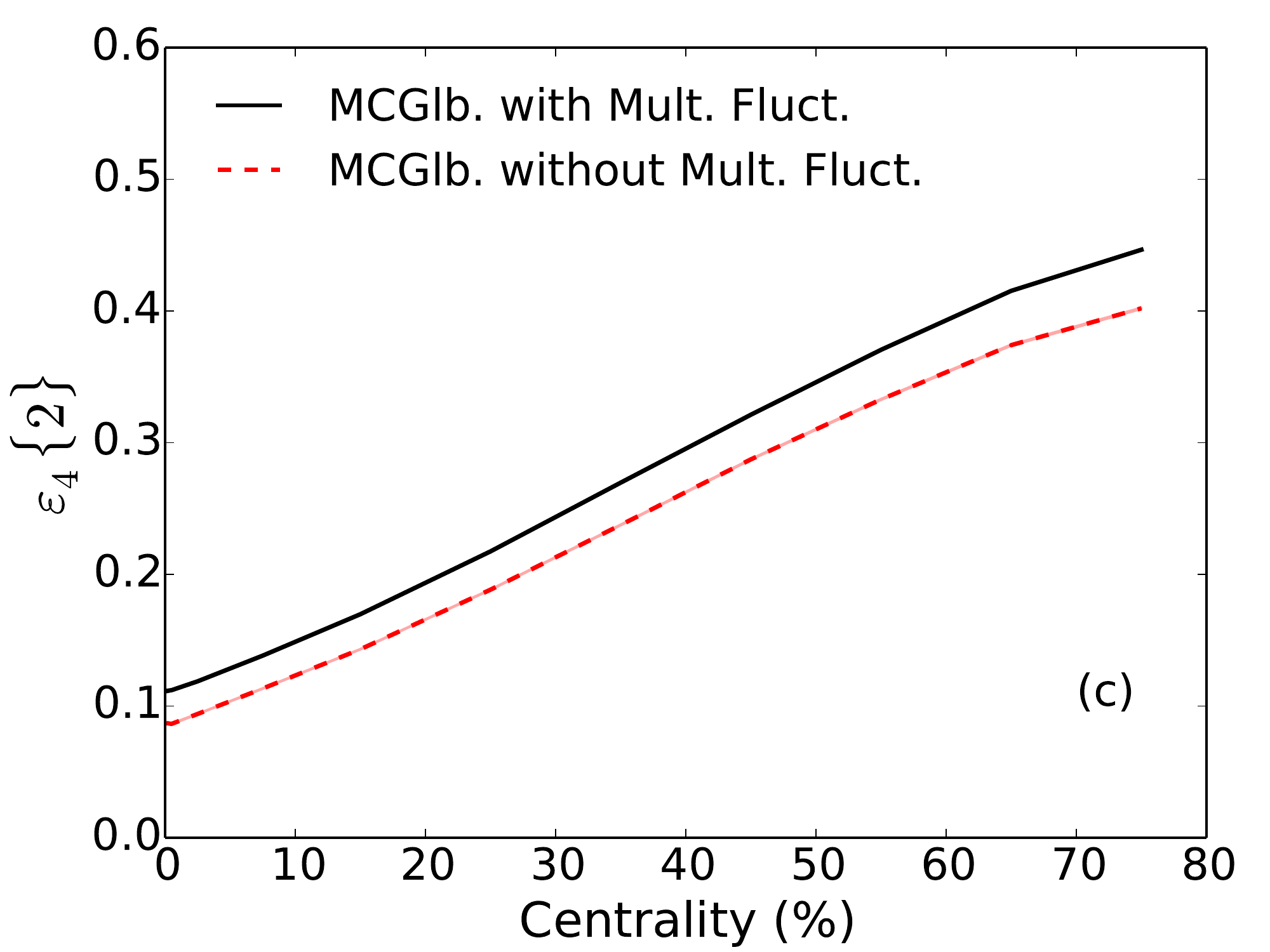} &
  \includegraphics[width=0.45\linewidth]{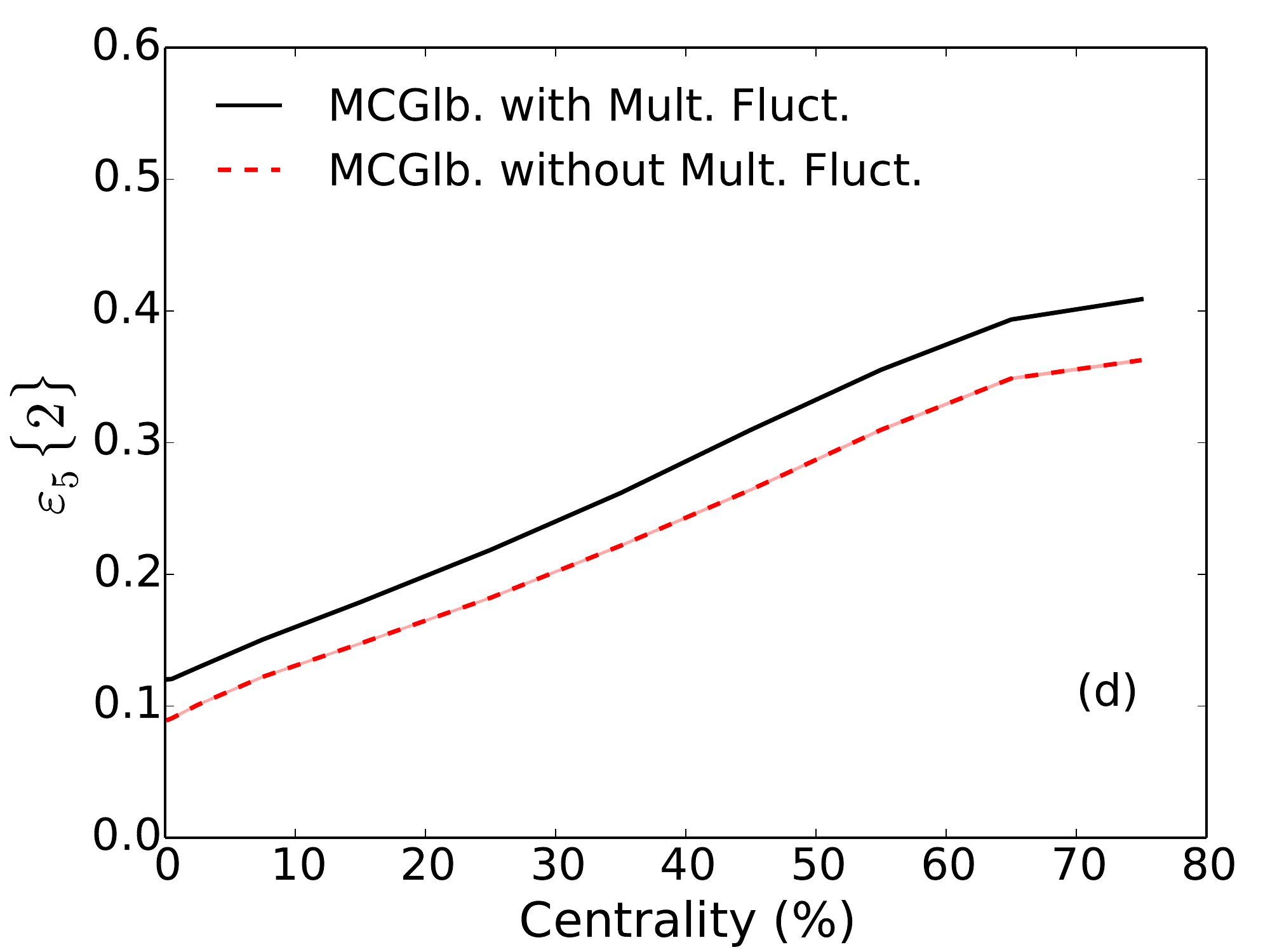}
  \end{tabular}
  \caption{Centrality dependence of the root mean square initial spatial eccentricities $\varepsilon_{2,3,4,5}\{2\}$.}
  \label{eccn_vs_centrality}
\end{figure*}
%=======================================

In Fig. \ref{eccn_vs_centrality}, we show $\varepsilon_2$ to  $\varepsilon_5$ as functions of the collision centrality. We find that collision-by-collision multiplicity fluctuations are not only important in ultra-central collisions, but that they increase the spatial eccentricities at all collision centralities. 

\subsection{Centrality cuts in theoretical calculations}
\label{chap2.centralitycut}

Centrality is a key quantity that links the theoretical calculations with the experimental measurements. It is introduced to characterize the collision geometry in nucleus-nucleus collisions. Experimentally, the centrality is typically defined by sorting the recorded events according to their measured charged hadron multiplicity at mid-rapidity, $dN^\mathrm{ch}/d\eta \vert_{\vert \eta \vert < 0.5}$. Applying the same procedure theoretically is computationally expensive since, due to viscous heating, the final charged hadron multiplicity can not be determined directly from the initially produced entropy, but requires the calculation of the full viscous hydrodynamic evolution, event by event. 

\begin{figure}[h!]
  \centering
  \begin{tabular}{cc}
  \includegraphics[width=0.48\linewidth]{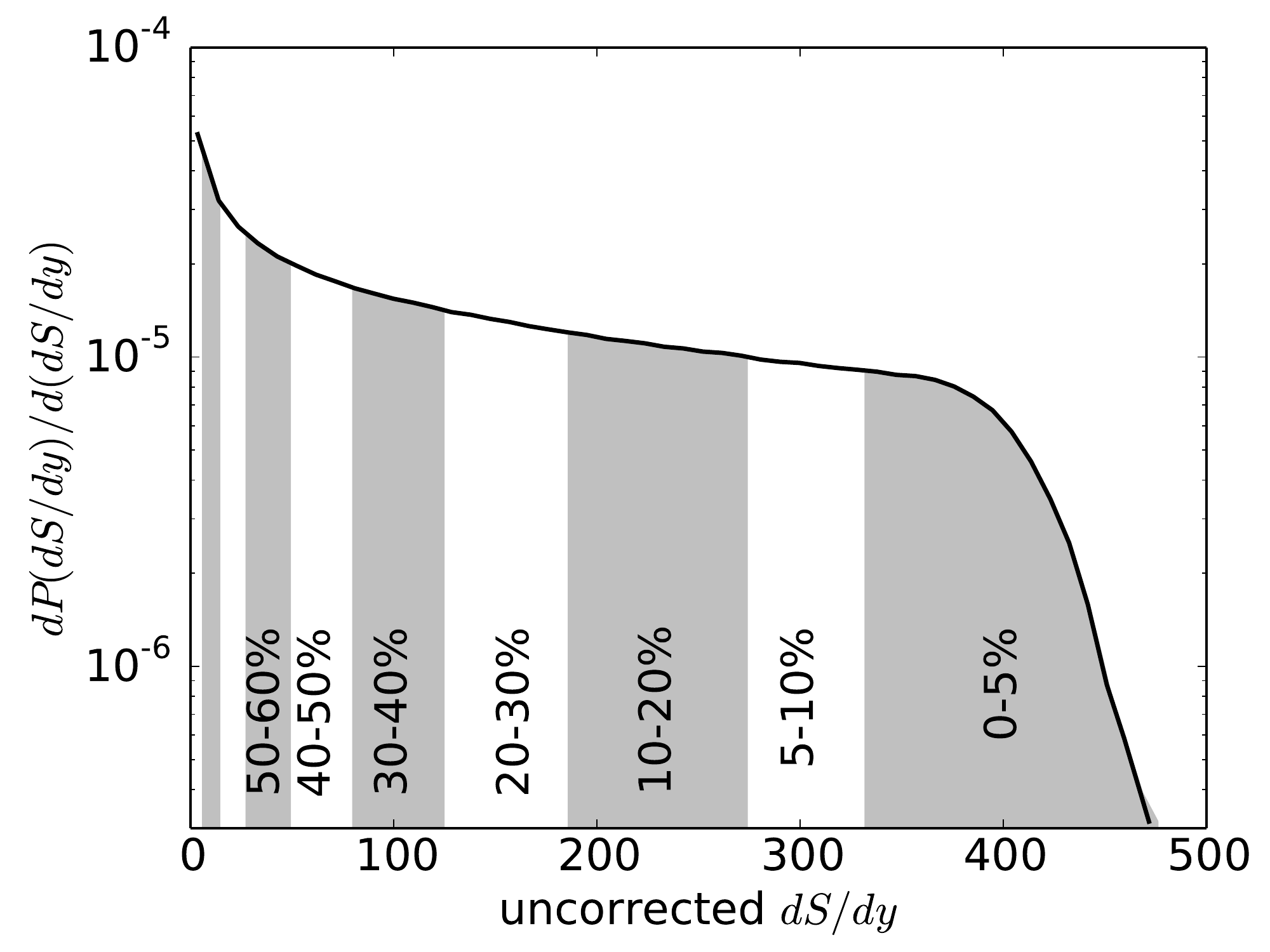} &
  \includegraphics[width=0.48\linewidth]{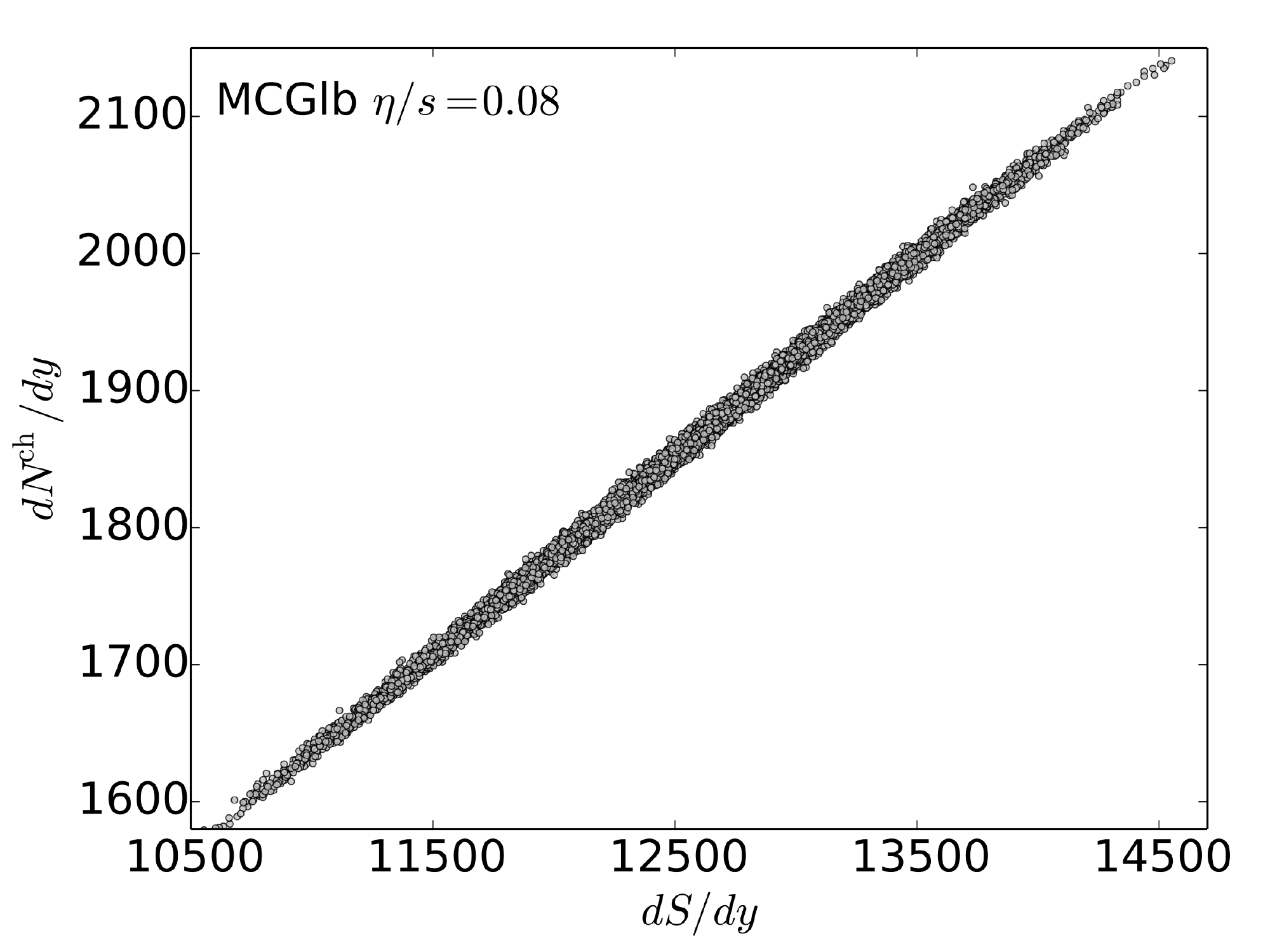}
  \end{tabular}
  \caption{Left Panel: Probability distribution of the total entropy density $dS/dy \vert_{y=0}$ from MC-Glauber model for Pb + Pb at $\sqrt{s} = 2.76 A$ TeV. Right Panel: Correlation between initial $dS/dy$ and final measured $dN^\mathrm{ch}/dy$ at 0-5\% most central collisions for MC-Glauber with $\eta/s = 0.08$.  }
  \label{centralityCut_dSdy}
\end{figure}

However, we can use the following approximation to save simulation time: we select centrality on the initially produced total entropy in the transverse plane, $dS/dy\vert_{y=0}$, assuming that, on average, the final charged hadron multiplicity, $dN^\mathrm{ch}/d\eta$, is monotonically related to $dS/dy\vert_{y=0}$. This procedure is illustrated in Fig. \ref{centralityCut_dSdy}. 

%=======================================
%=======================================

After having fixed the normalization constant $\kappa$ in Eq. (\ref{superMC.eq10}) such that, on average, the measured charged multiplicity in central collisions is correctly reproduced, we first sort the minimum bias events generated by the MC-Glauber model according to the initial entropy $dS/dy$. Then, we can classify their collision centrality through their relative positions in the sorted array. The events with largest total entropy define the most central collisions where the two nuclei completely overlap each other. The 0-10\% centrality bin includes the 10\% of all events with the largest initial $dS/dy$, 90-100\% centrality the 10\% of all events with the smallest $dS/dy$ values.

This procedure ignores event-by-event fluctuations in the fractional increase of the entropy due to viscous heating during the hydrodynamic evolution. This extra entropy production depends on the actual shape of the initial density profile as well as on the chosen value for the specific shear viscosity, $\eta/s$. Event-by-event fluctuation of the viscous entropy production will de-correlate the one-to-one correspondence between the initial total entropy, $dS/dy$ and the final measured charged hadron multiplicity, $dN^\mathrm{ch}/dy$. However, as shown in the right panel of the Fig. \ref{centralityCut_dSdy}, this decorrelation is weak: For given $dS/dy$, the spread in final $dN_\mathrm{ch}/dy$ is very small ($2-3$\%). Therefore, our procedure of defining collision centrality by cutting on the initial total $dS/dy$ is a pretty good theoretical approximation to the experimental centrality definition using final state charged hadron multiplicities.

%%%%%%%%%%%%%%%%%%%%%%%%%%%%%%%%%%%%%%%%%%%%%
\section{(2+1)-d viscous hydrodynamics {\tt VISHNew}}
\label{sec4}
%%%%%%%%%%%%%%%%%%%%%%%%%%%%%%%%%%%%%%%%%%%%%
\subsection{Solving the hydrodynamic equations}
The module {\tt VISHNew} is an improved version of {\tt VISH2+1}, the (2+1)-d longitudinally boost-invariant viscous hydrodynamic algorithm developed by H. Song \cite{Song:2007ux, Song:2008si, Heinz:2008qm}. It includes several improvements for efficiency and stability which will be discussed in this section. We solve the following equation of motion for second order viscous hydrodynamics (``Israel-Stewart equations''):

\begin{equation}
d_\mu T^{\mu\nu} = 0, \mbox{   } T^{\mu\nu} = e u^\mu u^\nu - (p + \Pi) \Delta^{\mu\nu} + \pi^{\mu\nu}.
\label{VISH.eq1}
\end{equation}
The shear stress tensor $\pi^{\mu\nu}$ and bulk pressure $\Pi$ satisfy the following transport equations,
\begin{eqnarray}
\Delta^{\mu\alpha} \Delta^{\nu \beta} D \pi_{\alpha \beta} &=& -\frac{1}{\tau_\pi} (\pi^{\mu\nu} - 2 \eta \sigma^{\mu\nu}) - \frac{1}{2} \pi^{\mu\nu} \frac{\eta T}{\tau_\pi} d_\lambda \left(\frac{\tau_\pi}{\eta T} u^\lambda \right), \label{VISH.eq2}\\
D\Pi &=& - \frac{1}{\tau_\Pi} (\Pi + \zeta \theta) - \frac{1}{2} \Pi \frac{\zeta T}{\tau_\Pi} d_\lambda \left(\frac{\tau_\Pi}{\zeta T} u^\lambda \right),
\label{VISH.eq3}
\end{eqnarray}
where $D = u^\mu d_\mu$. The hydrodynamic equations need to be solved together with a given equation of state (EOS). {\tt VISHNew} supports three versions of the lattice-based equation of state, {\tt s95p-v0-PCE}, {\tt s95p-v1}, and {\tt s95p-v1-PCE} \cite{Huovinen:2009yb}. The differences among these three EOS are different implementations of partial chemical equilibrium in the hadronic phase \cite{Huovinen:2009yb,Shen:2010uy}. In general, Eq. (\ref{VISH.eq1}) must be supplemented by an evolution equation (conservation law) for the baryon current $j^\mu = n u^\mu$. We start with the case $n = 0$.

\subsubsection{Without baryon current}
The hydrodynamic code evolves the components of energy stress tensor. In  order to use the EOS for determining the pressure in the liquid, we first need to
solve for the local energy density and velocity of the fluid cell. In the (2+1)-d case, we define a
vector $M^{\mu}  =  ( M^{0} ,M^{x},M^{y} ) = ( T^{\tau \tau} -
\pi^{\tau \tau},T^{\tau x} - \pi^{\tau x} ,T^{\tau y} - \pi^{\tau y}
)$. Using the decomposition Eq. (\ref{VISH.eq1}) for $T^{\mu\nu}$, we find
\begin{equation}
  M^{0} =  ( e+P+ \Pi ) ( u^{0} )^{2} -P- \Pi,
\end{equation}
\begin{equation}
  M^{1} = ( e+P+ \Pi ) u^{0} u^{1},
\end{equation}
\begin{equation}
  M^{2} = ( e+P+ \Pi ) u^{0} u^{2},
\end{equation}
The local energy density thus satisfies the following equation:
\begin{equation}
  e=M^{0} - \frac{( M^{1} )^{2} + ( M^{2} )^{2}}{M^{0} +P+ \Pi}.
  \label{VISH.energyEq}
\end{equation}
To solve Eq. (\ref{VISH.energyEq}) we define
\begin{equation}
  f ( e )  =  ( M^{0} -e ) ( M^{0} +P+ \Pi ) - ( ( M^{1} )^{2} + ( M^{2} )^{2}).
\label{VISH.quardraticf_e}
\end{equation}
We first observe that $f ( M^{0} ) =- ( ( M^{1} )^{2} + ( M^{2}
)^{2} ) \leq 0$. In order for Eq. (\ref{VISH.energyEq}) to have an odd number of positive
solutions, we need to require $f ( 0 )   \geqslant  0$. With a non-zero bulk viscous pressure, this leads to the condition,
\begin{equation}
  f ( 0 ) = ( M^{0} )^{2} - ( M^{1} )^{2} - ( M^{2} )^{2} +M^{0} \Pi \geqslant
  0.
\end{equation}
When this requirement is not fulfilled because $\Pi$ (which is negative) is too large, we regulate $\Pi$ such that $f ( 0 )$ = 0. In this special situation one can further compute
\begin{equation}
  \frac{d f}{d e}_{} ( e=0 ) = ( c^{2}_{s} -1 ) M^{0} - \Pi.
\end{equation}
If $\frac{d f}{d e}\vert_{e=0} \le 0$, $e = 0$ is the solution. For $\frac{d f}{d e}\vert_{e=0} >0
$, there will be a positive energy density solution. Without $\Pi$,
$\frac{d f}{d e}\vert_{e=0}$ is always less than 0  because the square of the speed of sound is always smaller
than 1.

Once these two boundary conditions are set up, it is guaranteed that there will
be at least one solution of Eq.
(\ref{VISH.energyEq}) with positive energy density. Newton's root finding method is a very efficient in finding this solution with a minimal number of iterations. To ensure numerical stability and optimal efficiency, we use
the fact that to fairly good approximation the pressure is roughly proportional to the energy density. We rewrite Eq. (\ref{VISH.quardraticf_e}) as,
\[ f ( e )  =  ( M^{0} -e ) \left( M^{0} + \frac{P}{e}  e+ \Pi \right) - ( (
   M^{1} )^{2} + ( M^{2} )^{2} )  \]
and use that $\tilde{c}_{s}^{2} = \frac{P}{e}$ has a very weak dependence
on $e$. This turns the condition $f ( e ) =0$ into approximately a quadratic equation with solution
\begin{equation}
 e= \frac{- ( M^{0} ( 1- \tilde{c}_{s}^{2} ) + \Pi ) \pm \sqrt{( M^{0} ( 1-
   \tilde{c}_{s}^{2} ) + \Pi )^{2} +4 \tilde{c}_{s}^{2} ( M^{0} ( M^{0} + \Pi
   ) -M )}}{2 \tilde{c}_{s}^{2}} .
\end{equation}
To identify the correct sign, we note that for $M=0$ we must recover $e = M^{0}$. Therefore,
\begin{equation}
e= \frac{- ( M^{0} ( 1- \tilde{c}_{s}^{2} ) + \Pi ) + \sqrt{( M^{0} ( 1-
   \tilde{c}_{s}^{2} ) + \Pi )^{2} +4 \tilde{c}_{s}^{2} ( M^{0} ( M^{0} + \Pi
   ) -M )}}{2 \tilde{c}_{s}^{2}} . 
\end{equation}
This equation is the most efficient satisfying form for applying Newton's method, and it
is  implemented in {\tt{VISHNew}}.

Once Eq. (\ref{VISH.eq1}) has been solved for $e$, the flow velocity can be calculated from
\begin{equation}
  u^{0} = \left( \frac{M^{0} +P+ \Pi}{e+P+ \Pi} \right)^{1/2}
  \label{VISH.eq37}
\end{equation}
where $P = P(e)$ is obtained from the EOS. 
Please note that calculating $u^{0}$ instead of $v$ is numerically more stable
when $v \rightarrow 1$. Since $u^{0}   \geqslant 1$, this requires $M^{0}  
\geqslant e$. So $M^{0}$ should be set as an upper limit for $e$ when intreating $e$ using
Newton's root finding routine. Similarly,
\begin{equation}
  u^{i} = \frac{M^{i}}{\sqrt{M^{0} +P+ \Pi} \sqrt{e+P+ \Pi}} \mbox{   } (i = 1,2).
   \label{VISH.eq38}
\end{equation}
One can check that if $e$ is the exact solution of Eq. (\ref{VISH.energyEq}), the
flow velocity components Eq. (\ref{VISH.eq37}) and Eq. (\ref{VISH.eq38}) satisfy the normalization constraint 
\begin{equation}
  ( u^{0} )^{2} - ( u^{1} )^{2} - ( u^{2} )^{2} =1.
\end{equation}

\subsubsection{With baryon density current}

The derivations above assumed zero net baryon density where the pressure is
only a function of the local energy density. In order to deal with the  more general
cases of non-zero conserved charge current in the future,  we now consider the situation where the baryon current
is not zero.

In this case the pressure is a function of both the local energy density
and the local net baryon density: $P = P ( e,n )$. For the baryon current, we have the 
additional hydrodynamic equation
\begin{equation}
  \partial_{\mu} j^{\mu} =0
\end{equation}
where ($V^\mu$ is the heat flow vector)
\begin{equation}
  j^{\mu} =n u^{\mu} + V^\mu.
\end{equation}
Now, the problem of implementing the EOS presents itself as follows: knowing $j^{0}$, $T^{00}$, $T^{01}$, $T^{02}$, $T^{03}$ and the EOS, we would like to solve for 5 unknowns $n,e,u^{\mu}$. We have the following 5
equations:
\begin{equation}
  M^{0} =  ( e+P+ \Pi ) ( u^{0} )^{2} -P- \Pi,
\end{equation}
\begin{equation}
  M^{i} = ( e+P+ \Pi ) u^{0} u^{i}, \mbox{   } (i = 1,2,3),
\end{equation}
\begin{equation}
  j^{0} =n u^{0} + V^0.
\end{equation}
We can no longer solve for $e$ easily, because the pressure now depends on both $e$ and
$n$. The equations for $e$ and $n$ are coupled with each other. To decouple these two equations, we need to
know the actual functional dependence for $P ( e,n )$. In such a situation, it is easilier to solve for $v$ or $u^{0}$
first. For $v$, we have solve the following equation:
\begin{equation}
  v =  \frac{M}{M^{0} +P+ \Pi} 
\label{VISH.eq_v}
\end{equation}
where $M= \sqrt{( M^{1} )^{2} + ( M^{2} )^{2} + ( M^{3} )^{2}}$. For the pressure from the EOS, we need to work out
\begin{equation}
  e = M^{0} -v M 
  \label{VISH.e_v}
\end{equation}
\begin{equation}
  n = (j^{0} - V^0)\sqrt{1-v^{2}}
 \label{VISH.n_v}
\end{equation}
To solve Eq. (\ref{VISH.eq_v}) we define
\begin{equation}
  f ( v )  = v ( M^{0} +P+ \Pi )  - M.
\end{equation}
We have the boundary conditions
\begin{equation}
  f ( 0 ) =-M \leqslant 0
\end{equation}
and
\begin{equation}
  f ( 1 )  = M^{0} +P+ \Pi -M.
\end{equation}
Imposing $f ( 1 ) \geqslant 0$ will ensure an odd number of solutions.
From Eqs. (\ref{VISH.e_v}) and Eq. (\ref{VISH.n_v}) we see that $e $ and
$n$ are roughly linear in $v$, which means that $P$ is also roughly
linear in $v$. So we expect to have only one solution. Please note that
since $v$ is bounded between 0 and 1, we need to ensure high precision of the
solution, otherwise $u^{\mu}$ will not be accurate, especially when $v \rightarrow 1$. 

Once $v$ is solved and thus $e$ and $n$ are known from Eqs. (\ref{VISH.e_v}) and (\ref{VISH.n_v}), $v_{x} ,v_{y} ,v_{z}$ can be solved easily from $M^{1}
,M^{2} ,M^{3}$,
\begin{equation}
  v^{i} = \frac{M^{i}}{M^{0} +P(e, n) + \Pi} , \mbox{   } (i = 1,2,3).
\end{equation}

In order to use Newton's method to find the root of the key equation (\ref{VISH.eq_v}), we can reorganize it as follows:
\begin{equation}
  f ( v ) =v ( M^{0} + \tilde{c}_{s}^{2} ( M^{0} -v M ) + \Pi ) -M.
  \label{VISH.quadraticf_v}
\end{equation}
Eq. (\ref{VISH.quadraticf_v}) can be considered as an approximatly quadratic equation for $v$. The condition $f (
v ) =0$ it has the solutions
\begin{equation}
  v= \frac{( M^{0} ( 1+ \tilde{c}_{s}^{2} ) + \Pi ) \pm \sqrt{( M^{0} ( 1+
  \tilde{c}_{s}^{2} ) + \Pi )^{2} -4 \tilde{c}_{s}^{2} M^{2}}}{2
  \tilde{c}_{s}^{2} M} .
 \label{VISH.eq53}
\end{equation}
The correct sign is found by checking the limit $M \rightarrow 0$, when $v$ approaches to zero. This selects the negative sign in Eq. (\ref{VISH.eq53}), which can thus be rewritten as
\begin{equation}
  v= \frac{2M}{( M^{0} ( 1+ \tilde{c}_{s}^{2} ) + \Pi ) + \sqrt{( M^{0} ( 1+
  \tilde{c}_{s}^{2} ) + \Pi )^{2} -4 \tilde{c}_{s}^{2} M^{2}}} .
\label{VISH.solution_v}
\end{equation}
The advantage of Eq. (\ref{VISH.solution_v}) is that the right hand side is only very weakly
dependent on $v$ as long as $\tilde{c}_{s}^{2}$ is approximately a constant
which is true over a very wide range of energy densities for {\tt s95p} EOS. Additionally, it is numerically stable in the limit
$M \rightarrow 0$. Similarly, we can find a solution for $u^{0}$:
\begin{equation}
  u^{0} = \frac{1}{\sqrt{1-v^{2}}} 
  %= \frac{1}{\sqrt{1- \left[ \frac{2M}{(M^{0} ( 1+ \tilde{c}_{s}^{2} ) + \Pi ) + \sqrt{( M^{0} ( 1+ \tilde{c}_{s}^{2} ) + \Pi )^{2} -4 \tilde{c}_{s}^{2} M^{2}}} \right]^{2}}} .
\label{VISH.solution_u0}
\end{equation}
Eqs. (\ref{VISH.solution_v}) and (\ref{VISH.solution_u0}) in principle give consistent solutions for $v$ and
$u^{0}$. In practice,  inevitable numerical errors render the use of Eq. (\ref{VISH.solution_v}) preferable for small velocities $v \rightarrow 0$, while Eq. (\ref{VISH.solution_u0}) should be used for $v \rightarrow 1$. Let us see why this is the case:

If we solve $u^{0}$ from Eq. (\ref{VISH.solution_u0}) and write the numerical solution as
$\tilde{u}^{0} =u^{0} + \Delta u$ where $u^{0}$ is the exact
solution and $\Delta u$ is the numerical error, the numerical error for
$v$ can be estimated as,
\begin{equation}
  \Delta v= \frac{d v}{d u^{0}}   \Delta u^{0} = \frac{\Delta u^{0}}{(
  \tilde{u}^{0} )^{2} \sqrt{( \tilde{u}^{0} )^{2} -1}}(1 +O ( \Delta u^{0} )) .
\end{equation}
In this situation, $\Delta v$ becomes small, $\Delta v \ll \Delta u^0$, for large flow velocity, $\tilde{u}^{0} \rightarrow + \infty$.
On the other hand, when $v \rightarrow 0$ and
$\tilde{u}^{0} \rightarrow 1$, the numerical error for $v$ is amplified by a factor
$\frac{1}{\sqrt{( \tilde{u}^{0} )^{2} -1}}$ compared to $\Delta u^{0}$, which is not good. Therefore Eq. (\ref{VISH.solution_u0}) is numerically stable for $v \rightarrow 1$ and unstable for $v \rightarrow 0$. 

The opposite is true for Eq. (\ref{VISH.solution_v}), Writing the numerical
solution of Eq. (\ref{VISH.solution_v}) as $\tilde{v} =v+ \Delta v$, we find,
\begin{equation}
  \Delta u= \frac{d u^{0}}{d v} \Delta v= \frac{\tilde{v}}{\left( \sqrt{1-
  \tilde{v}^{2}} \right)^{3}} \Delta v.
\end{equation}
In this case, $\Delta u^0 \ll \Delta v$ for small $\tilde{v} \rightarrow 0$, $\Delta u^{0}
\ll \Delta v$, which is favorable. On the other hand, for large velocity, $\tilde{v} \rightarrow 1$,
the error in $u^0$, $\Delta u^{0} \sim \frac{1}{\left( \sqrt{1-\tilde{v}^{2}} \right)^{3}} \Delta v$,
is amplified by a factor$\frac{1}{\left( \sqrt{1-\tilde{v}^{2}} \right)^{3}}$ relative to $\Delta v$, making $u^0$ numerically unstable. 

In the actual numerical
implementation, we solve both Eq. (\ref{VISH.solution_v}) and Eq. (\ref{VISH.solution_u0}), but we then select the
preferred solution according to the magnitude of the resulting velocity. The transition
point from one choice to the other happens at
\begin{equation}
  \frac{\tilde{v}}{\left( \sqrt{1- \tilde{v}^{2}} \right)^{3}} = \frac{1}{(
  \tilde{u}^{0} )^{2} \sqrt{( \tilde{u}^{0} )^{2} -1}} ,
\label{VISH.vtransition}
\end{equation}
with the relation $\tilde{u} = \frac{1}{\sqrt{1- \tilde{v}^{2}}}$.The
numerical solution of Eq. (\ref{VISH.vtransition}) is $\tilde{v}  = 0.563624$ or $\tilde{u}^{0}
=1.21061$. For velocities smaller than this critical value, we use the solution for $v$ from 
Eq. (\ref{VISH.solution_v}), while for larger velocities, we should select the solution for $u^0$ from Eq. (\ref{VISH.solution_u0}) as the more reliable one.

\subsection{Numerical check for {\tt VISHNew} using semi-analytic solutions}

\begin{figure*}[h!]
  \begin{center}
  \begin{tabular}{cc}
  \includegraphics[width=0.45\linewidth]{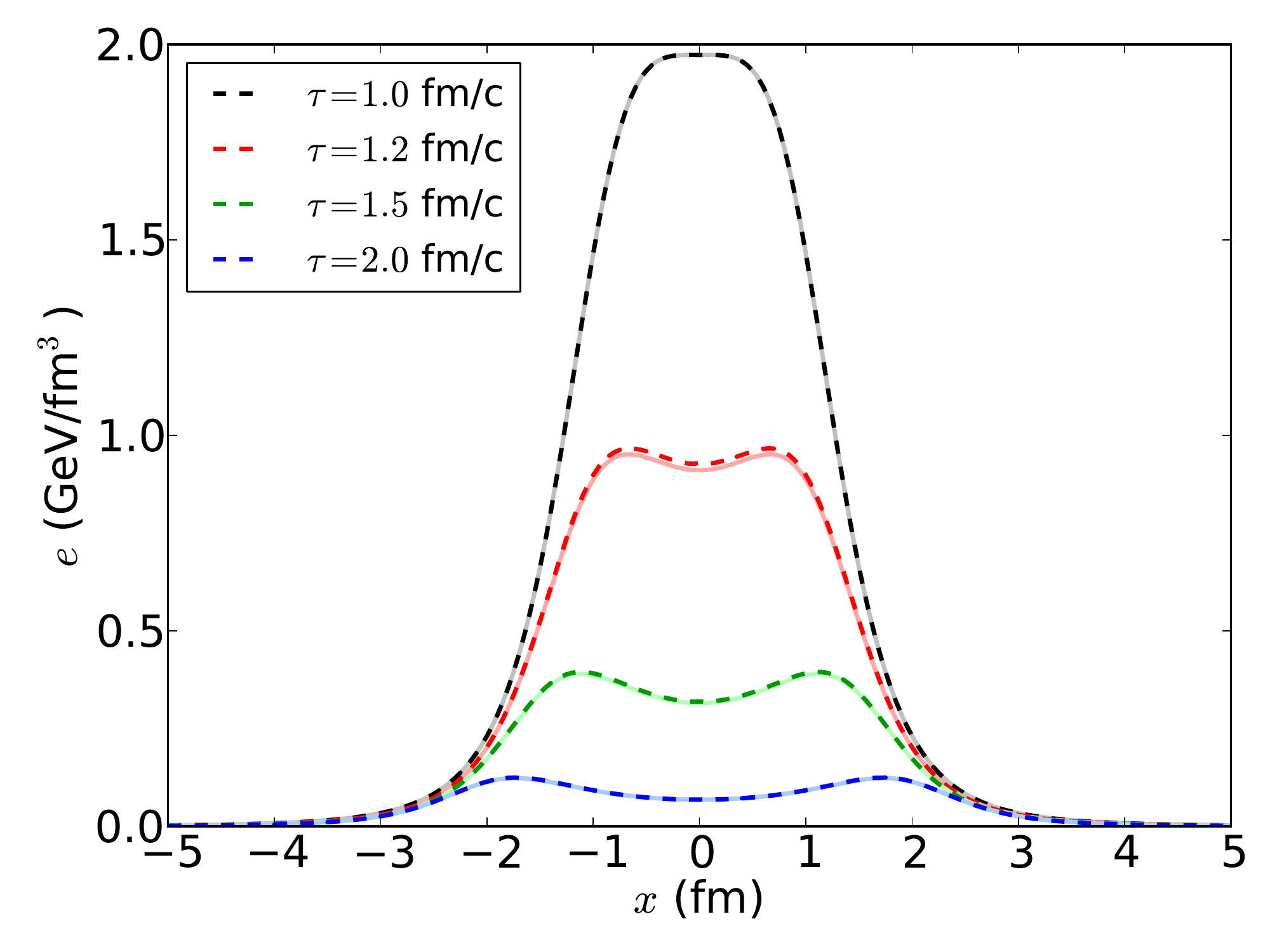} &
  \includegraphics[width=0.45\linewidth]{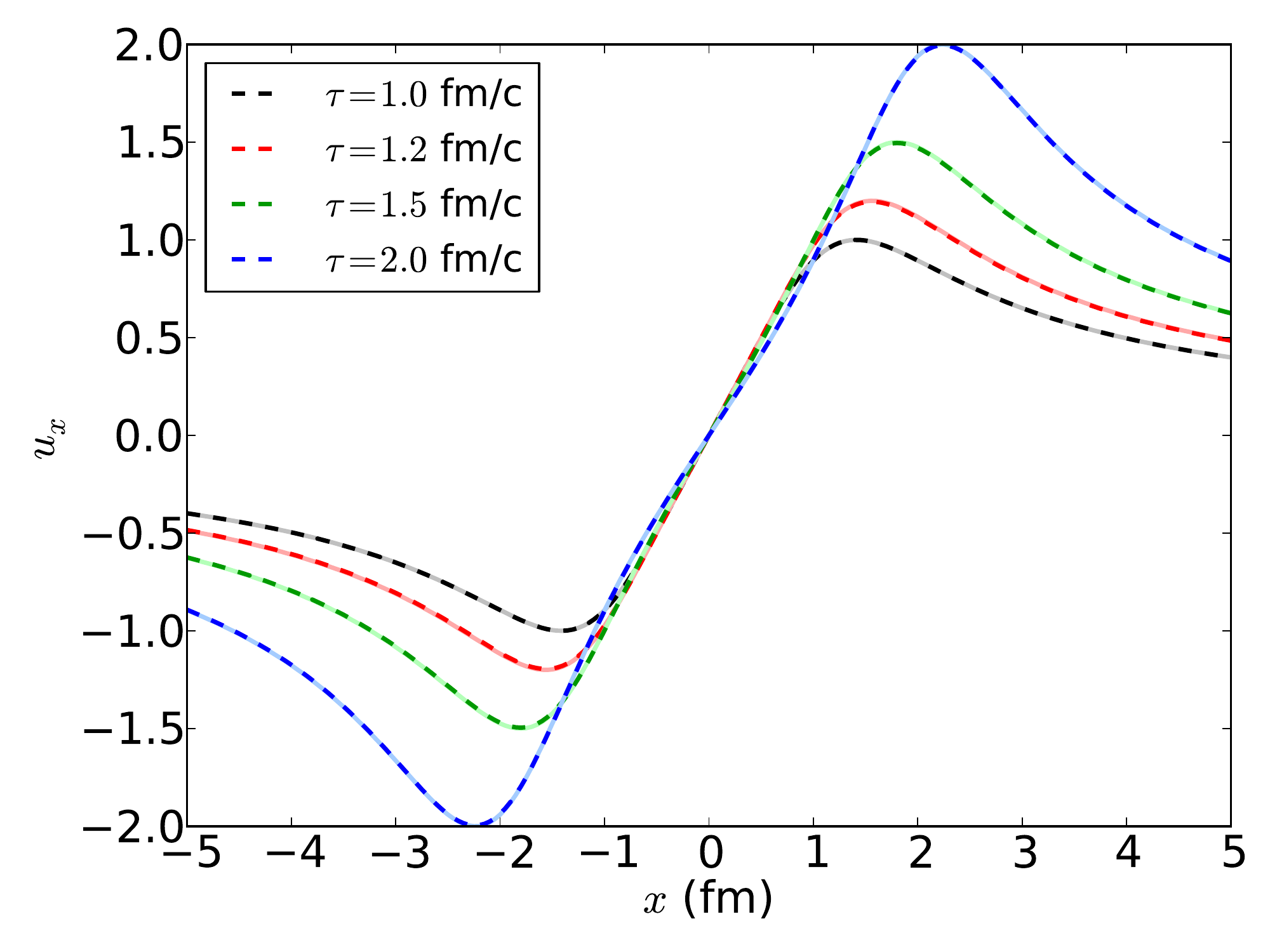}
  \end{tabular}
  \end{center}
  \caption{Comparison of the numerical solution (dark dashed lines) for (1+1)-d ideal fluid dynamical evolution with Gubser symmetry with Gubser's analytical results (light solid lines).}
  \label{Gubserflow.fig1}
\end{figure*}

In \cite{Gubser:2010ze,Gubser:2010ui}, the authors derived $SO(3) \otimes SU(1,1) \otimes Z_2$ invariant (``Gubser symmetric'') solutions of ideal relativistic conformal fluid dynamics which couple boost-invariant longitudinal expansion to an azimuthally symmetric transverse expansion. We first use this (1+1)-d solution to check the ideal hydrodynamic mode in {\tt VISHNew}. We start our ideal hydrodynamic simulation with Gubser's solution for the energy density and flow velocity at $\tau = 1.0$ fm/c and compare results at later proper time with Gubser's analytic solution. Fig. \ref{Gubserflow.fig1} shows excellent agreement between our simulations and the analytical solution.

%=======================================
%=======================================

For viscous hydrodynamics, Marrochio et al.  \cite{Marrochio:2013wla} have used the same symmetry argument developed by Gubser and constructed a nontrivial semi-analytic solution of the Israel-Stewart equations for (1+1)-d expansion with Gubser symmetry. In order to use this solution as a check of the {\tt VISHNew} code, we have to change the source term in the transport equation of shear stress tensor as specified in \cite{Marrochio:2013wla},
\begin{equation}
\Delta^{\mu\alpha} \Delta^{\nu \beta} D \pi_{\alpha \beta} = -\frac{1}{\tau_\pi} (\pi^{\mu\nu} - 2 \eta \sigma^{\mu\nu}) - \frac{4}{3} \pi^{\mu\nu} \theta.
\label{VISH.eq4}
\end{equation}
The difference between Eqs. (\ref{VISH.eq2}) and  (\ref{VISH.eq4}) only appears in third and higher orders in velocity gradients. However, since these gradients are large for the Gubser profile, the difference is noticeable and would be visible even if {\tt VISHNew} were a perfect numerical algorithm. We use the same parameters as in \cite{Marrochio:2013wla} (described below) to test our viscous hydrodynamic simulations. We start the simulation at $\tau = 1.0$ fm/c and use the semi-analytical solutions from \cite{Marrochio:2013wla} at $\tau = 1.0$ fm/c as the initial conditions for our simulations. We use an ideal massless gas equation of state $e = 3P$, with 
\begin{equation}
e = N_c \times \left(16 + \frac{7}{2}\times 3 N_f\right) \times \frac{\pi^2}{90} T^4,
\end{equation}
using $N_c = 3$ for the number of colors and $N_f = 2.5$ for the number of flavors. We set the specific shear viscosity to $\eta/s = 0.2$ and its corresponding relaxation time to $\tau_\pi = 5 \eta/(Ts)$. 

%=======================================
\begin{figure*}[h!]
  \centering
  \begin{tabular}{cc}
  \includegraphics[width=0.45\linewidth]{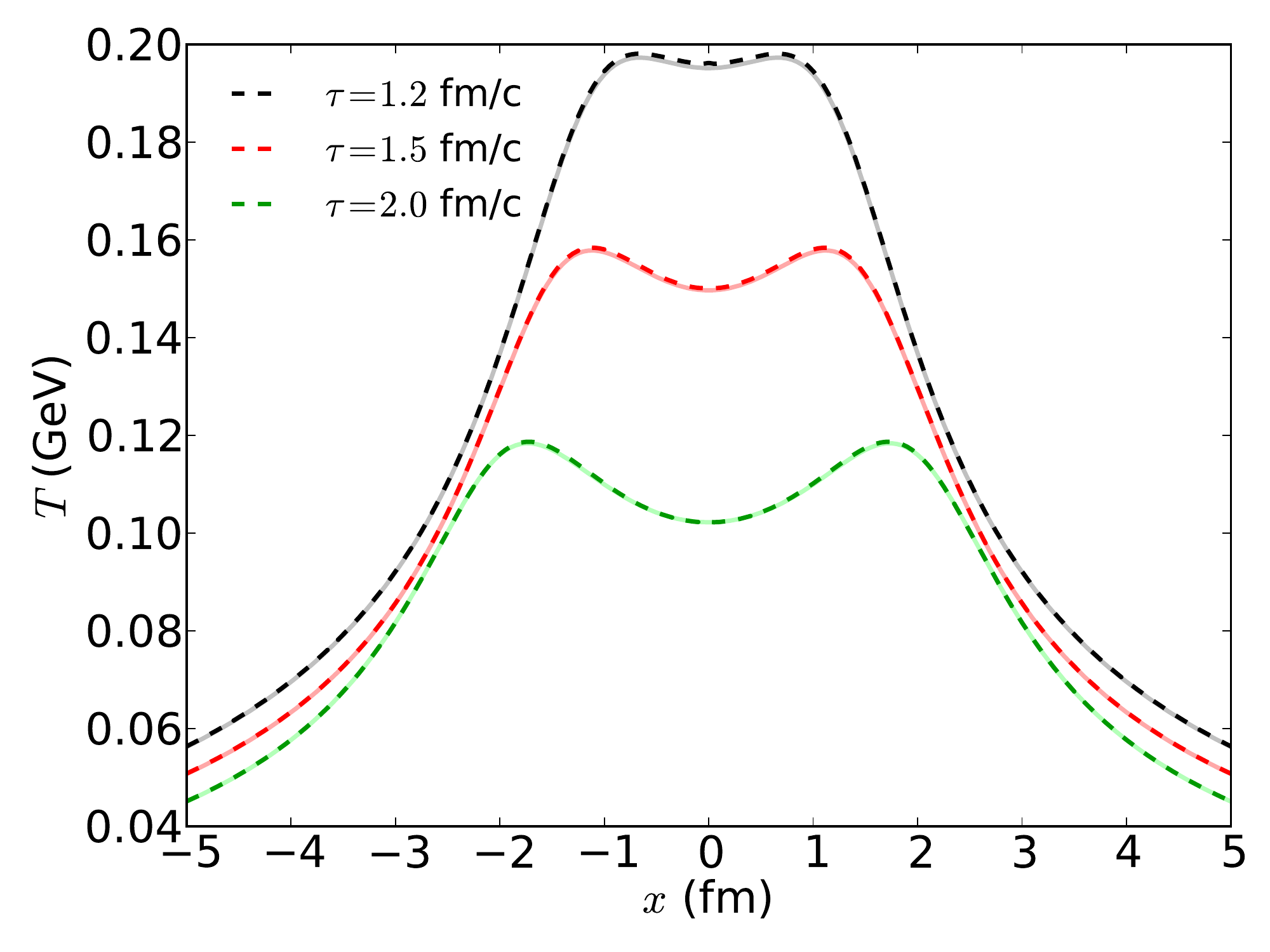} &
  \includegraphics[width=0.45\linewidth]{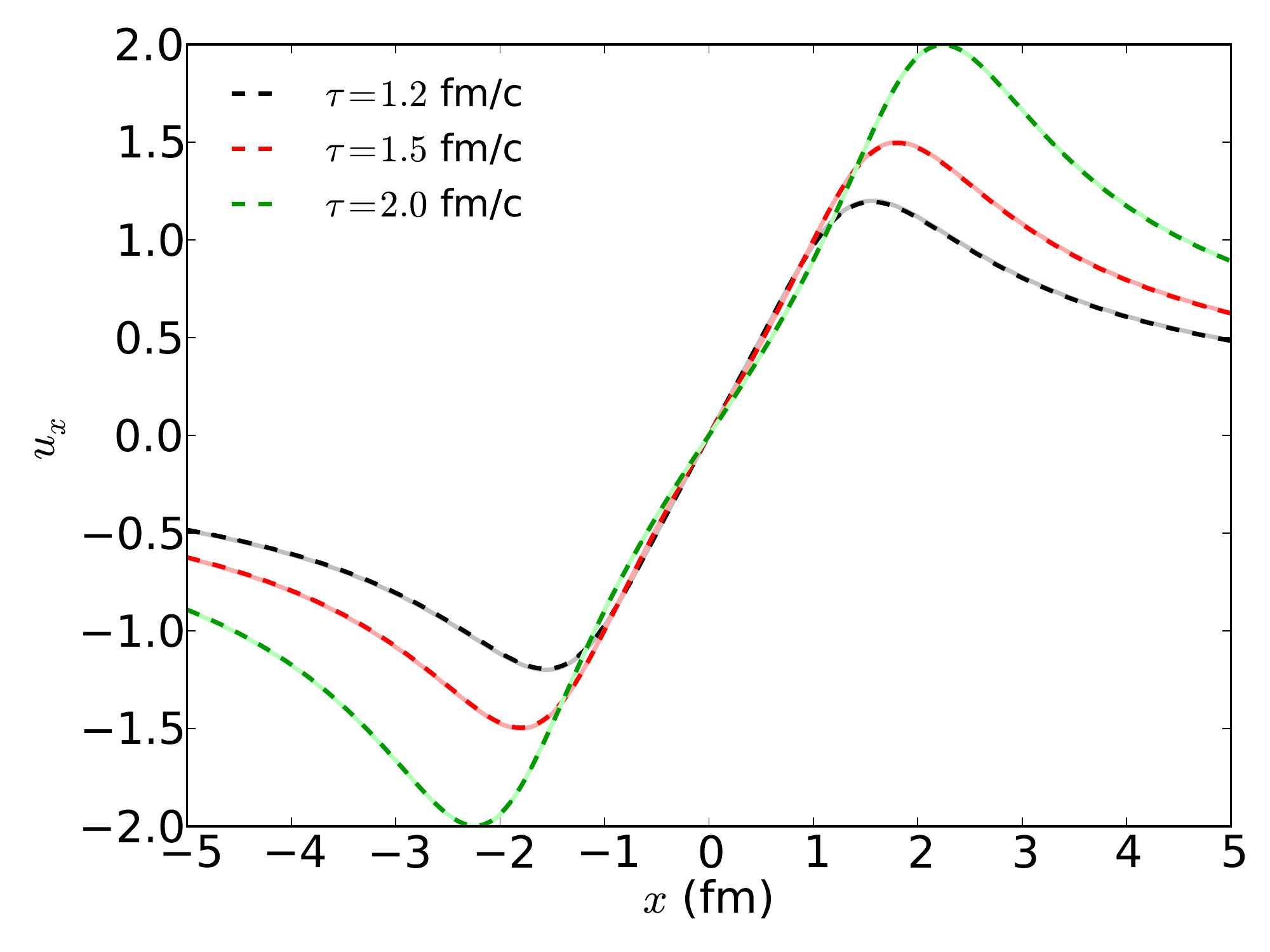}
  \end{tabular}
  \caption{Comparison of the temperature and flow velocity evolution from {\tt VISHNew} (dark dashed) with the semi-analytical solutions from \cite{Marrochio:2013wla} (light solid).}
  \label{Gubserflow.fig2}
\end{figure*}
%=======================================

%=======================================
\begin{figure*}[h!]
  \centering
  \begin{tabular}{cc}
  \includegraphics[width=0.45\linewidth]{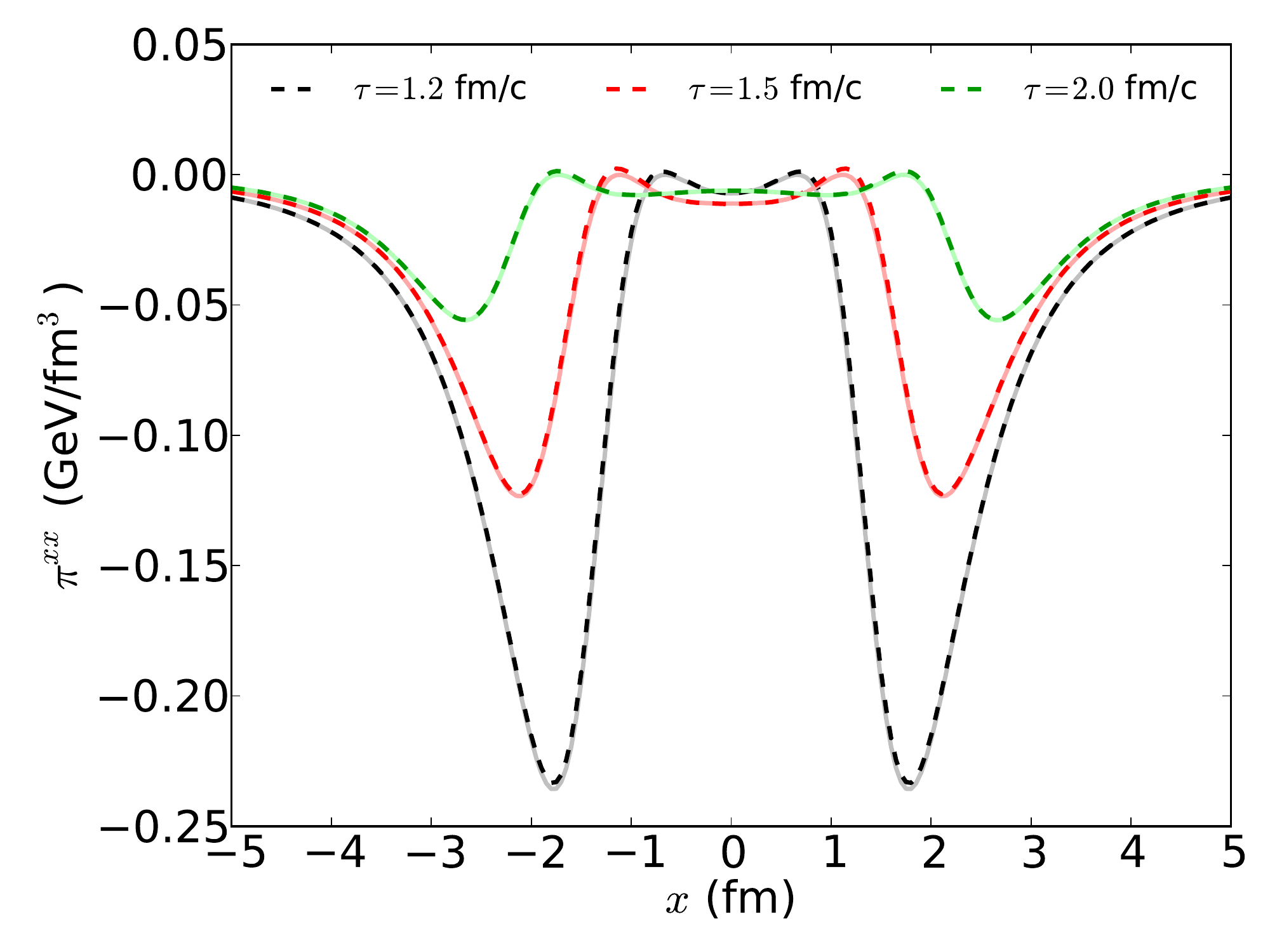} &
  \includegraphics[width=0.45\linewidth]{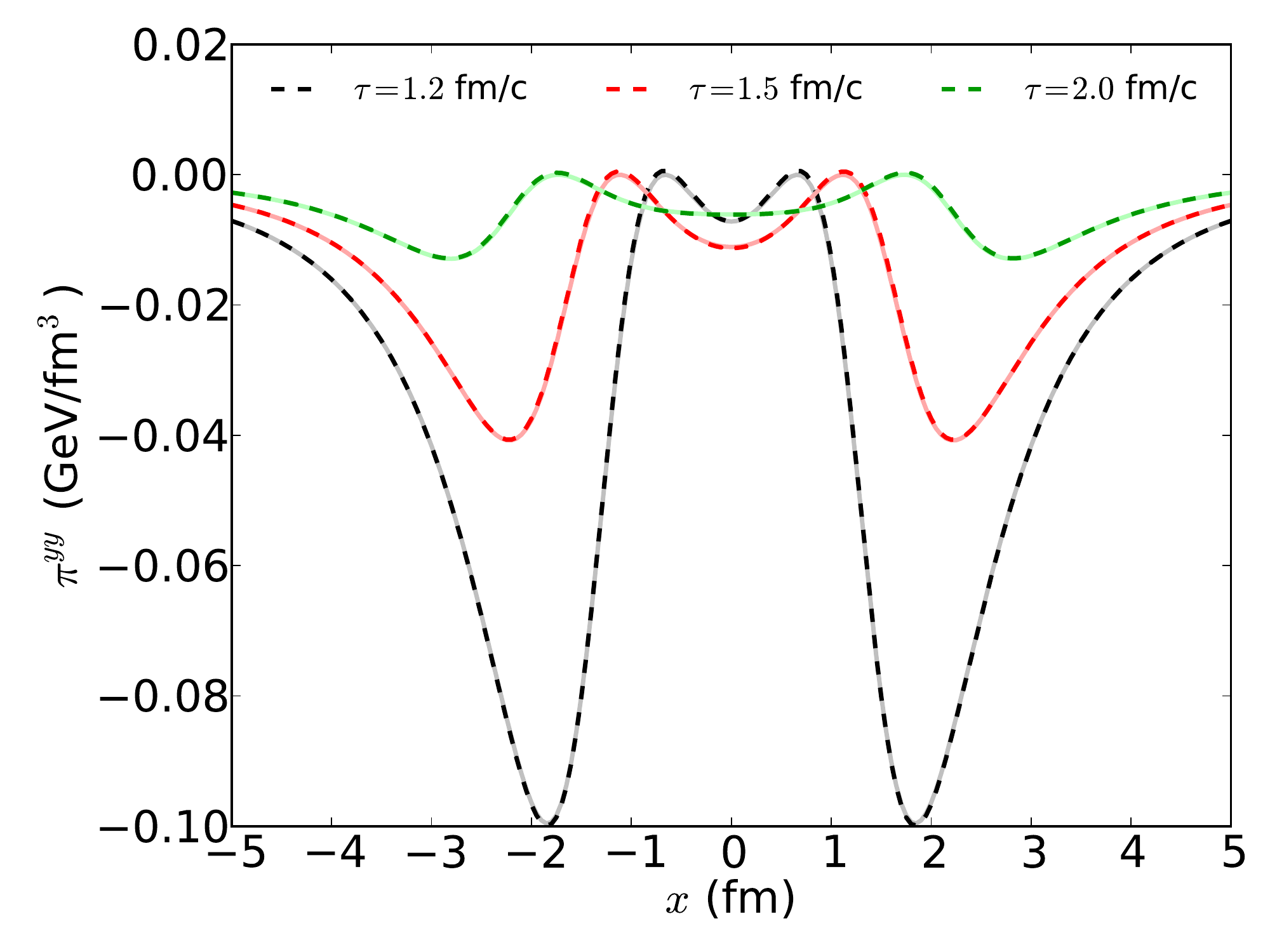} \\
  \includegraphics[width=0.45\linewidth]{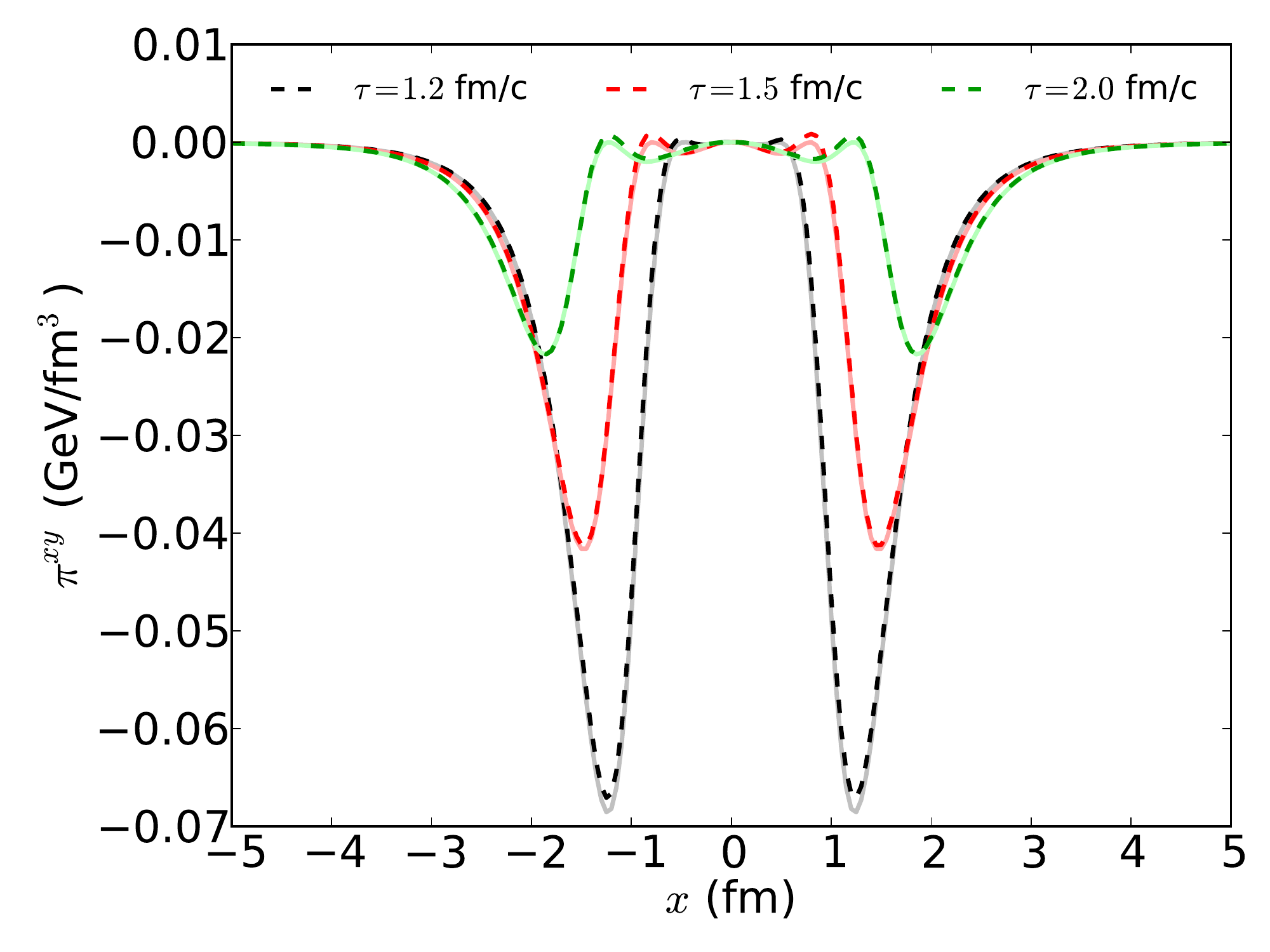} &
  \includegraphics[width=0.45\linewidth]{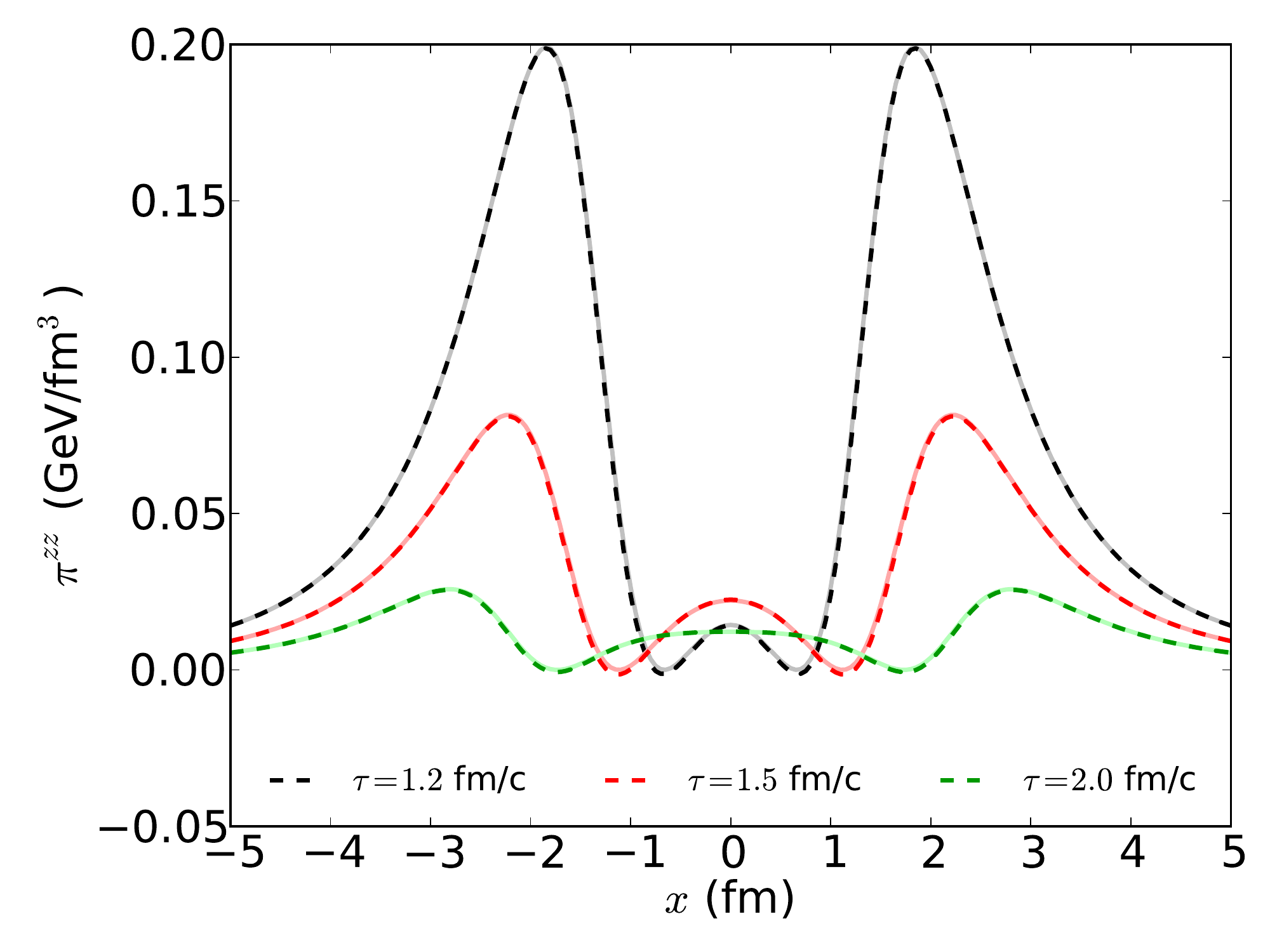}
  \end{tabular}
  \caption{The evolution of individual components of the shear stress tensor from {\tt VISHNew} (dark dashed) compared with the semi-analytical solutions from \cite{Marrochio:2013wla} (light solid). }
  \label{Gubserflow.fig3}
\end{figure*}
%=======================================

In Figs. \ref{Gubserflow.fig2} and \ref{Gubserflow.fig3}, we compare our numerical calculations with the semi-analytical solutions from \cite{Marrochio:2013wla} for the evolution of the local temperature, flow velocity, and shear stress tensor. For all hydrodynamic quantities we find very good agreement of our {\tt VISHNew} simulations with the semi-analytical results. 

\subsection{Stabilizing {\tt VISHNew} against numerical fluctuations in the viscous shear}

{\tt VISHNew} solves the minimum set of second order viscous hydrodynamic equations. The shear stress tensor is evolved according to Eq. (\ref{VISH.eq2}), which only includes spatial gradients up to second order. Such a truncation of the gradient expansion converges and gives good approximations only when higher order gradient terms are negligible. When we perform event-by-event hydrodynamic simulations, the fluctuating initial conditions usually feature large spatial gradients in the transverse plane. Under such conditions, the missing higher order gradient corrections to Eq. (\ref{VISH.eq2}) have the potential to grow large during the hydrodynamic evolution, and not including them in the code may eventually drive the whole numerical simulation into instability. However including all the higher order gradient terms in the transport equation for $\pi^{\mu\nu}$ is not practical. It would require the knowledge of the corresponding higher order transport coefficients, which are poorly constrained both theoretically and experimentally.

Therefore, staying within the framework of second order viscous hydrodynamics, we apply a regulation to the shear stress tensor that aims to suppress numerical instabilities caused by large spatial gradients. Similar regulation procedures are also performed in Refs. \cite{Bozek:2011ua,Schenke:2010rr}. In general, for second order viscous hydrodynamics to be valid, $\pi^{\mu\nu}$ must to satisfy the following criteria:

\begin{enumerate}
\item $\pi^{\mu\nu}$ should be smaller than the ideal part of the energy momentum tensor, $T_{0}^{\mu\nu} = e u^\mu u^\nu - P \Delta^{\mu\nu}$. To implement this we compare the following Lorentz invariant quantities,
\begin{equation}
\mathrm{Tr}(\pi^2) = \pi^{\mu\nu}\pi_{\mu\nu} \mbox{ and } T_{0}^{\mu\nu} {T_0}_{\mu\nu} = e^2+3P^2. \notag
\end{equation}
	
Consistency for our theoretical framework requires
\begin{equation}
	\pi^{\mu\nu}\pi_{\mu\nu} \ll e^2+3P^2 .
\label{VISH.eq6}
\end{equation}

\item $\pi^{\mu\nu}$ should be traceless:
	\begin{equation}
	\pi^{\mu}_{\ \mu} = 0
	\label{VISH.eq7}
	\end{equation}
	
\item $\pi^{\mu\nu}$ should be perpendicular to $u^\mu$:
	\begin{equation}
	\pi^{\mu\nu} u_\nu=0
	\label{VISH.eq8}
	\end{equation}
\end{enumerate}

{\tt VISHNew} evolves all seven non-vanishing components of $\pi^{\mu \nu}$, $\pi^{\alpha \beta}$ (where $\alpha, \beta = \tau, x, y$) and $\pi^{\eta \eta}$, independently without enforcing the conditions 2 and 3. Checking the validity of Eqs. (\ref{VISH.eq7}) and (\ref{VISH.eq8}) for the numerically evolved $\pi^{\mu\nu}$ thus amounts to a check of the numerical accuracy of our code. 
In actual calculations, there are limits to the numerical accuracy of $\pi^{\mu\nu}$ so we choose a small number $\xi_0\ll 1$ as the ``relative numerical zero" and replace conditions 2 and 3 by
\begin{equation} 
	\pi^{\mu}_{\ \mu} \le \xi_0 \sqrt{\pi^{\mu\nu}\pi_{\mu\nu}} \mbox{ and } \pi^{\mu\nu} u_\nu \le \xi_0 \sqrt{\pi^{\mu\nu}\pi_{\mu\nu}}, \forall \mu
\label{VISH.eq9}
\end{equation}
The vector $\pi^{\mu\nu} u_\nu$ should be component-wise zero (in any frame), therefore all its components should be compared to the ``relative numerical zero" multiplied by $\sqrt{\pi^{\mu\nu}\pi_{\mu\nu}}$(for dimensional reasons). Here we use the scalar $\sqrt{\pi^{\mu\nu}\pi_{\mu\nu}}$ as a measure for the magnitude of the $\pi^{\mu\nu}$ tensor that sets the scale (via the factor $\xi_0$) for how close the numerical result is to zero.

In practice, to ensure that Eq. (\ref{VISH.eq6}) is satisfied, we choose a number $\rho_\mathrm{max} \ll 1$ and require that\footnote{$\rho_\mathrm{max}\ll 1$ corresponds to the required ``$\ll$" condition in Eq. (\ref{VISH.eq9}); $\rho_\mathrm{max}=\infty$ corresponds to no constraint at all.}:
\begin{equation} 
	\sqrt{\pi^{\mu\nu}\pi_{\mu\nu}} \leq \rho_\mathrm{max}\sqrt{e^2+3p^2}.
\label{VISH.eq10}
\end{equation}

In our simulations, we found, that this condition is sometimes violated during the early stage and/or in the dilute regions outside the freeze-out surface (see at Fig. \ref{VISH.inverseReynold} for an example). The violation of Eq. (\ref{VISH.eq6}) in these regions do not have much influence on the dynamical behavior of the QGP in the physical region inside the freeze-out surface; however, if left untreated, such violations lead to accumulating numerical errors that eventually cause the evolution code to break down at later times. For these reasons, in the following we develop a systematic treatment that suppresses large viscous terms. This stabilizes the code with negligible effects on the physics and negligible extra numerical cost.

We enforce a continuous systematic regulation on $\pi^{\mu\nu}$ in each time step on the whole lattice by replacing $\pi^{\mu\nu}$ by $\hat{\pi}^{\mu\nu}$:
\begin{equation}
 \pi^{\mu\nu} \rightarrow \hat{\pi}^{\mu\nu} \equiv \pi^{\mu\nu} \frac{\tanh(\rho)}{\rho},
\label{VISH.eq12}
\end{equation}
where $\rho$ is the largest quantity at each lattice point among the following:
\begin{equation}
\frac{\sqrt{\pi^{\mu\nu}\pi_{\mu\nu}}}{\rho_\mathrm{max}\sqrt{e^2+3p^2}}, \frac{\pi^{\mu}_{\ \mu}}{\xi_0\,\rho_\mathrm{max}\,\sqrt{\pi^{\mu\nu}\pi_{\mu\nu}}}, \mbox{ or } \frac{\pi^{\mu\nu} u_\nu}{\xi_0 \rho_\mathrm{max}\,\sqrt{\pi^{\mu\nu}\pi_{\mu\nu}}}, \forall \mu \notag
\end{equation}
It is easy to check that $\hat{\pi}^{\mu\nu}$ satisfies Eq. (\ref{VISH.eq10}), and that it is close to $\pi^{\mu\nu}$ where no modifications are needed; that is, when the left hand side of the inequality in Eq. (\ref{VISH.eq10}) is small compared to the right hand side, Only at those grid points where $\pi^{\mu\nu}$ violates or is close to violating the inequality (\ref{VISH.eq10}) will it be strongly modified; if this is the case, all components of $\pi^{\mu\nu}$ are suppressed by the same factor. 

Because smoother flow velocity profiles give smaller $\pi^{\mu\nu}$, the systematic suppression of $\pi^{\mu\nu}$ can be understood as locally replacing sharp jumps in the flow profile by smoother pieces; the regulation process is therefore an implicit and automatic way of smoothing profiles. This treatment allows us to perform hydrodynamic calculations using very bumpy initial conditions, including those using disk-like nucleons that have density discontinuities. Without this regularization {\tt VISHNew} breaks down for such initial conditions.We note that typically no regulations are required inside the freeze-out surface at later times; shear viscosity leads to dynamical smoothing of initial fluctuation by dissipation, suppressing sharp velocity gradients and large values of $\pi^{\mu\nu}$ as time proceeds. Regulation remains necessary in the dilute region outside the freeze out surface where $e$ and $P$ (which for massless degree of freedom both fall like $T^4$) fall faster than $\pi^{\mu\nu} $ (which falls only like $T^3$).

In our calculations, we take $\xi_0=0.1$. If we choose smaller $\xi_0$ in the simulations, we overkill the physical viscous damping effects. In the following section, we show tests invoking several choices of $\xi_0$ and their influence on the final observables. During our tests we found that $\rho_\mathrm{max}$ is best chosen to be a value between $1-10$. By choosing $\rho_\mathrm{max}$ of order unity or larger, we reduce the regulation strength in each step to the point where the code is numerically stable with minimum modification. (Note that this implies that the code may run in a domain where the strong inequality Eq. (\ref{VISH.eq6}) is not satisfied, i.e. second order viscous hydrodynamics may not be strictly valid (see \cite{Niemi:2014wta} for a related study).)

\subsection{Hydrodynamic evolution with regulation}

In this section, we study the sensitivity of final hadronic observables on the choice of the $\xi_0$ parameter used in the $\pi$ regulation routine.  For these tests, we choose MC-Glauber initial conditions for  Pb + Pb collisions at $\sqrt{s} = 2760$\,$A$\,GeV at 20-30\% centrality, using $\eta/s = 0.20$. We simulate 200 events for every choice of $\xi_0$. 

%=======================================
\begin{figure}[h!]
  \centering
  \includegraphics[width=0.55\linewidth]{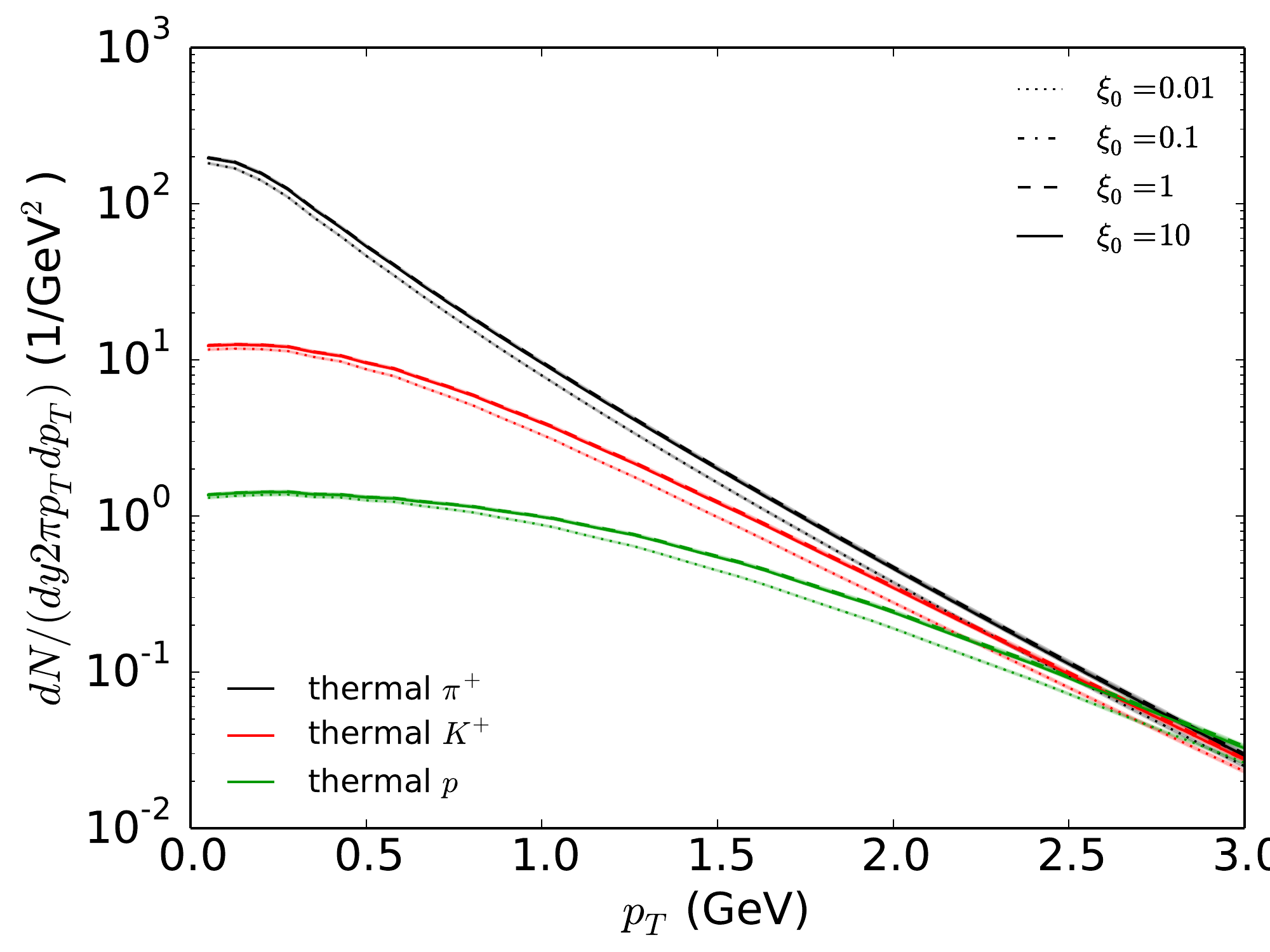}
  \caption{Thermal particles' $p_T$ spectra for different choice of $\xi_0$ used in the regulation trigger routine.  }
  \label{VISH.xi0Spdepdence}
\end{figure}
%=======================================

In Fig. \ref{VISH.xi0Spdepdence}, we show the $p_T$-spectra for thermal $\pi^+$, $K^+$, and protons, with different values of $\xi_0$ used in the trigger routine for the $\pi^{\mu\nu}$ regulation. We find that for $0.1 \le \xi_0 \le 10$,  there is no noticeable difference between different choices of $\xi_0$ used in the simulations. Only for the very small value $\xi = 0.01$ we see an effect: particle spectra get steeper, and the yield decreases. This means that the system generates less entropy and radial flow during the evolution, which indicates that the shear viscous effects in the simulations are suppressed too strongly by the regulations.   

%=======================================
\begin{figure*}[h!]
  \centering
  \begin{tabular}{cc}
  \includegraphics[width=0.48\linewidth]{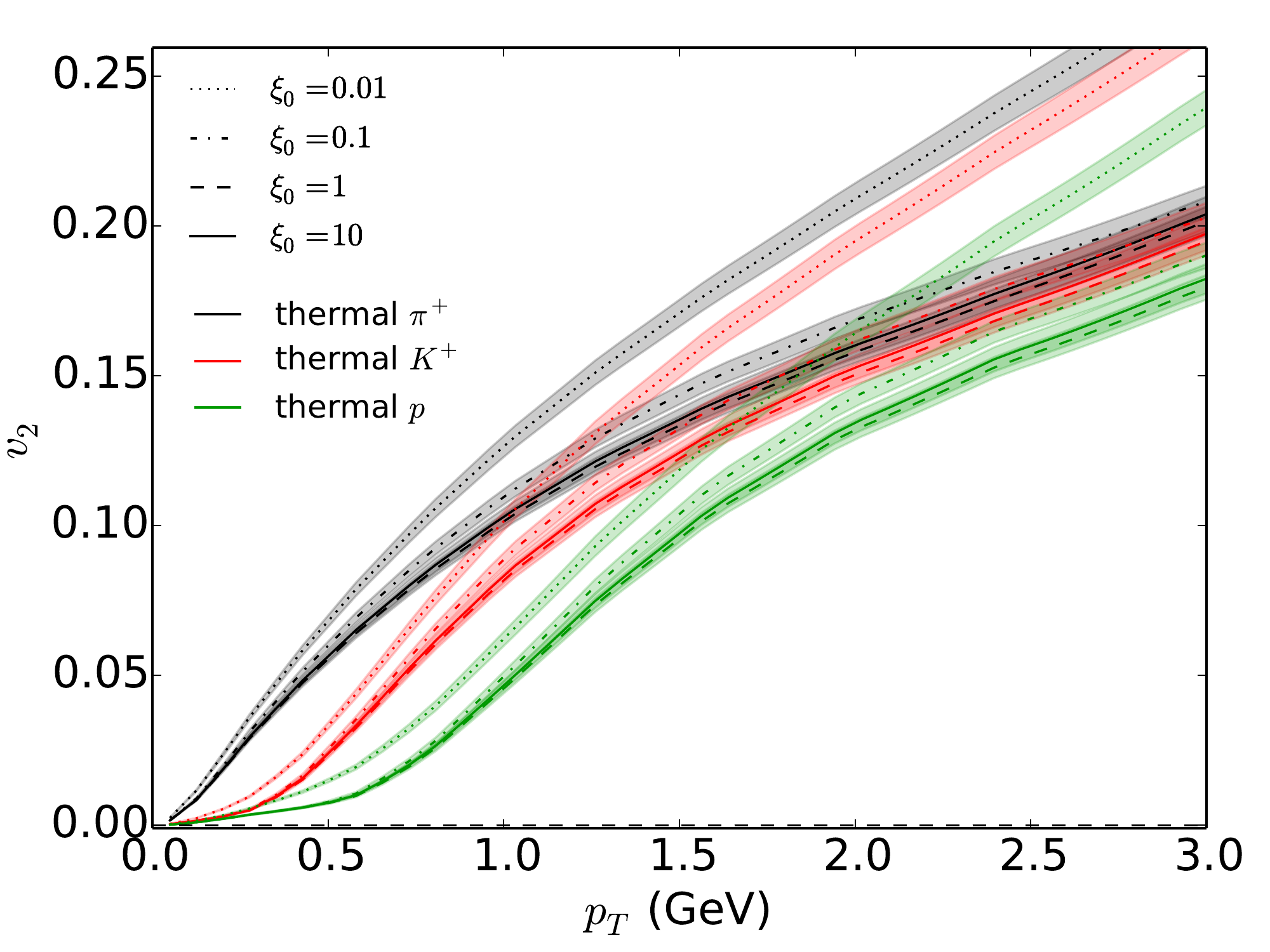} &
  \includegraphics[width=0.48\linewidth]{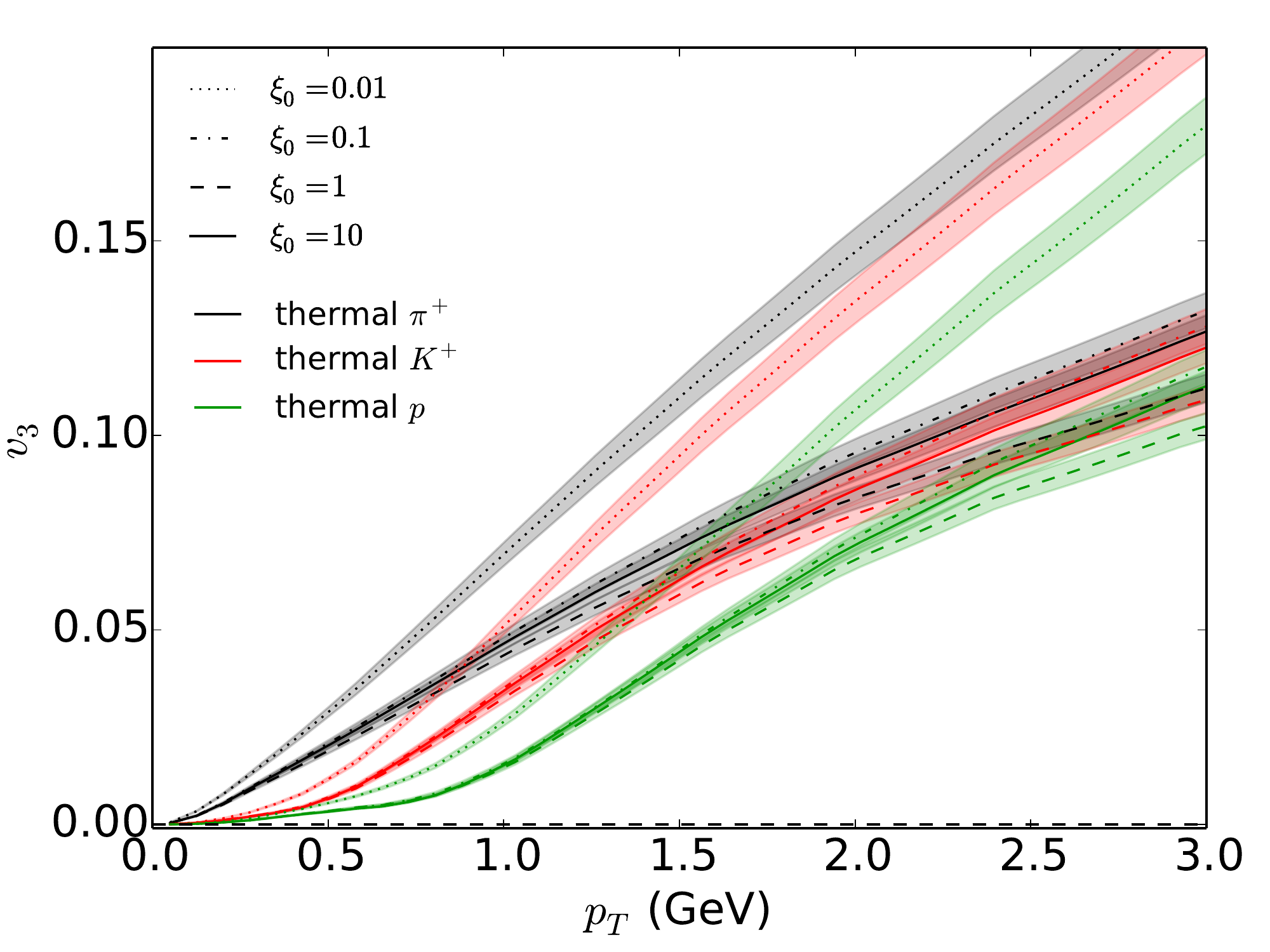} \\
   \includegraphics[width=0.48\linewidth]{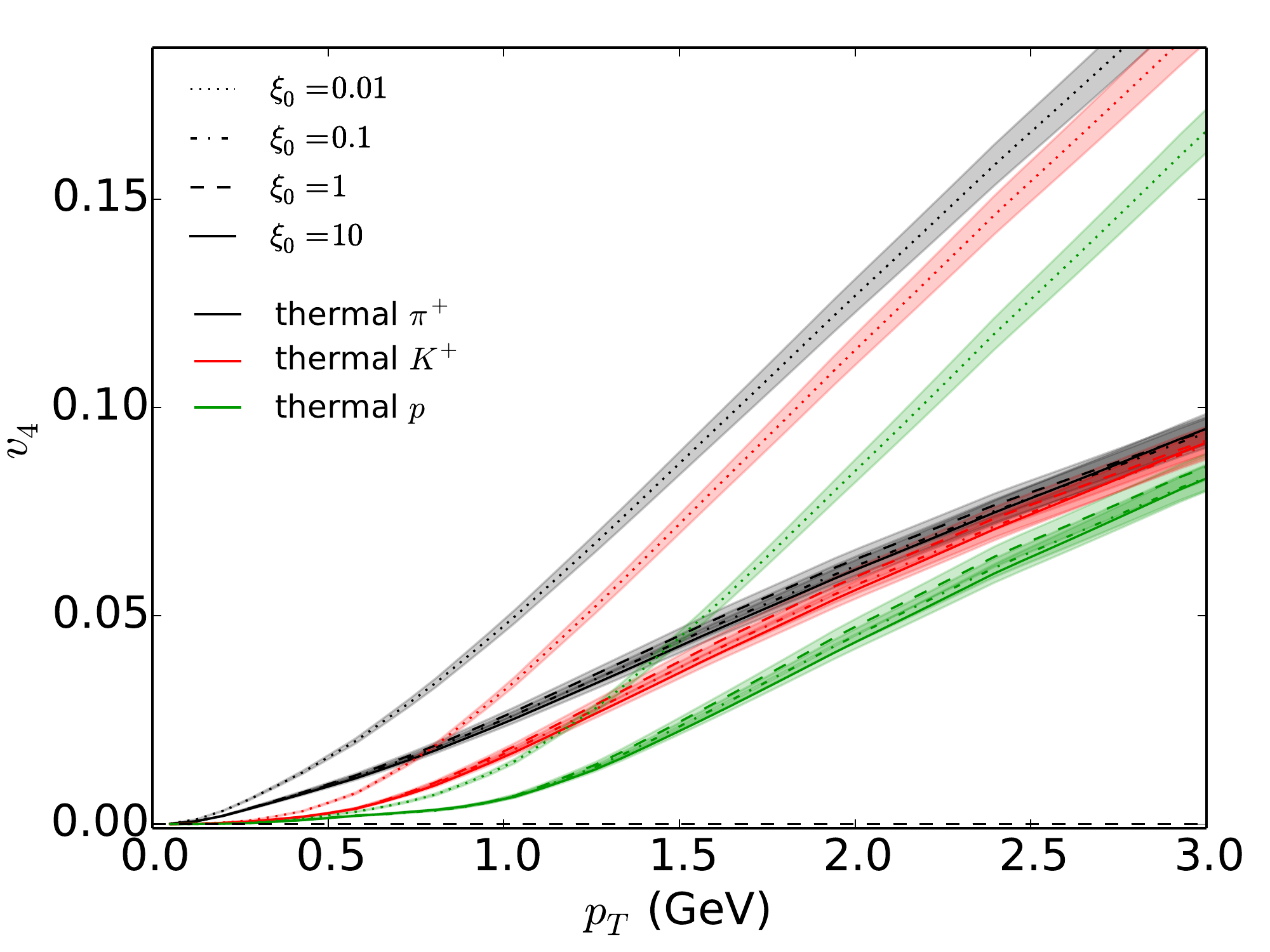} &
  \includegraphics[width=0.48\linewidth]{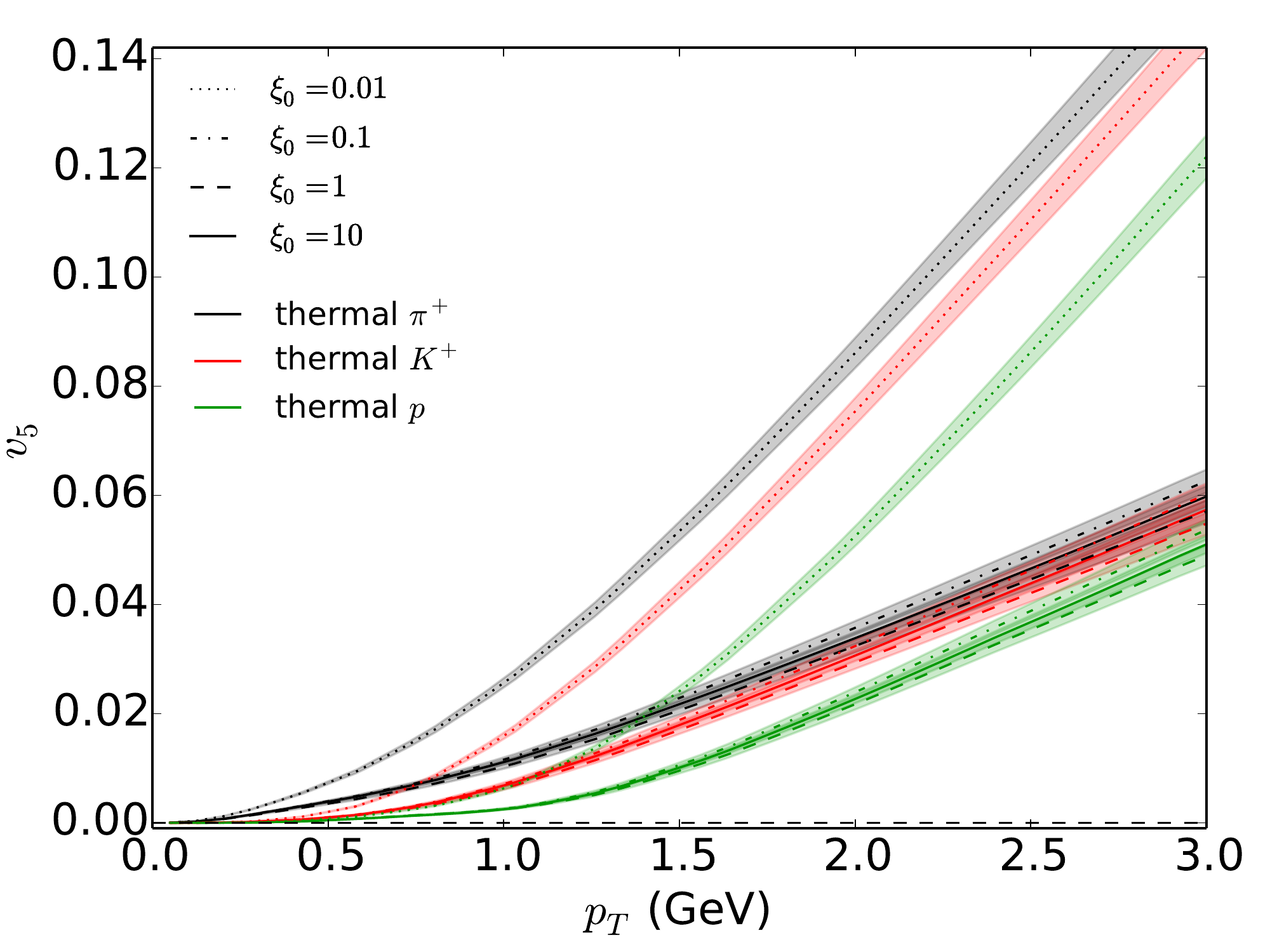}
  \end{tabular}
  \caption{Thermal particles' $p_T$-differential $v_n$ for different choice of $\xi_0$ used in the regulation trigger routine.  Statistical errors are indicate as shaded bands in the plots. }
  \label{VISH.xi0vndepdence}
\end{figure*}
%=======================================

In Fig. \ref{VISH.xi0vndepdence}, we show results for the $p_T$-differential anisotropic flows $v_2$ to $v_5$. They show a larger sensitivity to the choice of $\xi_0$ than the single particle spectra. For $0.1 \le \xi_0 \le 10$, the $v_n$ of thermal particles agree reasonably well with each other within the statistical error bands. But for $\xi_0 = 0.01$, the regulation again over-suppresses the viscous effects, which damp the anisotropic flow of the system. 

From this parameter study we conclude that the final hadronic observables are not sensitive to the choice of $\xi_0$ as long as we keep it in the range $0.1 \le \xi_0 \le 10$. For $\xi_0$ larger than 10, the code becomes numerically unstable due to too strong violations of the criteria that ensure validity of the second order viscous hydrodynamic description. For $\xi_0$ smaller than 0.1, the regulation routine seems to over kill the shear viscous effects in the system, thereby altering the physics by using an effective shear viscosity that is much smaller than the input value. In our application of {\tt VISHNew}, we therefore always use $\xi_0$ in the range $ 0.1 \le \xi_0 \le 10$. 

%=======================================
\begin{figure*}[h!]
  \centering
  \begin{tabular}{cc}
  \includegraphics[width=0.48\linewidth]{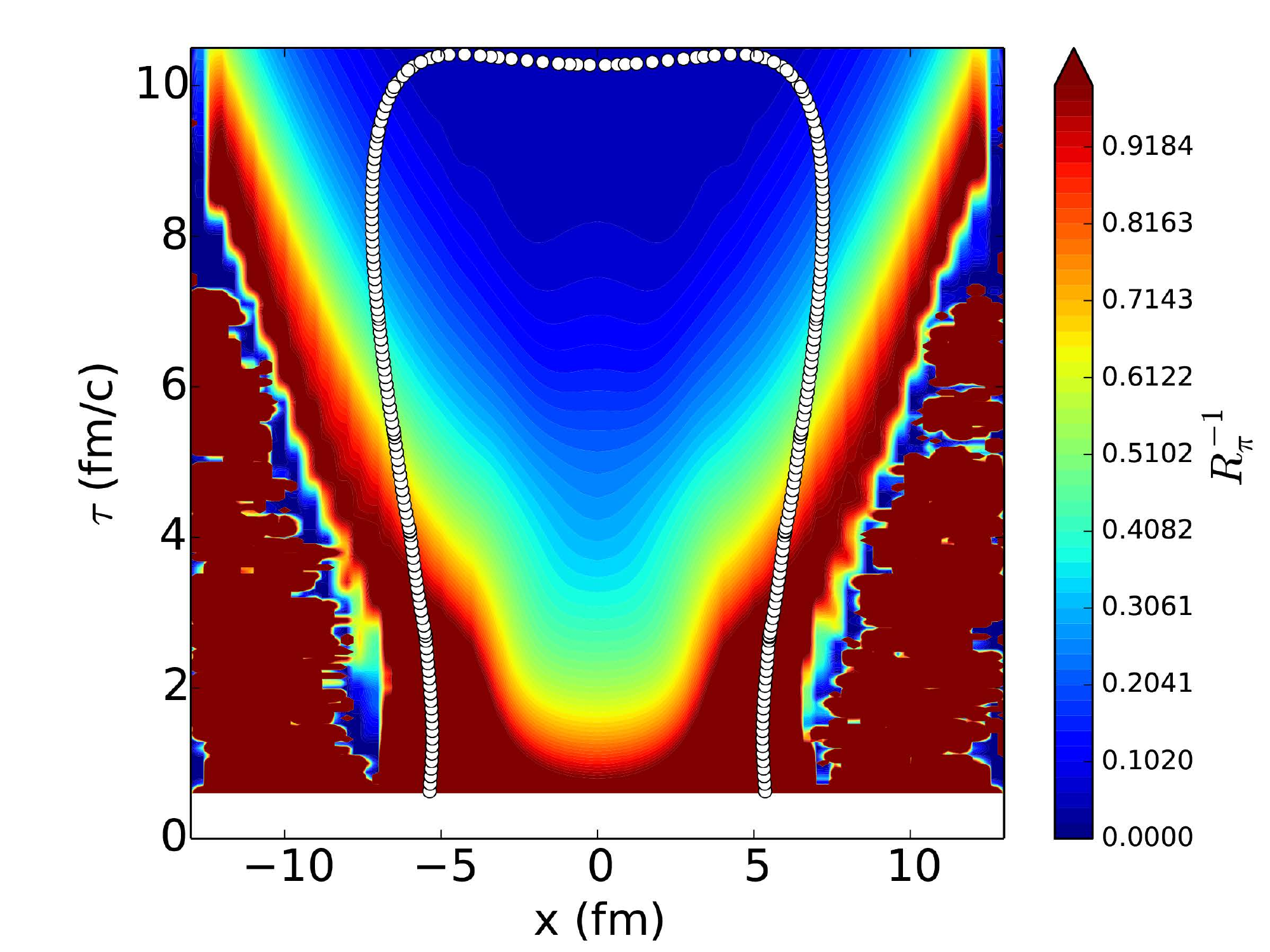} &
  \includegraphics[width=0.48\linewidth]{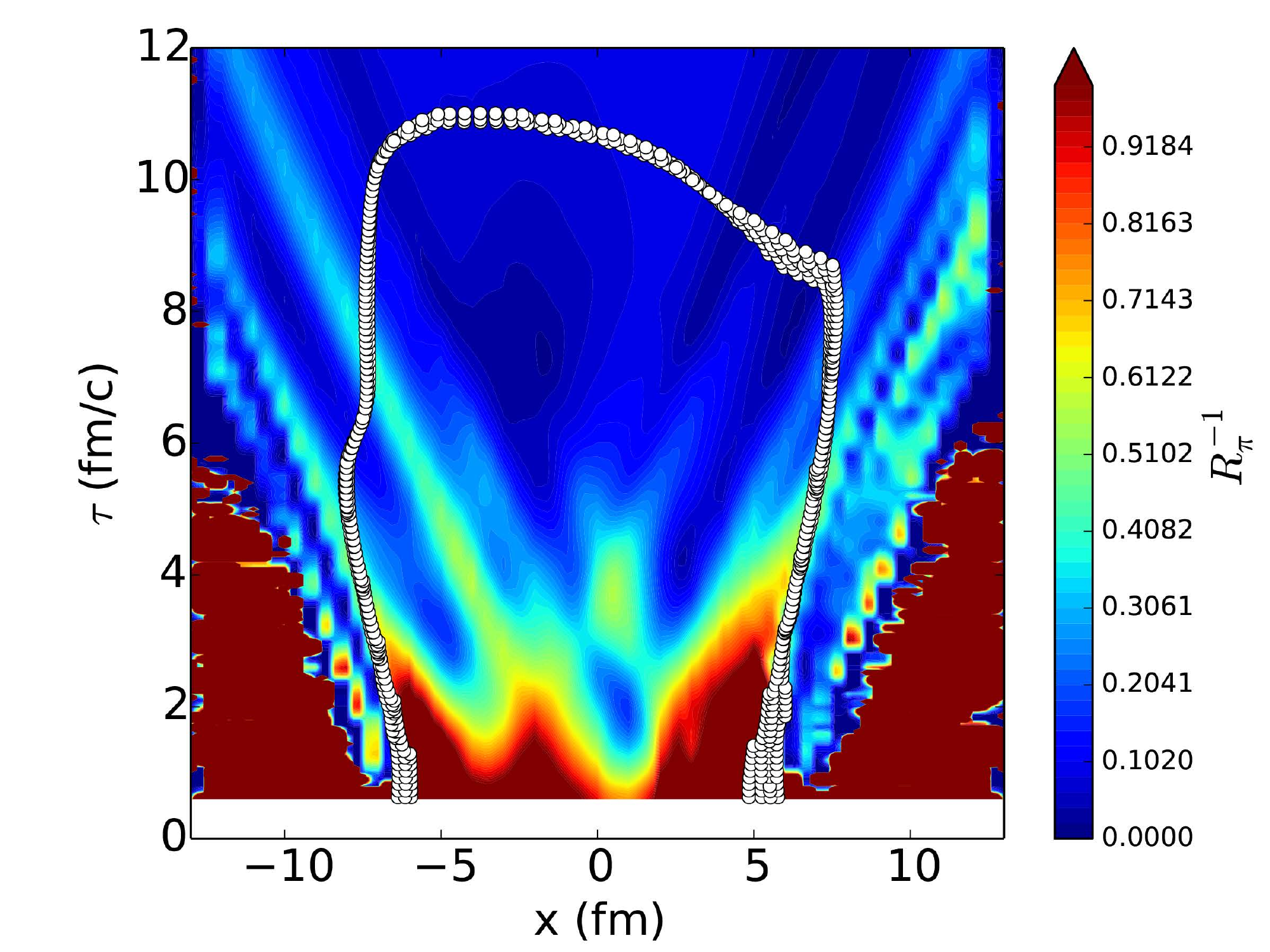} \\
\includegraphics[width=0.48\linewidth]{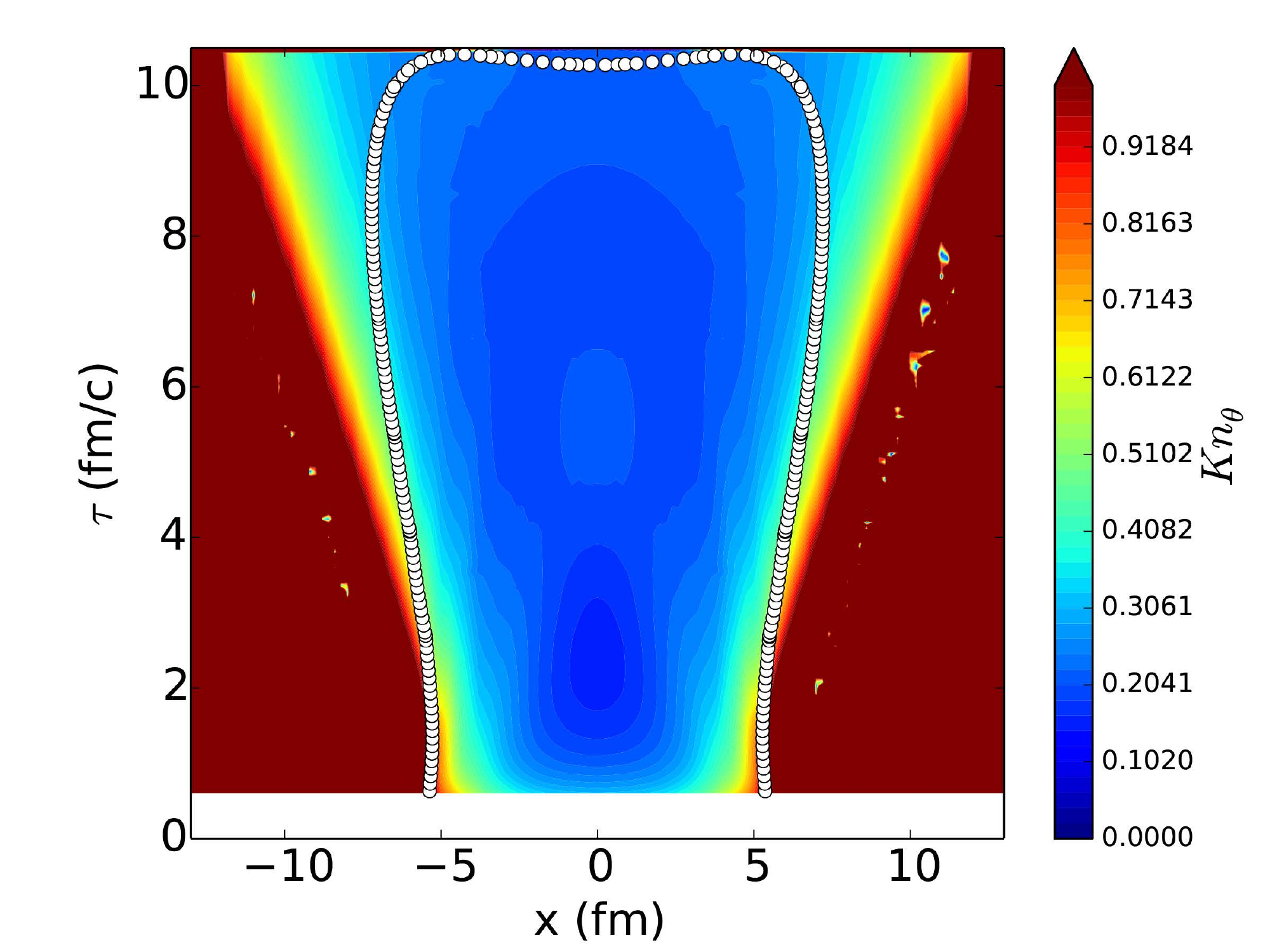} &
  \includegraphics[width=0.48\linewidth]{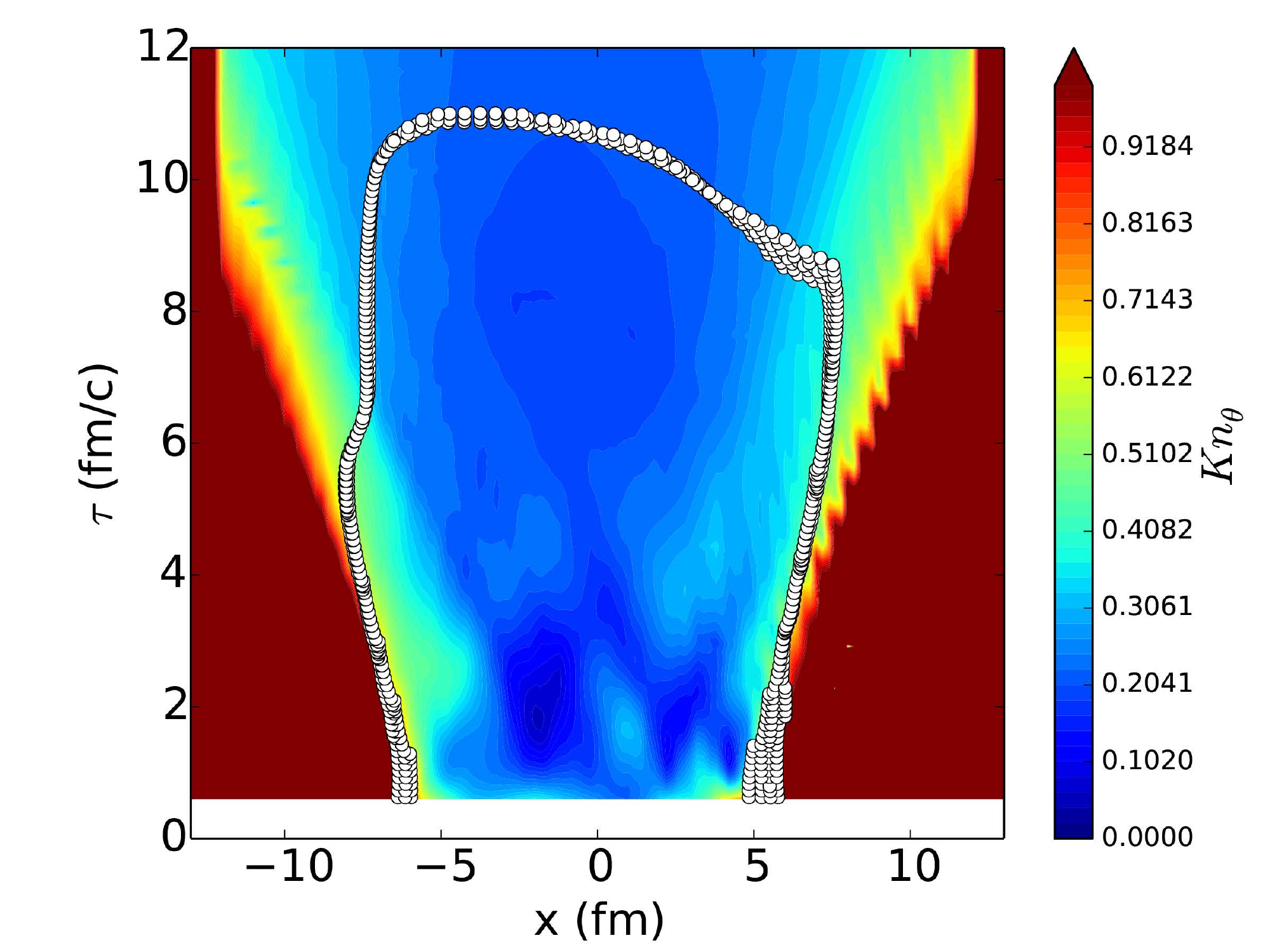}
  \end{tabular}
  \caption{Contour plot for the evolution of the inverse Reynolds number (upper panels) and Knudsen number (lower panels) in viscous hydrodynamic simulations with $\eta/s = 0.20$ at 20-30\% LHC energy. The white points indicate the kinetic freeze surface at $T_\mathrm{dec} = 120 $ MeV. }
  \label{VISH.inverseReynold}
\end{figure*}
%=======================================

In order to quantify the quality of reliability of our second order viscous hydrodynamic approach we can monitor the inverse Reynolds number and Knudsen number associated with shear stress, defined as \cite{Niemi:2014wta}:
\begin{equation}
R^{-1}_\pi = \frac{\sqrt{\pi^{\mu\nu} \pi_{\mu\nu}}}{\mathcal{P}}
\label{VISH.inverseReynolddef}
\end{equation}
and
\begin{equation}
Kn_\theta = \frac{\lambda_\mathrm{mfp}}{L_\mathrm{hydro}} = \tau_\pi \theta = 5 \frac{\eta \theta}{s T}. 
\label{VISH.Knudsendef}
\end{equation}
Here $\mathcal{P}$ is the thermal pressure, $\tau_\pi$ is the shear relaxation time, and $\theta$ is system expansion rate. As long as $R^{-1}_\pi$ and $Kn_\theta$ are smaller than 1, the system behaves like a fluid with low viscosity. For $R_\pi^{-1} \gg 1$, the behavior is more dissipative and viscous hydrodynamics is no long a good description. And for $Kn_\theta > 1$, the collision rate is not high enough to keep the system to stay near thermal equilibrium. In Fig. \ref{VISH.inverseReynold}, we show a contour plot of the evolution of the inverse Reynolds number and Knudsen number in our hydrodynamic simulation at $y=0$ in the transverse plane. In the left panels we start with a smooth event-averaged initial condition; in the right panels we show results for a fluctuating initial profile.  The white points indicate the position of the kinetic freeze-out surface. We notice that the largest inverse Reynolds numbers are encountered at early times of the hydrodynamic evolution or outside the freeze-out surface for both smooth and fluctuating initial conditions. And we find similar situation for the Knudsen number evolution. As time goes on, the magnitude of the shear stress tensor decreases, and the relativistic hydrodynamic modeling becomes more and more reliable. At very early times the use of viscous hydrodynamics becomes questionable, especially for fluctuating initial profiles. Please note that with a constant $\eta/s$, our estimation of the Knudsen number $Kn_\theta = 5 \frac{\eta}{sT} \theta $ might not be reliable in dilute hadronic regions. In principle the value of $\eta/s$ will increases as temperature decreases. So the Knudsen number should increase in the low density region in the hadron gas phase.

%%%%%%%%%%%%%%%%%%%%%%%%%%%%%%%%%%%%%%%%%%%%%
\section{Cooper-Frye freezeout using {\tt iS} and particle sampler {\tt iSS}}
\label{sec5}
%%%%%%%%%%%%%%%%%%%%%%%%%%%%%%%%%%%%%%%%%%%%%

The name ``{\tt iS}'' stands for ``{\tt iSpectra}"; {\tt iS} is a fast Cooper-Frye particle momentum distribution calculator along the conversion surface. Its output is a continuous function, evaluated at discrete momenta provided by the user, for the invariant momentum distributions of the desired hadron species. The code ``{\tt iSS}'', whose name stands for ``{\tt iSpectraSampler}", goes one step further to generate individual particles samples, using the calculated particle momentum distributions as the relative emission probability. {\tt iSS} is an ``event generator" which generates a complete collision event of emitted hadrons, similar to the events created in the experiment. Both codes are written keeping the following factors in mind:

\begin{itemize}
 \item {\bf Readability and extendability.} The most important goal is to create a cleanly written framework that calculates particle momentum distributions and performs sampling, whose components and output can be used easily for further physics analyses and tests of new physical ideas. To achieve this, the entire program is divided into modules according to their functionalities, the structures and the algorithms are documented with comments, and long but informative names are chosen for variables and function names.
 \item {\bf Efficiency.} Both the {\tt iS} and {\tt iSS} codes are written aiming for intensive event-by-event calculations where every CPU cycle counts. To achieve the necessary degree of efficiency, much effort is put into optimizing the algorithms at different levels of the calculations.
 \item {\bf Easy maintainability and re-usability.} The framework is divided into different carefully chosen functionality modules, for better interoperability and to maximize re-usability.
\end{itemize}

\subsection{Cooper-Frye freeze-out}

The particle emission function that implements sudden decoupling from a surface element $d^3\sigma$ located on a freeze-out hyper-surface $\Sigma(x^\mu)$ is given by the Cooper-Frye formula,
\begin{equation}
E\frac{d N}{d^3 p}(x^\mu, p^\mu) = \frac{g}{(2\pi)^3} p^\mu d^3 \sigma_\mu \left(f_0(x^\mu, p) + \delta f(x^\mu, p^\mu)\right),
\label{iSS.eq1}
\end{equation}
where $g$ is the spin degeneracy, $d^3 \sigma_\mu = (\cosh \eta_s, -\partial \tau/\partial x, -\partial \tau/\partial y, -\sinh \eta_s) \tau dx dy d\eta_s$ is the infinitesimal surface element on $\Sigma(x^\mu)$ for systems with longitudinal boost-invariance, and $f_0(x^\mu, p)$ is local thermal equilibrium distribution function. $\delta f(x^\mu, p^\mu)$ represents the deviation from local thermal equilibrium due to viscous effect and takes the following form,
\begin{equation}
\delta f(x^\mu, p^\mu) = f_0(x^\mu, p) (1 \pm f_0(x^\mu, p)) \frac{\pi_{\mu\nu} \hat{p}^\mu \hat{p}^\nu}{2(e+P)} \chi \left(\frac{p \cdot u}{T} \right),
\label{iSS.eq2}
\end{equation}
where $\hat{p}^\mu = p^\mu/(p \cdot u)$ and $\chi (p \cdot u/T) = (p \cdot u/T)^\alpha$ with $1 \le \alpha \le 2$. Integrating the emission function over the freeze-out surface we obtain particle momentum distribution
\begin{equation}
\frac{dN}{dy p_T dp_T d\phi_p} = \int_\Sigma \frac{g}{(2\pi)^3} p^\mu d^3 \sigma_\mu (f_0(x^\mu, p) + \delta f(x^\mu, p^\mu)).
\label{iSS.eq3}
\end{equation}
The azimuthally averaged $p_T$-spectrum is given by,
\begin{equation}
\frac{dN}{2\pi dy p_T dp_T} = \int \frac{d\phi_p}{2\pi}\frac{dN}{dy p_T dp_T d\phi}
\label{iSS.eq4}
\end{equation}
while the anisotropic flow coefficients are computed from,
\begin{equation}
V_n \equiv v_n e^{i n \Psi_n} = \frac{\int p_T dp_T d\phi_p e^{i n\phi_p} dN/(dy p_T dp_T d\phi_p)}{\int p_T dp_T d\phi_p dN/(dy p_T dp_T d\phi_p)},
\label{iSS.eq5}
\end{equation}
\begin{equation}
V_n(p_T) \equiv v_n(p_T) e^{i n \Psi_n} = \frac{\int d\phi_p e^{i n\phi_p} dN/(dy p_T dp_T d\phi_p)}{\int d\phi_p dN/(dy p_T dp_T d\phi_p)}.
\label{iSS.eq5b}
\end{equation}
To optimize the efficiency of the numerical calculations, gaussian quadrature points are used for the variables $p_T, \phi_p$, and $\eta_s$. Further optimization for performing the numerical integral in Eq. (\ref{iSS.eq3}) involves adjusting the order of the integration loops, using local variables, pre-tabulating mathematical functions, etc. The resulting code {\tt iS} is $\sim$ 7 times faster compared to its ancestor {\tt AZSpectra} \cite{Kolb:2003dz}. 

\subsection{Methodology for particle sampling}

The particle emission function from the Cooper-Frye formula Eq.(\ref{iSS.eq1}) can be regarded as the probability of emitting particle from a given freeze-out fluid cell with specified momentum. The program {\tt iSS} uses this probability to generate sets of momenta and positions for actual particles emitted at the end of the hydrodynamic simulation. This information is then used as input for the following microscopic hadron cascade simulation. In the sampling procedure, we employ two well-known statistical sampling methods, the inverse cumulative distribution function (CDF) method and the direct probability distribution function (PDF) method, the latter is also known as the acceptance and rejection method.
 
\subsubsection{Purely numerical approach}

The straightforward (although not necessarily the fastest) approach is to compute all the required quantities numerically.

For a given particle species, the average total number of particles per unit rapidity, $dN/dy$, is calculated by numerically integrating Eq. (\ref{iSS.eq1}) over all freeze-out fluid cells and all particle transverse momenta $\vec{p}_T$. During the numerical integration, an inverse CDF can be built up  with negligible numerical cost for latter efficient sampling. However, in practice, the inverse CDF for a full set of spatial and momentum variables is memory demanding. In order to sample such a multi-dimensional probability distribution function, we divide the random variables into two groups and use efficient specific sampling methods to handle each of them. It is natural to group the spatial information $(\tau, \vec{x}_\perp, \eta_s)$ for the sampled particles into one set of random variables, and their momenta $(p_\perp, \phi_p, y)$ into the other. Dividing the random variables into two groups allows us to perform the sampling in different order and with different methods. 

One way to proceed is to first sample the spatial information, $(\tau, \vec{x}_\perp, \eta_s)$, using the inverse CDF method. Along with calculating the particle yield $dN/dy$, (see above) we build up the inverse CDF for the particle's spatial variables, $(\tau, \vec{x}_\perp, \eta_s)$, by integrating Eq. (\ref{iSS.eq1}) over the transverse momentum, $(p_\perp, \phi_p)$. For a collision event at top RHIC energy, the typical size of the array to store the inverse CDF is about 30,000 freeze-out fluid cells in the transverse plane times 40 points along the $\eta_s$ direction. Once we have the particles' spatial information, we can evaluate Eq. (\ref{iSS.eq1}) at any given point $(\tau, \vec{x}_\perp, \eta_s)$ for the particle's probability distribution in momentum space. To sample the particle's transverse momentum $(p_T, \phi_p)$ from this distribution we use the  the direct PDF method. In the end, since we assume longitudinal boost-invariance, the particle's rapidity can be sampled uniformly within given rapidity range. By sampling particles in this order, we optimize the sampling of the particle's spatial coordinates since the inverse CDF method has zero rejection rate. The direct PDF method used in momentum space, on the other hand, allows us to use continuous random variables for $p_\perp$ and $\phi_p$ instead of sampling them at some discrete lattice points. 

A second way to proceed is to first sample the particle's momentum information with the inverse CDF method. To this end we first build the inverse CDF for the particle's momentum variables, $(p_\perp, \phi_p)$. Using 15 points in $p_\perp$ and 48 points in $\phi_p$. Once we have $(p_\perp, \phi_p)$, Eq.(\ref{iSS.eq1}) is used as a probability distribution for the particle's spatial coordinates $(\tau, \vec{x}_\perp, \eta_s)$ which is then sampled with the direct PDF method.

\subsubsection{Semi-analytic approach}

In a given collision event the number of particles of species $a$ being emitted from a given fluid cell at $x^\mu$ can be calculated analytically as follows:
\begin{equation}
\Delta N_a (\tau_f, \vec{x}_\perp, \eta_s) = \frac{g_a}{(2\pi)^3} \Delta^3 \sigma_\mu \int \frac{d^3 p}{E} p^\mu (f_0(p) + \delta f(p)).
\label{iSS.eq6}
\end{equation}
Here the surface element of the given fluid cell is $\Delta^3 \sigma_\mu = \sigma_\mu \Delta^2 x_\perp \tau \Delta \eta_s$ with $\sigma_\mu = (\cosh \eta_s$, $ -\partial \tau/\partial x, -\partial \tau/\partial y, -\sinh \eta_s)$. 
The off-equilibrium correction $\delta f$ originating from the shear stress tensor does not contribute to the total particle yield, due to the properties that $\pi^{\mu\nu}$ is traceless and orthogonal to the flow velocity.
\begin{equation}
\int \frac{d^3 p}{E} p^\mu \delta f(p) = \int \frac{d^3 p}{E} p^\mu f_0(p)(1 \pm f_0(p)) \frac{\pi_{\alpha\beta}\hat{p}^\alpha \hat{p}^\beta}{2(e+P)} \chi \left(\frac{p}{T}\right) = A u^\mu,
\label{iSS.eq7}
\end{equation}
where $A = u_\mu \int \frac{d^3 p}{E} p^\mu \delta f(p) = \frac{\pi_{\alpha \beta}}{2(e+P)} \int \frac{d^3 p}{E} (u \cdot p) \frac{p^\alpha p^\beta}{(u \cdot p)^2} \chi(\frac{p}{T}) f_0(p) ( 1\pm f_0(p)).$
In the local rest frame of the fluid cell, it is easy to see that the integrand is proportional to $\delta^{\alpha \beta}$, hence
\begin{equation}
A = \int d^3 p f_0(p)(1 \pm f_0(p)) \frac{p^2}{3E^2} \frac{\pi^\alpha\,_\alpha}{2(e+P)} \chi \left(\frac{p}{T}\right) = 0.
\label{iSS.eq8}
\end{equation}
Thus the particle yield is totally determined by its equilibrium distribution, 
\begin{eqnarray}
\Delta N_a (\tau_f, \vec{x}_\perp, \eta_s) &=& \frac{g_a}{(2 \pi)^3} \Delta^3 \sigma_\mu \int \frac{d^3 p}{E} p^\mu f_0(p) \notag \\
&=& \frac{g_a}{(2 \pi)^3} \Delta^3 \sigma_\mu u^\mu \int p^2 dp \, d\phi\, d\cos\theta \frac{1}{e^{\beta (E - \mu_a)} \pm 1} \notag \\
&=& \frac{g_a}{2 \pi^2} \Delta^3 \sigma_\mu u^\mu \frac{m_a^2}{\beta} \sum_{n=1}^{\infty}\frac{(\mp 1)^{n-1}}{n} e^{n \beta \mu_a} K_2(n\beta m_a).
\label{iSS.eq9}
\end{eqnarray}

With the assumption of boost invariance, the particle's rapidity $y$ and its space-time rapidity $\eta_s$ only enters in the combination $y-\eta_s$, and therefore $\frac{dN}{d\eta_s} = \frac{dN}{dy}$. This leads to the following relation:
\begin{eqnarray}
\Delta N_a (\tau_f, \vec{x}_\perp, \eta_s) &=& \frac{g_a}{(2\pi)^3} \Delta^3 \sigma_\mu \int d y \int d^2 p_\perp p^\mu f_0(p) \notag \\
&=& \frac{g_a}{(2\pi)^3} \int d y \Delta^2 x_\perp \tau \Delta \eta_s \int d^2 p_\perp (m_\perp \cosh(y-\eta_s) - \vec{p}_\perp \cdot \vec{\nabla}_\perp \tau) f_0(p) \notag \\
&=& \Delta \eta_s \frac{g_a}{(2\pi)^3} \int \tau d \tilde{y} \Delta^2 x_\perp \int d^2 p_\perp (m_\perp \cosh(\tilde{y}) - \vec{p}_\perp \cdot \vec{\nabla}_\perp \tau) f_0(p)
\end{eqnarray}
This integral is independent of $\eta_s$, so
\begin{equation}
\Delta N_a (\tau_f, \vec{x}_\perp, \eta_s) = \Delta \eta_s \frac{\Delta N}{\Delta y} (\tau, \vec{x}_\perp). 
\label{iSS.eq10}
\end{equation}
In the numerical sampling procedure, we first consider all freeze-out fluid cells $(\tau, \vec{x}_\perp)$ in the transverse plane and use (\ref{iSS.eq9}) (together with (\ref{iSS.eq10})) to compute the total particle yield per unit rapidity for particle species $a$, $\Delta N/\Delta y$, for each cell. If freeze-out occurs on a surface of constant inverse temperature $\beta$ and chemical potential $\mu_a$, as will be the assumed in the rest of this thesis, Eq. (\ref{iSS.eq9}) can be written as,
\begin{equation}
\Delta N_a  (\tau_f, \vec{x}_\perp, \eta_s) = n_a u_\mu (\tau_f, \vec{x}_\perp, \eta_s) \Delta^3\sigma_\mu(\tau_f, \vec{x}_\perp, \eta_s). 
\end{equation}
where $n_a = \frac{g_a}{2\pi^2} \frac{m_a^2}{\beta} \sum_{n=1}^\infty e^{n \beta \mu_a} K_2(n \beta m_a)$ is the freeze-out density of particle species $a$, which is the same for all freeze-out cells. In this case, $\Delta N_a/\Delta y$ depends on the position of the fluid cell only through its freeze-out volume,
\begin{equation}
\Delta V (\tau_f, \vec{x}_\perp, \eta_s) = u^\mu \Delta^3 \sigma_\mu (\tau_f, \vec{x}_\perp, \eta_s).
\end{equation}
The we use $(\Delta N_a/\Delta y) (\tau_f, \vec{x}_\perp) $ to build up an inverse CDF for the spatial variables  $(\tau_f, \vec{x}_\perp)$. Their sum over all  $(\tau_f, \vec{x}_\perp)$  points gives the total rapidity density $\frac{dN_a}{dy}$ of particle species $a$ in a given collision event. The constructed inverse CDF is then used to sample the positions  $(\tau_f, \vec{x}_\perp)$ of the particles of species $a$. Finally, we use the Cooper-Frye formula Eq. (\ref{iSS.eq1}) at these sampled positions $ (\tau_f, \vec{x}_\perp)$ as the relative probability distribution for sampling the particle's momentum $(p_T, \phi_p, y-\eta_s)$ using the direct PDF method:
\begin{eqnarray}
P (p_\perp, \phi_p, y-\eta_s;\tau_f, \vec{x}_\perp) &=& \frac{g_i}{(2\pi)^3} \Delta^3 \sigma_\mu p^\mu (f_0(p) + \delta f(p)) \notag \\
&=& \frac{g_i}{(2\pi)^3} \tau_f \Delta^2 x_\perp \Delta \eta_s (m_\perp \cosh(y-\eta_s) - \vec{p}_\perp \cdot \vec{\nabla}_\perp \tau)  \notag \\
&& \times (f_0(p) + \delta f(p)).
\label{iSS.eq11}
\end{eqnarray}
Having obtained $(y - \eta_s)$ by sampling Eq. (\ref{iSS.eq11}), we use boost-invariance and sample $y$ uniformly from a given range specified by the user (e.g., -4 to 4) and then obtain $\eta_s$ from the previously determined $y - \eta_s$. 

Since for every $(\tau, \vec{x}_\perp)$, the probability Eq. (\ref{iSS.eq11}) for $(p_T, \phi_p, y-\eta_s)$ is only sampled once, building an inverse CDF for Eq. (\ref{iSS.eq11}) would be excessively expensive, which is why we choose to use the direct PDF method to sample $(p_\perp, \phi_p, y-\eta_s)$. However, the direct PDF method requires one to estimate the maximum value of the probability distribution function given in Eq. (\ref{iSS.eq11}) which is closely related to the function
\begin{equation} 
  G(E; A) = \frac{E^A}{e^{\beta(E-\mu)} \pm 1},\,A>0.
\label{iSS.eq12}
\end{equation}
By setting its derivative to zero, the extrema can be found by solving
\begin{equation} 
  (1 \mp f_0) = \frac{A}{\beta E} \Longleftrightarrow
  \left\{
  \begin{aligned}
    & x e^x = y;\,x=\beta E-A,\,y=A e^{\beta\mu-A}, \mbox{fermions (upper),} \\
    & x e^{-x} = y;\,x=A-\beta E,\,y=A e^{\beta\mu-A}, \mbox{bosons (lower).} \\
  \end{aligned}
  \right.
\label{iSS.eq13}
\end{equation}
This equation is transcendental and cannot be solved algebraically; however, the solutions to the equations $x e^{\pm x} = y$ in Eq. (\ref{iSS.eq13}) can be pre-calculated and tabulated. For fermions (upper sign), a solution always exists and it is expressed by the Lambert W-function; for bosons (lower sign) the equation  has real solutions only when $y<1/e$, and the it yields two solutions; the physical solution must satisfy $x\in[0,1]$. In the following, the solution to Eq. (\ref{iSS.eq13}) will be denoted as $E_\mathrm{max}^\pm$ when it exists.

The maximum of $G(E;A)$ with constraint $E\geq m$ will be denoted as $G_\mathrm{max}^{(A)}$. It depends on several conditions:
\begin{enumerate}
  \item For fermions (upper sign), $G(E)$ has a single peak at $E_\mathrm{max}^+$ and the constraint maximum is taken as $G(E_\mathrm{max}^+)$ if $E_\mathrm{max}^+>m$ and as $G(m)$ otherwise.
  \item For bosons (lower sign) with $A e^{\beta\mu-A} > 1/e$, Eq. (\ref{iSS.eq13}) has no solution and the maximum takes $G(m)$.
  \item For bosons (lower sign) with $A e^{\beta\mu-A} \leq 1/e$, $G(E)$ has two extrema in $(\mu,\infty)$, with the larger one being the maximum and given by $E_\mathrm{max}^-$. If $E_\mathrm{max}^-<m$ then the maximum is taken as $G(m)$; otherwise the maximum is taken as the larger one of the two numbers $G(m)$ and $G(E_\mathrm{max}^-)$.
\end{enumerate}

In Eq. (\ref{iSS.eq11}), an upper limit for the factor $p^\mu \Delta^3 \sigma_\mu$ can obtained using  the H\"{o}lder inequality, 
\begin{equation}
p^\mu \Delta^3 \sigma_\mu = E \Delta^3 \sigma_0 + p^i \Delta \sigma_i \le (p\cdot u) (\vert \Delta^3 \sigma_\mu u^\mu \vert + \sqrt { \vert \Delta^3 \sigma_\mu \Delta^3 \sigma_\nu \Delta^{\mu\nu} \vert}).
\end{equation}
For the equilibrium contribution, it is clear that the remaining part is to calculate the maximum of the function
\begin{equation} 
  E f_0 = \frac{E}{e^{(E-\mu)/T} \pm 1} = G(E; 1);
  \label{iSS.eq14}
\end{equation}
the solution to this problem is $G_\mathrm{max}^{(1)}$.

For the off-equilibrium correction, it is convenient to estimate its maximum in the local rest frame of the fluid cell. We can further rotate the shear stress tensor in the transverse plane such that $\pi^{xy} = 0$. In such a coordinate system
\begin{eqnarray}
p^\mu p^\nu \pi_{\mu\nu} &=& (p^x)^2 \pi_{xx} + (p^y)^2 \pi_{yy} + (p^z)^2 \pi_{zz} \le E(\vert p^x \pi_{xx} \vert + \vert p^y \pi_{yy} \vert + \vert p^z \pi_{zz} \vert) \notag \\
&\le& E^2 \sqrt{\pi_{xx}^2 + \pi_{yy}^2 + \pi_{zz}^2} = E^2 \sqrt{\pi^{\mu\nu}\pi_{\mu\nu}}.
\label{iSS.eq15}
\end{eqnarray}
In the last step, we rewrote the expression again in Lorentz invariant form such that it is now valid in any frame. With the form of $\delta f$ in Eq. (\ref{iSS.eq2}) and assuming $f_0 < 1$,
\begin{equation}
E^{\alpha + 1} f_0 (1 \mp f_0) \le \lambda E^{\alpha + 1} f_0  = \lambda G(E; \alpha + 1) \le \lambda G^{(\alpha + 1)}_\mathrm{max},
\label{iSS.eq16}
\end{equation}
where $\lambda = 1$ for fermions and $\lambda = 2$ for bosons.
To summarize, the maximum of the PDF Eq. (\ref{iSS.eq11}) for $(p_\perp, \phi_p, y-\eta_s)$ can be estimated as
\begin{equation}
P \le P_\mathrm{max} = \frac{g_a}{(2\pi)^3} \tau \left(\vert \Delta^3 \sigma_\mu u^\mu \vert + \sqrt {\Delta^3 \sigma_\mu \Delta^3 \sigma_\nu \Delta^{\mu\nu}}\right)\left(G^{(1)}_\mathrm{max} + \frac{\sqrt{\pi^{\mu\nu} \pi_{\mu\nu}}}{2(e+P)T^\alpha} \lambda G^{(\alpha + 1)}_\mathrm{max} \right).
\label{iSS.eq17}
\end{equation}

For light mesons, the validity of the assumption $f_0 < 1$ depends on the value of the freeze-out temperature and chemical potential. Especially, kinetic freeze-out at temperature much below the chemical decoupling temperature can lead to large non-equilibrium chemical potentials that can cause this assumption to break down in some of the fluid cells. We found that $f_0 < 1$ almost all the time, although there were some instances where it was violated. If a more rigorous result is desired, the inequality (\ref{iSS.eq16}) can be replaced by the following one:
\begin{eqnarray}
E^{\alpha + 1} f_0 (1 + f_0) &\le& \vert E^{\alpha + 1} f_0 \vert + \vert E^\gamma f_0 \vert \vert E^{\alpha + 1 - \gamma} f_0 \vert \le G^{(\alpha + 1)}_\mathrm{max} + G^{(\gamma)}_\mathrm{max}G^{(\alpha + 1 - \gamma)}_\mathrm{max},
\label{iSS.eq18}
\end{eqnarray}
where $0 \le \gamma \le \alpha + 1$. 

\subsubsection{The negative probability issue}

For hyper-surface of constant temperature, the Cooper-Frye formula in Eq. (\ref{iSS.eq1}) is not positive semi-definite. This is because on an isothermal hyper-surface $\Sigma$, $d^3 \sigma_\mu$ can be a space-like vector. So $p^\mu d^3 \sigma_\mu$ can be negative in certain regions. Physically, such regions represent parts of the switching surface through which more particles are flying into the fireball instead of being emitted.  These negative contributions to the Cooper-Frye integral are essential to ensure the conservation of energy across the hyper-surface. However, they become problematic when one wants to use Eq. (\ref{iSS.eq1}) as a probability distribution (which should always be positive). In the practical sampling procedure, we insert a $\theta$-function by hand to enforce positivity of the probability distribution function. Since we group the random variables differently in the different sampling approaches discussed above, insertion of the $\theta$-function will be done slightly differently in each case, with different consequences. In each case, a slight violation of energy-momentum conservation will occur. Let us therefore explore the implications of the $\theta$-function in some detail, we first sample particle's spatial information using the purely numerical approach, we use a $\theta$-function $\theta(u^\mu d\sigma_\mu)$ to enforce positivity of the $p_\perp$-integrated distribution function. This means that none of our sampled particles will come from the spatial regions where $u^\mu d\sigma_\mu < 0$. In the second step, when sampling the momenta we enforce the positivity of Eq. (\ref{iSS.eq1}) at the already sampled spatial coordinates. In this step, there are two possible quantities that can become negative. First, $p^\mu \sigma_\mu$ may be negative for some values of $p^\mu$. This represents the situation where a net number of particles with momentum $p^\mu$ flies into the fireball. Secondly, in the viscous case, when the off-equilibrium correction $\delta f$ becomes large, it may overwhelm the equilibrium term and turn the entire distribution function to negative. This situation represents a breakdown of the Chapman-Enskog expansion keeping only terms of first order in $\delta f$, Eq. (\ref{iSS.eq1}) should not be trusted in such regions of momentum space. With $\eta/s = 0.20$, we find that this problem usually occurs at high $p_T > 2.5$\,GeV. In the sampling procedure, we enforce both terms to be always positive, by inserting a product of theta functions, $\theta(f_0 + \delta f) \theta(p^\mu d^3 \sigma_\mu)$. $\theta(f_0 + \delta f)$ should be always kept in the calculation, even for the analytic results. The second factor $\theta(p^\mu d\sigma_\mu)$  causes a deviation of the sampled momentum distribution from the analytical result which will be studied below.

If we first sample the particle's momentum information, we enforce positivity of the momentum distribution $dN/(dy p_\perp dp_\perp d\phi_p) \ge 0$. In most cases, the total number of emitted particles with given transverse momentum $\vec{p}_\perp$, integrated over the entire freeze-out surface, is positive. The positivity constraint on $dN/(dy p_T dp_T d\phi_p)$ therefore has almost no effect at all. The set of momentum configurations obtained from this sampling procedure will reproduce momentum distributions and flow coefficients that agree most closely with the analytical Cooper-Frye formalism. In the second step, when we then additionally sample particle's spatial information, we need to enforce positivity of Eq. (\ref{iSS.eq1}) at a given momentum $\vec{p}_\perp$. Regions on the hypersurface where Eq. (\ref{iSS.eq1}) is negative will thus not contribute to particle emission at that $\vec{p}_\perp$. The sampled spatial distribution will therefore show some deviation from the analytic result. 

For the semi-analytic approach, the situation is similar to the purely numerical approach when sampling the positions first and the momenta second. 

\subsubsection{Multiplicity fluctuations at freeze-out}

The Cooper-Frye formula only yields the average number of particles emitted from a given hydrodynamic event. Each sampling of the Cooper-Fyre formula will, however, result in a number of emitted particle that fluctuates around that mean value. In principle, these sampling fluctuations are constrained by energy-momentum, baryon number and charge conservation. However, exact implementation of these constraints is non-trivial and will have to be left for future studies. 

We use an approximation based on the following procedure: We compute the integer value of the number of particles of species as predicted by Cooper-Frye, sample such particles until that number is exhausted, and then use the non-integer part of the predicted number to uniformly sample for one additional particle. 
This sampling procedure introduces minimum fluctuations in the total number of particles. In the current version of {\tt iSS}, there are options for users to instead fluctuate the particle number according to Poisson or negative Binomial distributions. 

\subsubsection{Performance}
To demonstrate the performance of the {\tt iSS} algorithm, we use an event-averaged hydrodynamically evolved Pb+Pb profile at 20-30\% centrality at LHC energy to obtain a rough estimate for the average running time of our particle sampler. The two sampling approaches have their individual advantages and disadvantages in dealing with different sampling requirements. 

\begin{table*}[htdp]
 \begin{center}
  \begin{tabular}{c|c|c}
   \hline \hline
   $100$ repeated samplings & purly numerical approach & semi-analytic approach \\
   \hline
   determining particle & $21.84$s for $\pi^+$ & negligible for $\pi^+$ \\ 
   yield $dN/dy$ & $\sim\times 100$ for rest of particles & $0.01$s for all particles\\
   \hline
   actual sampling & $1.25$s for $\pi^0$ & $2.65$s for $\pi^+$ \\
   & faster for heavier particles & faster for heavier particles \\
   \hline
   total & $2463.75$s & $15.16$s \\
   \hline \hline
   $50,000$ sampling & $88.73$s & $1327.99$s \\
    ($\pi^+$ only) & & \\
   \hline \hline
  \end{tabular}
 \end{center}
\caption{Efficiency comparison between pure numerical and semi-analytic methods. The test case has $32869$ conversion surface cells in the transverse plane and $dN/dy|_{\pi^+}\sim 144$. The test is done on a single core personal laptop.}
\label{iSS_performance}
\end{table*}

The numerical performance of the code is summarized in Table. \ref{iSS_performance}. The tests are done with the Intel C++ compiler with {\tt -O3} optimization. Our code runs about a factor of 6 faster with the Intel compiler compared to the GNU compiler (g++). 

The purely numerical approach is most suitable when a large number of repeated samplings of a single hydrodynamical event is desired. This is essential if one wants to study with good statistical precision rare multi-strange hadrons very few of which are emitted in a single event. On the other hand, the semi-analytic approach is extremely fast for small numbers of repeated samplings. This large gain in the numerical efficiency is due to the fact that it determines the particle yields analytically using, Eq. (\ref{iSS.eq9}). The drawback is that in this approach we need to sample one additional dimension (the rapidity direction) using the direct PDF method, which reduces the total sampling efficiency per simulation cycle.  

\subsection{Code verification}

In this section, we show some test results from our particle sampler. 

%=======================================
\begin{figure*}[h!]
  \centering
  \includegraphics[width=1.0\linewidth]{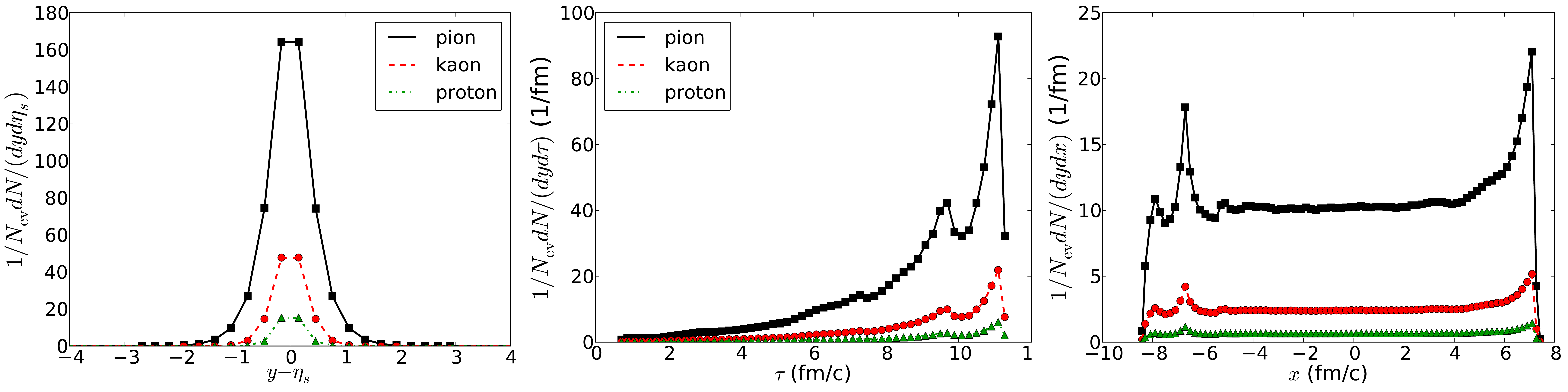} 
%  \begin{tabular}{ccc}
%  \includegraphics[width=0.3\linewidth]{figs/iSS_BenchMark_method_1_dNdeta.pdf} &
%  \includegraphics[width=0.3\linewidth]{figs/iSS_BenchMark_method_1_dNdtau.pdf} &
%  \includegraphics[width=0.3\linewidth]{figs/iSS_BenchMark_method_1_dNdx.pdf}
%  \end{tabular}
  \caption{For sampling method I in the purely numerical approach, the spatial distributions of the sampled thermal particles (solid dots) are compared to the emission function calculated from the Cooper-Frye Formula (lines). The left panel shows the particle distribution along the $\eta_s$ direction. The middle panel is the time emission function of the particles and the right panel shows the particle distribution along the x-axis in the transverse plane. All results are from a single hydrodynamic event with bumpy initial conditions.}
  \label{iSS.fig1}
\end{figure*}
%=======================================

%=======================================
\begin{figure*}[h!]
  \centering
  \begin{tabular}{cc}
  \includegraphics[width=0.48\linewidth]{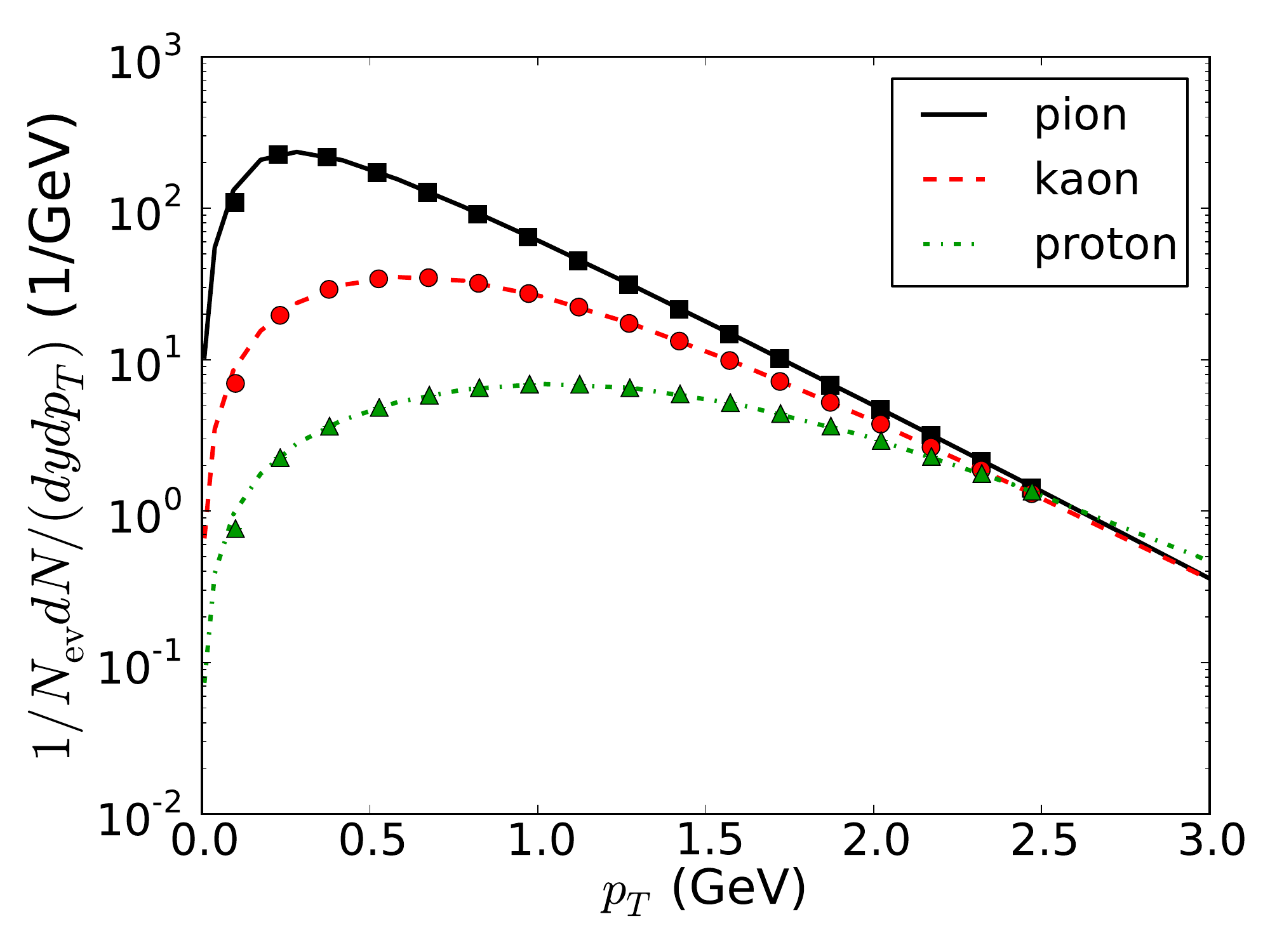} &
  \includegraphics[width=0.48\linewidth]{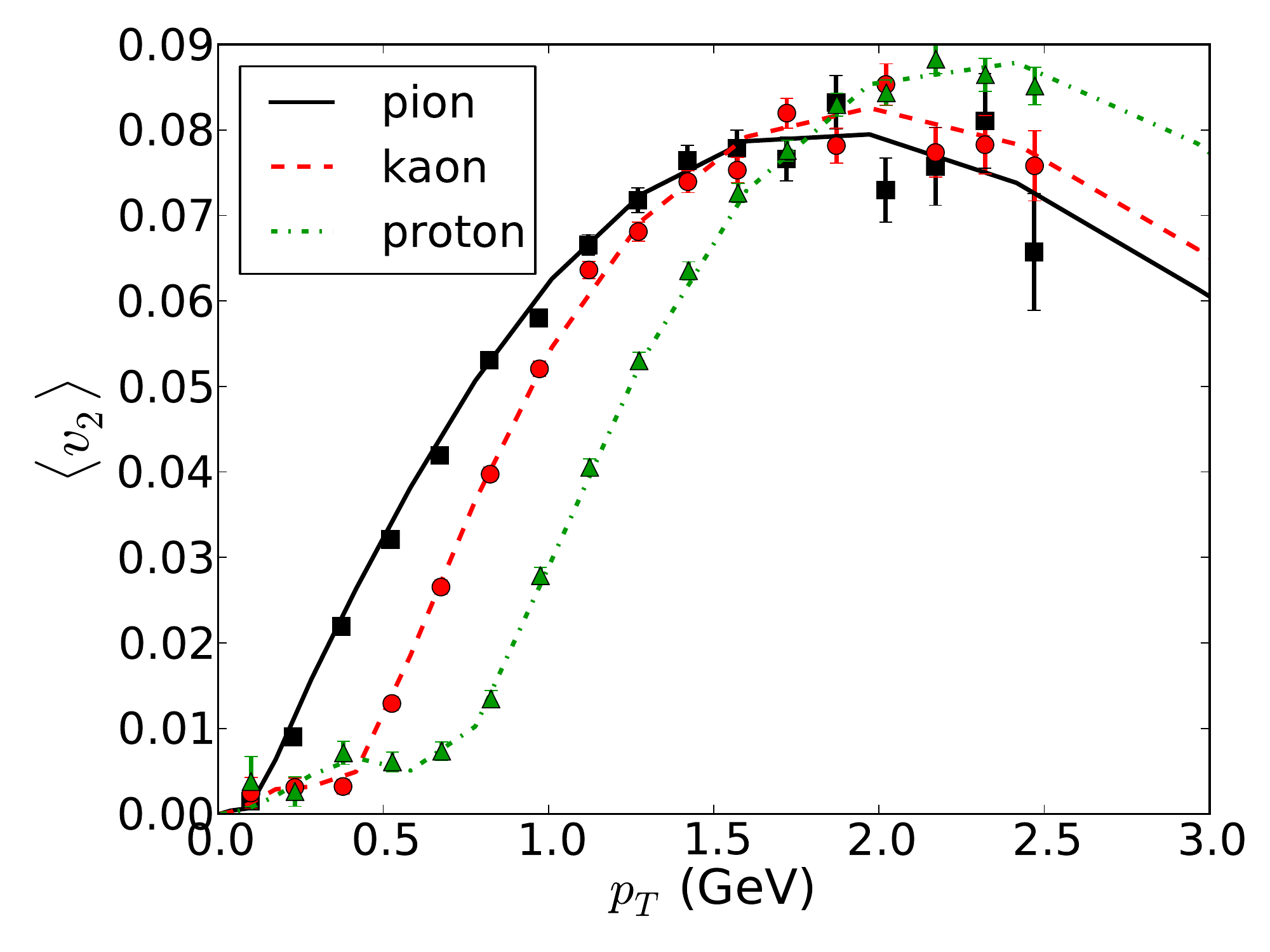} \\ 
  \includegraphics[width=0.48\linewidth]{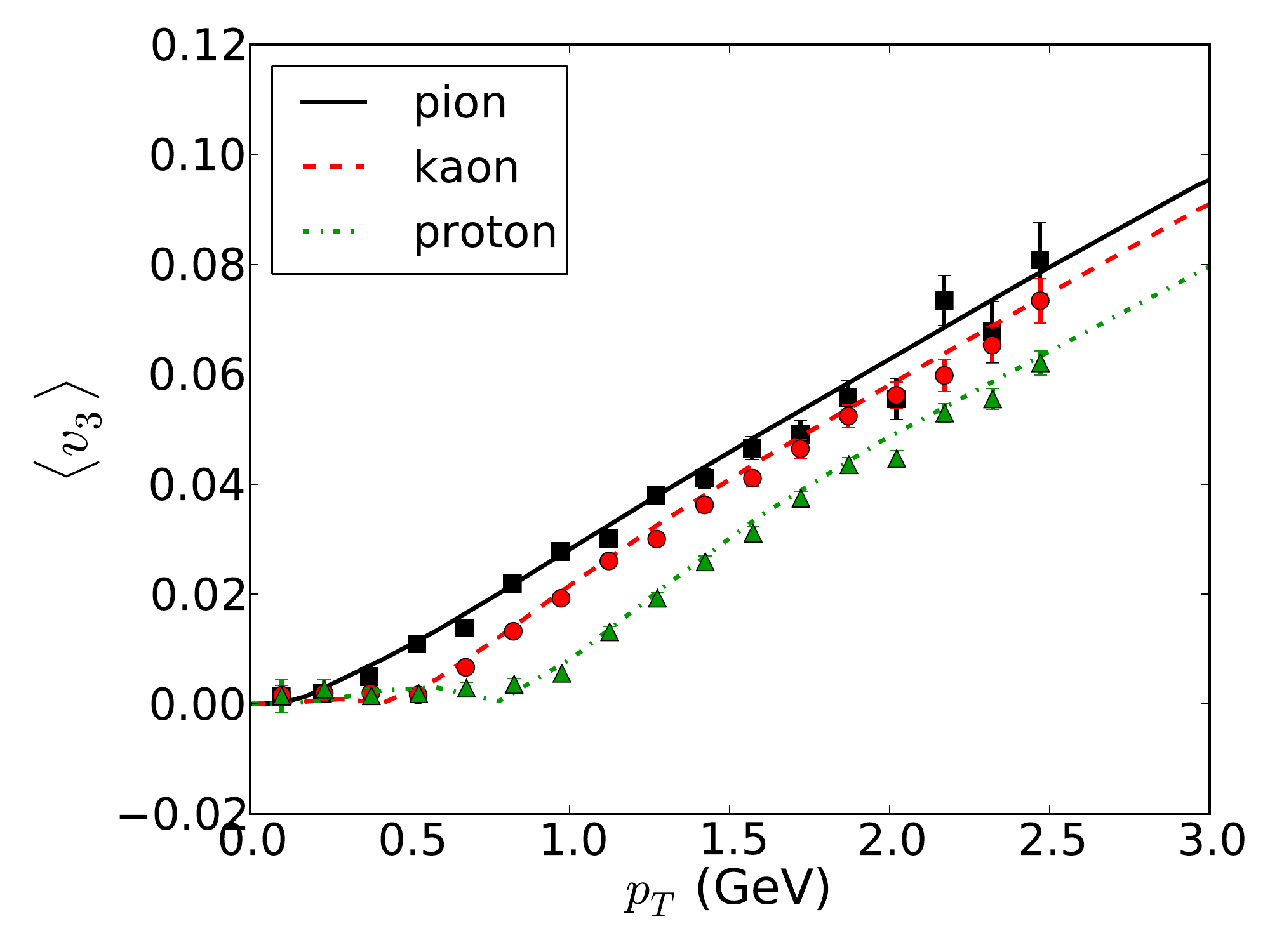} &
  \includegraphics[width=0.48\linewidth]{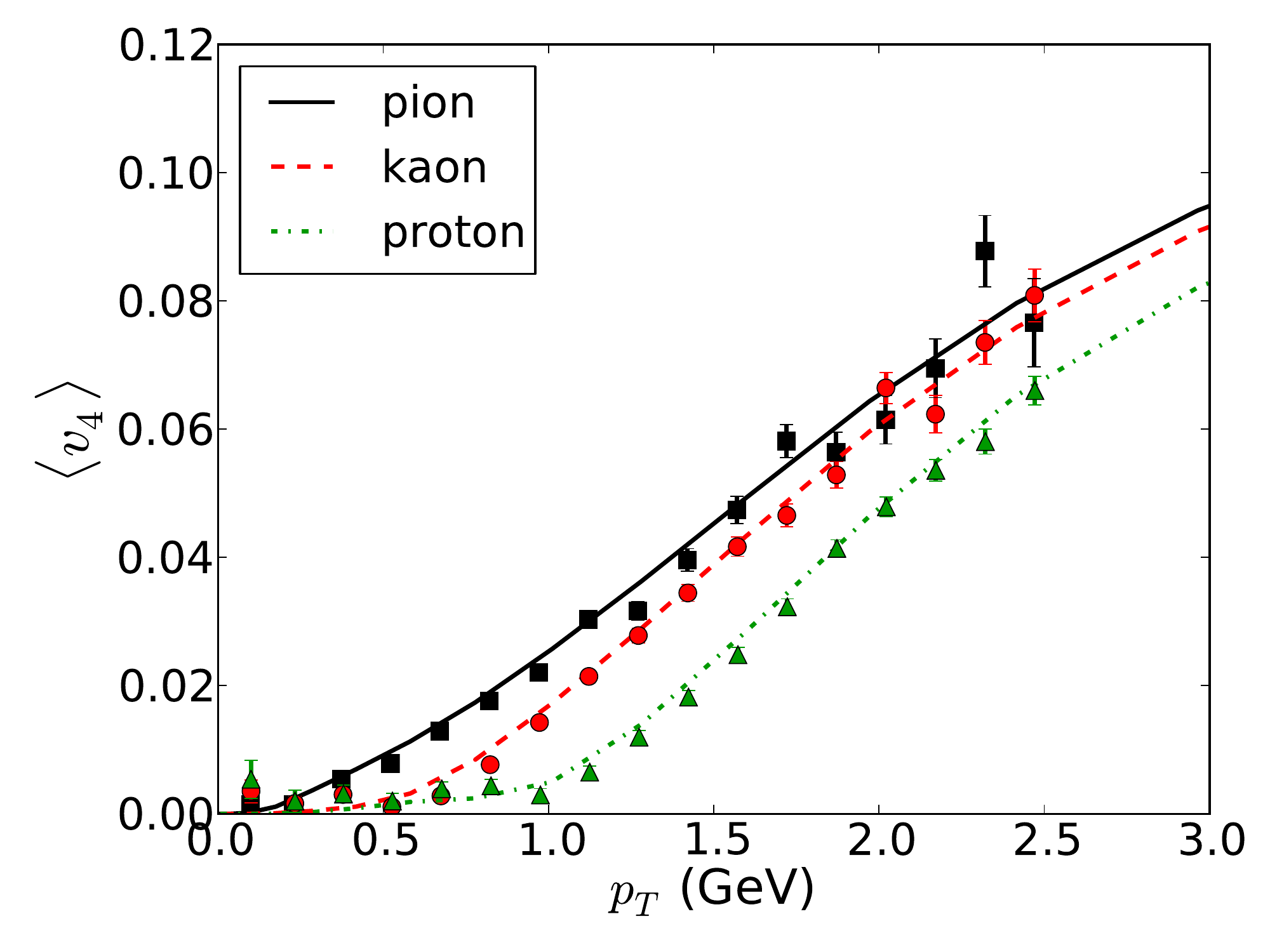}
  \end{tabular}
  \caption{Similar to Fig. \ref{iSS.fig1}, momentum distributions of the sampled particles are compared with results from the Cooper-Frye formula. Particles' $p_T$-differential spectra, $v_2$, $v_3$, and $v_4$ are presented. }
  \label{iSS.fig2}
\end{figure*}
%=======================================

In Figs. \ref{iSS.fig1} and \ref{iSS.fig2} we present the spatial and momentum distributions of thermally emitted particles (pions, kaons, and protons) and compare them against the emission function calculated directly from the Cooper-Frye Formula. We perform repeated samplings for a single hydrodynamic simulation with fluctuating initial conditions which emits about 172 positive pions, 40 positive kaons, and 11 protons. per unit rapidity thermally (i.e., not counting particles from resonance decays). To obtain sufficient statistics, we sample 50,000 events for thermal pions, 150,000 events for thermal kaons, and 500,000 events for thermal protons (these numbers account for the relative yields per event of their particle species.) Samples are generated using the purely numerical approach method I, which samples the spatial distributions first and then particle momenta. 

We find that the particle samples generated from this method reproduce very accurately the spatial distributions from the Cooper-Frye Formula. The regions where $u_\mu d^3 \sigma^\mu < 0$ do not affect the partially integrated emission functions shown in Fig. \ref{iSS.fig1}. In Fig. \ref{iSS.fig2}, we compare the particle momentum distributions against the results from the Cooper-Frye formula. We find very good agreement for the particle spectra as well as for the their anisotropy coefficients $v_2$, $v_3$, and $v_4$. 

%=======================================
\begin{figure*}[h!]
  \centering
  \includegraphics[width=1.0\linewidth]{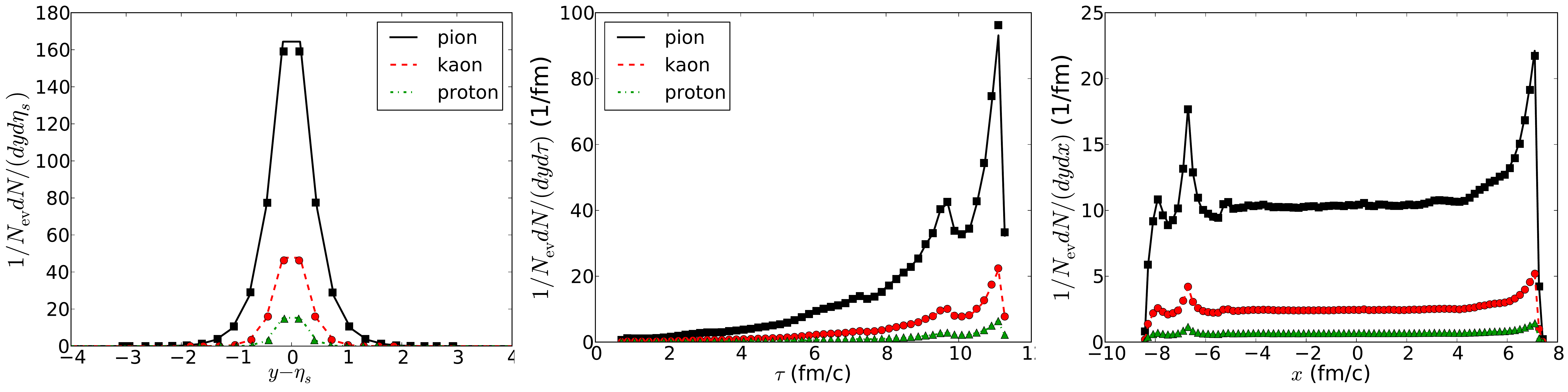} 
%  \begin{tabular}{ccc}
%  \includegraphics[width=0.3\linewidth]{figs/iSS_BenchMark_method_2_dNdeta.pdf} &
%  \includegraphics[width=0.3\linewidth]{figs/iSS_BenchMark_method_2_dNdtau.pdf} &
%  \includegraphics[width=0.3\linewidth]{figs/iSS_BenchMark_method_2_dNdx.pdf}
%  \end{tabular}
  \caption{Similar to Fig. \ref{iSS.fig1}, but for sampling method with the semi-analytic approach.}
  \label{iSS.fig3}
\end{figure*}
%=======================================

%=======================================
\begin{figure*}[h!]
  \centering
  \begin{tabular}{cc}
  \includegraphics[width=0.48\linewidth]{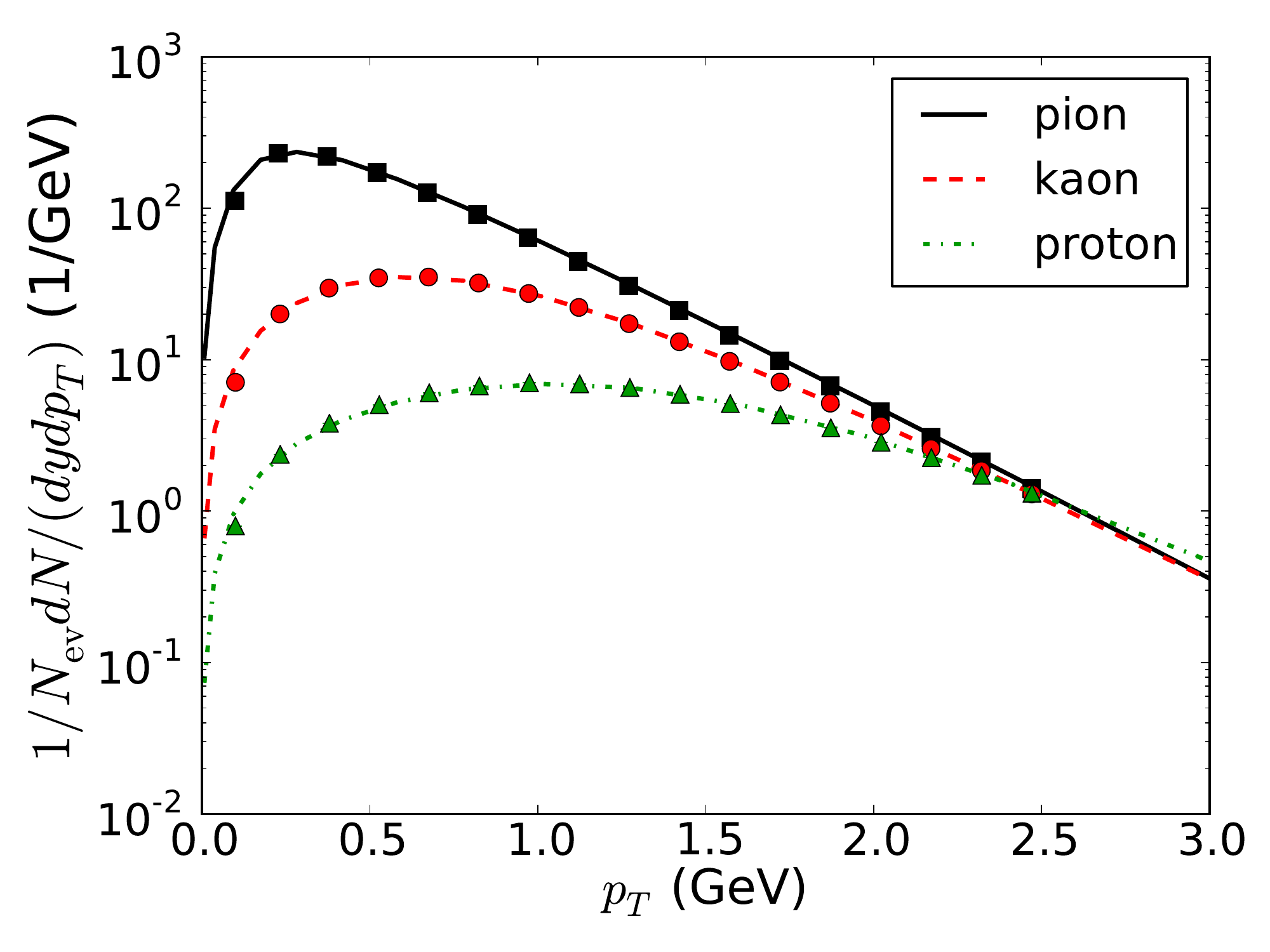} &
  \includegraphics[width=0.48\linewidth]{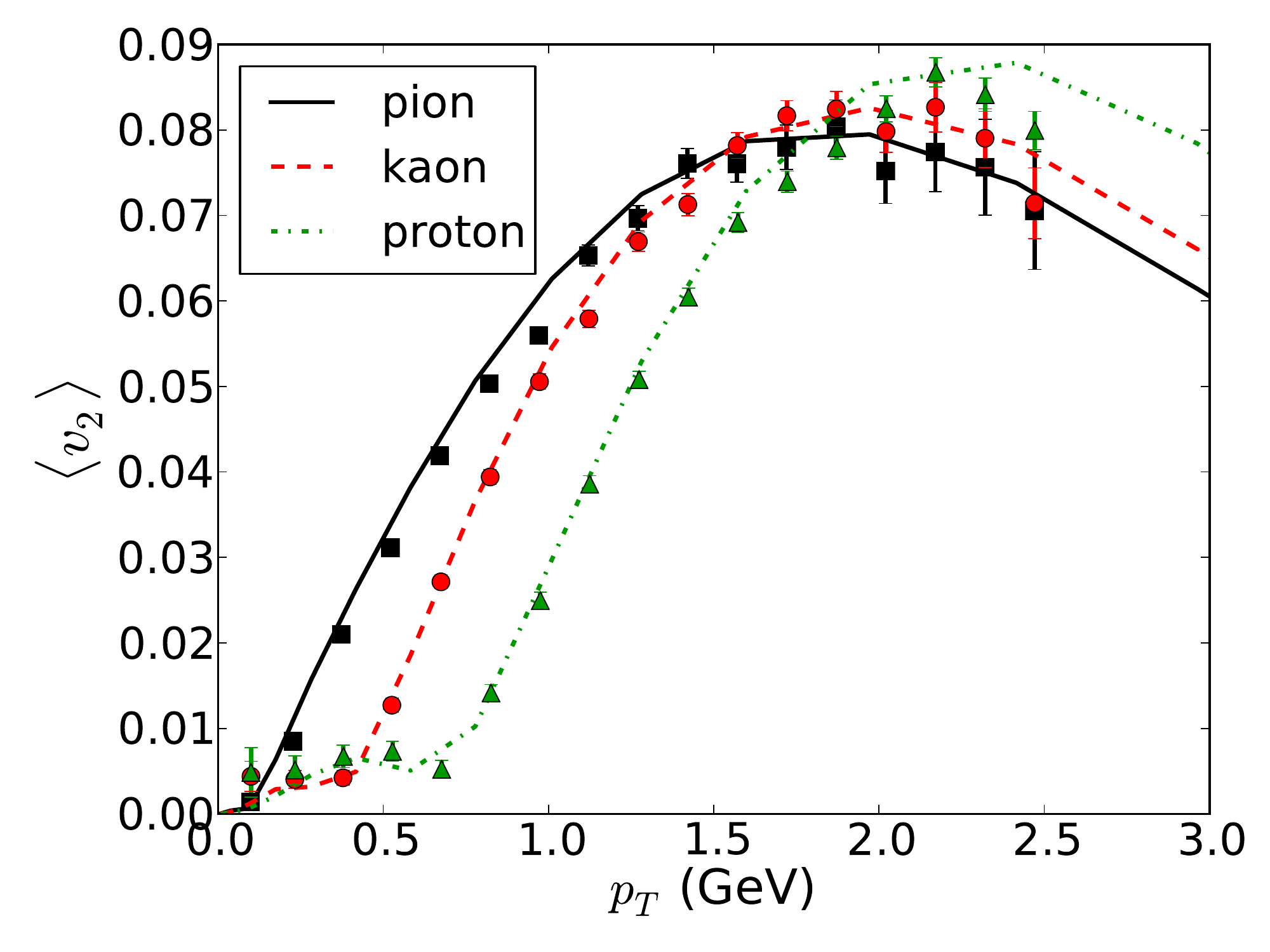} \\ 
  \includegraphics[width=0.48\linewidth]{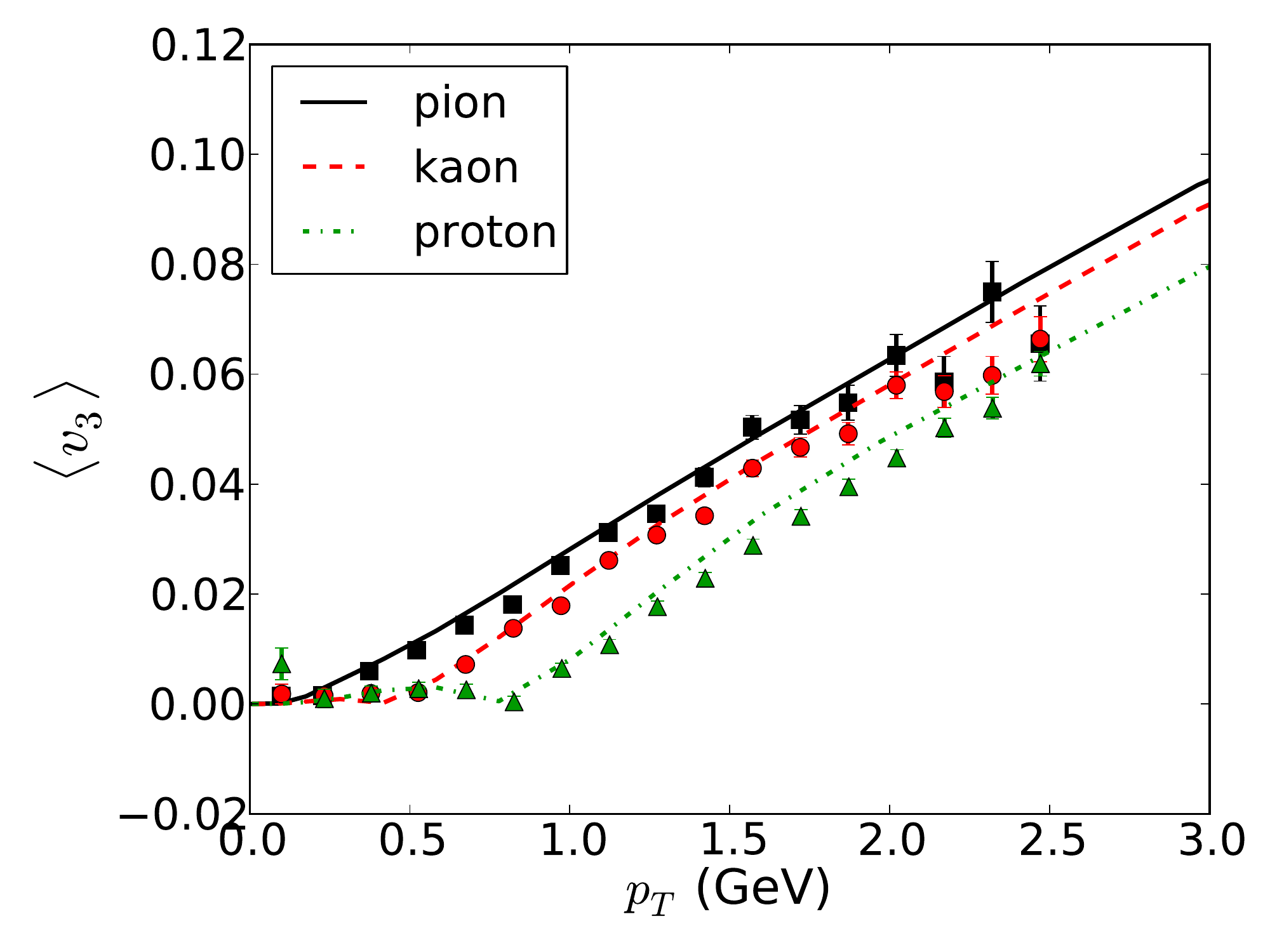} &
  \includegraphics[width=0.48\linewidth]{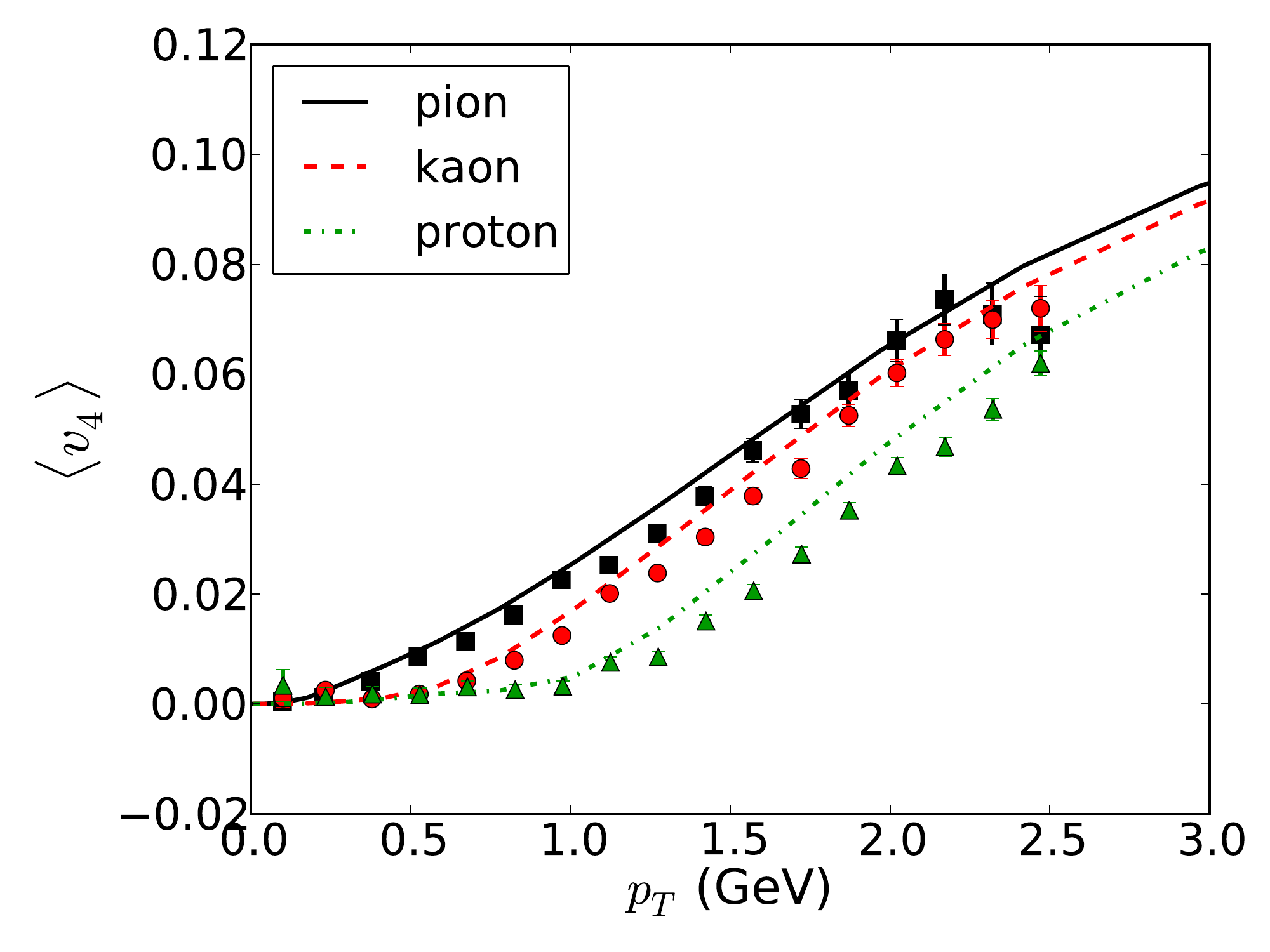}
  \end{tabular}
  \caption{Similar to Fig. \ref{iSS.fig2}, but for sampling with the semi-analytic approach.}
  \label{iSS.fig4}
\end{figure*}
%=======================================

In Figs. \ref{iSS.fig3} and \ref{iSS.fig4}, we show similar comparisons using the semi-analytic approach. Again the spatial distributions in the transverse plane from directly integrating the Cooper-Frye formula are very well reproduced but some slight differences are seen in the $\eta_s$ and $\tau$ distributions. Fig. \ref{iSS.fig4} shows that this method generates some noticeable disagreement in the higher order momentum anisotropies of the particle momentum distribution, caused by the particular way we remove negative contributions in this approach. 

%=======================================
\begin{figure*}[h!]
  \centering
  \includegraphics[width=1.0\linewidth]{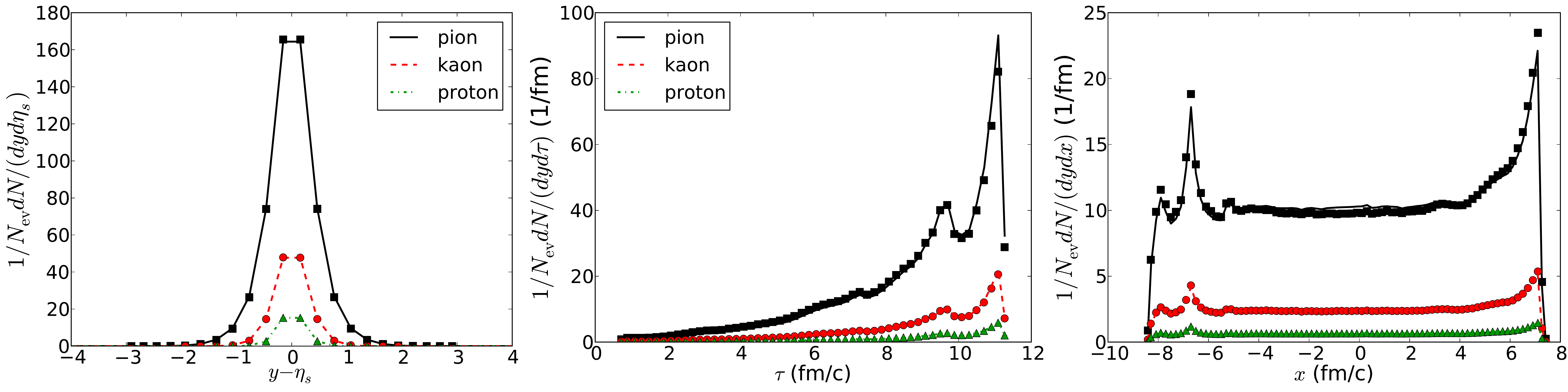} 
%  \begin{tabular}{ccc}
%  \includegraphics[width=0.3\linewidth]{figs/iSS_BenchMark_method_3_dNdeta.pdf} &
%  \includegraphics[width=0.3\linewidth]{figs/iSS_BenchMark_method_3_dNdtau.pdf} &
%  \includegraphics[width=0.3\linewidth]{figs/iSS_BenchMark_method_3_dNdx.pdf}
%  \end{tabular}
  \caption{Similar to Fig. \ref{iSS.fig1}, but for sampling method II of the purely numerical approach, which samples particle momenta first.}
  \label{iSS.fig5}
\end{figure*}
%=======================================

%=======================================
\begin{figure*}[h!]
  \centering
  \begin{tabular}{cc}
  \includegraphics[width=0.48\linewidth]{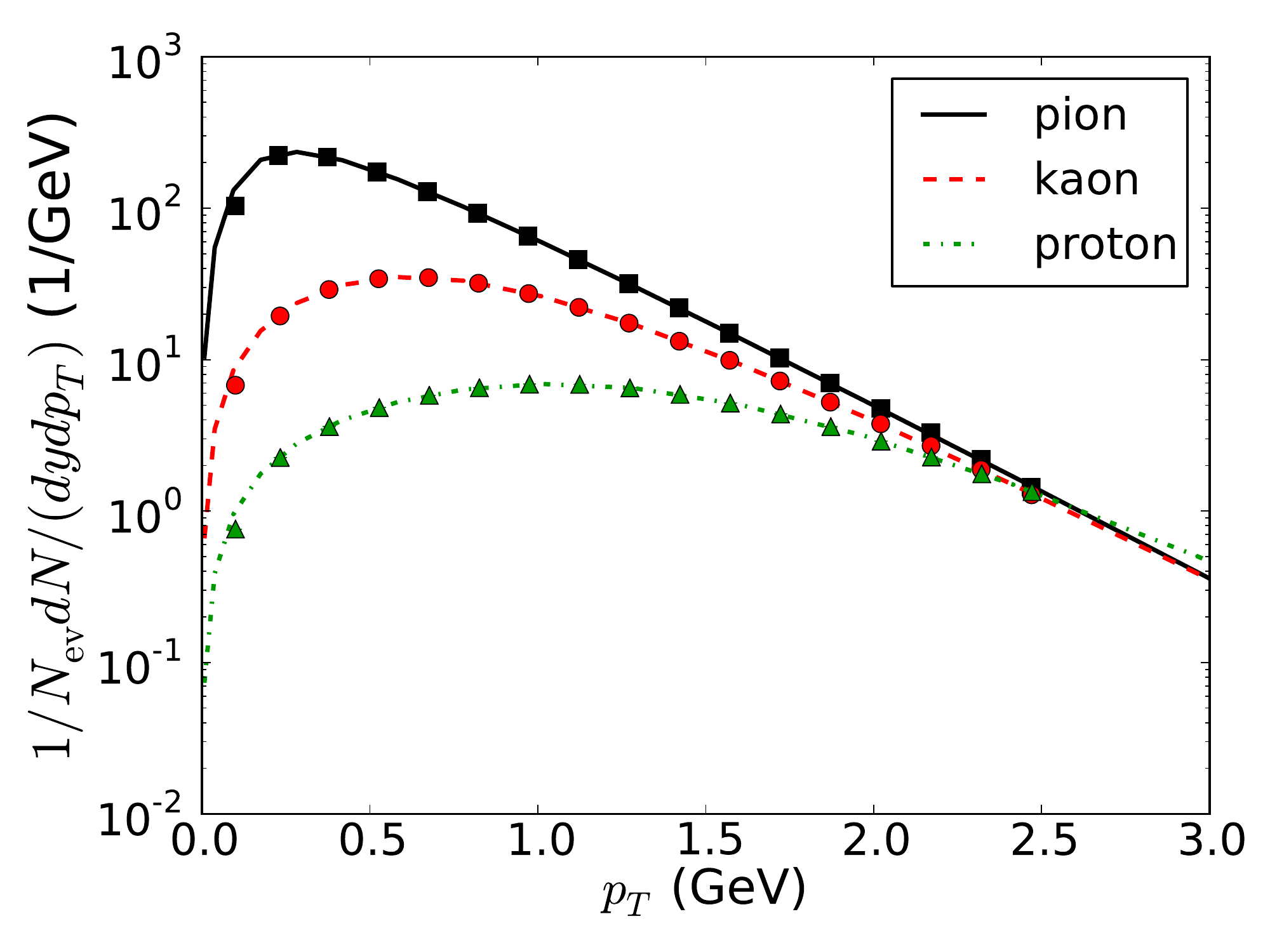} &
  \includegraphics[width=0.48\linewidth]{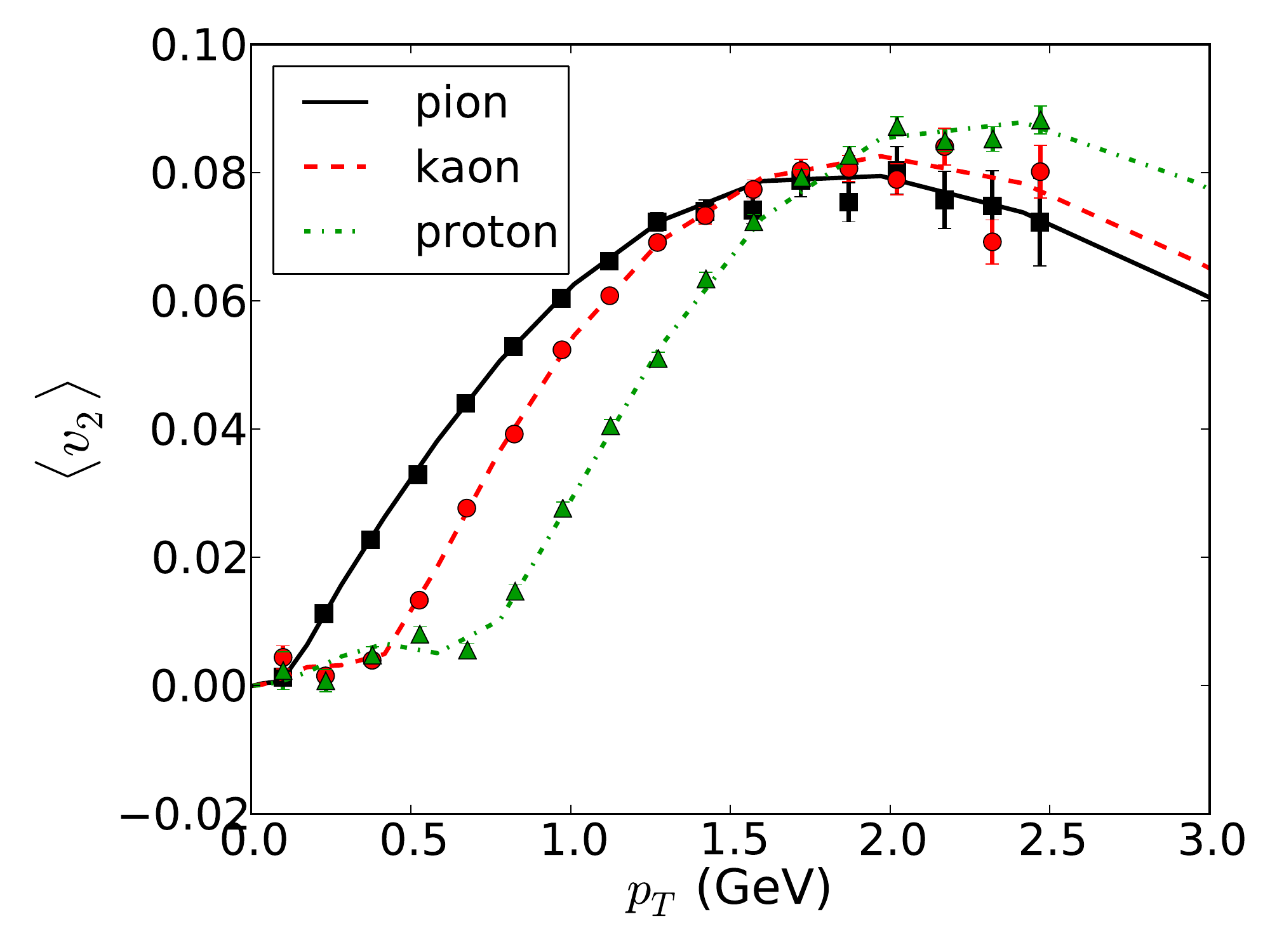} \\ 
  \includegraphics[width=0.48\linewidth]{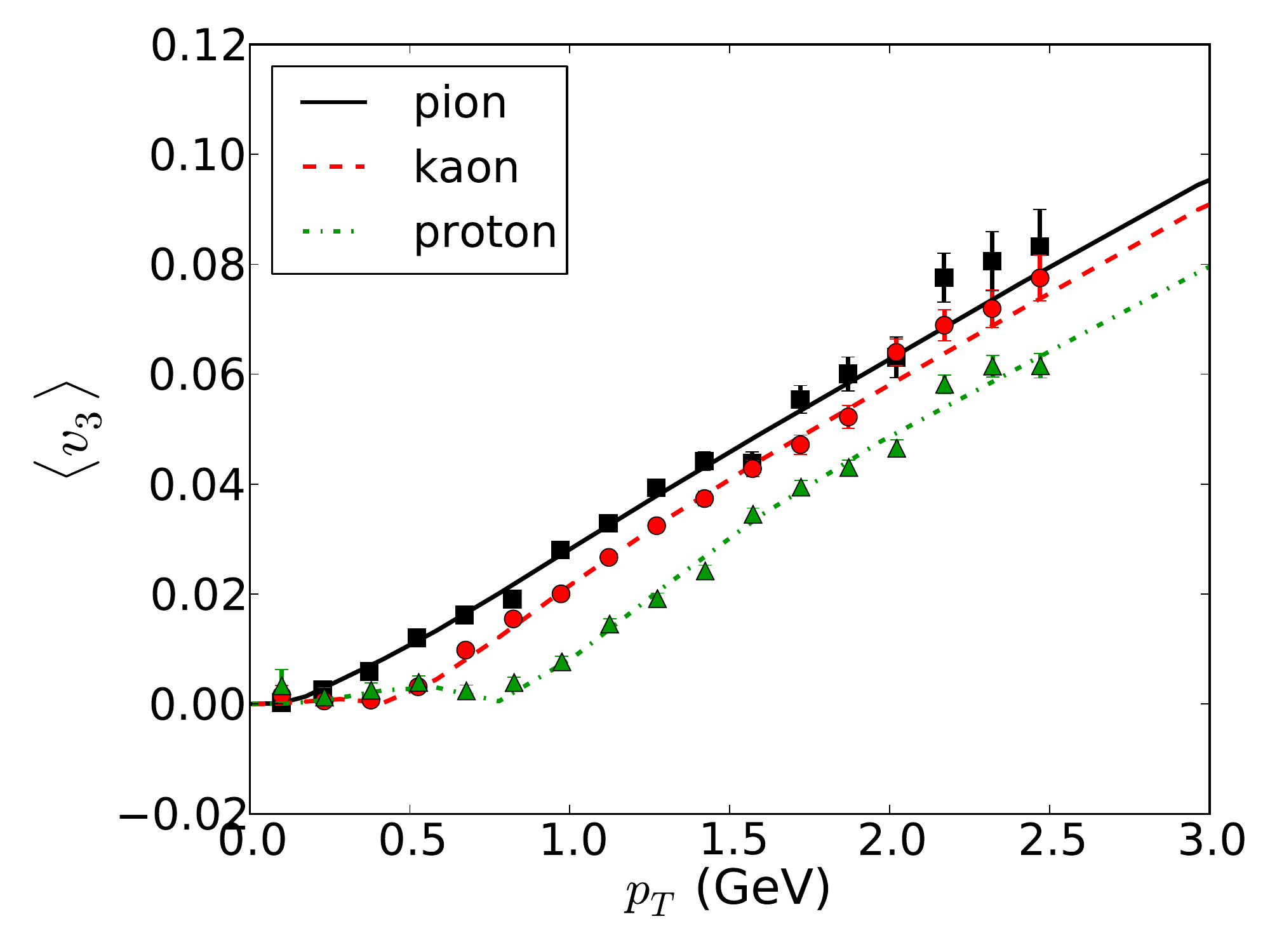} &
  \includegraphics[width=0.48\linewidth]{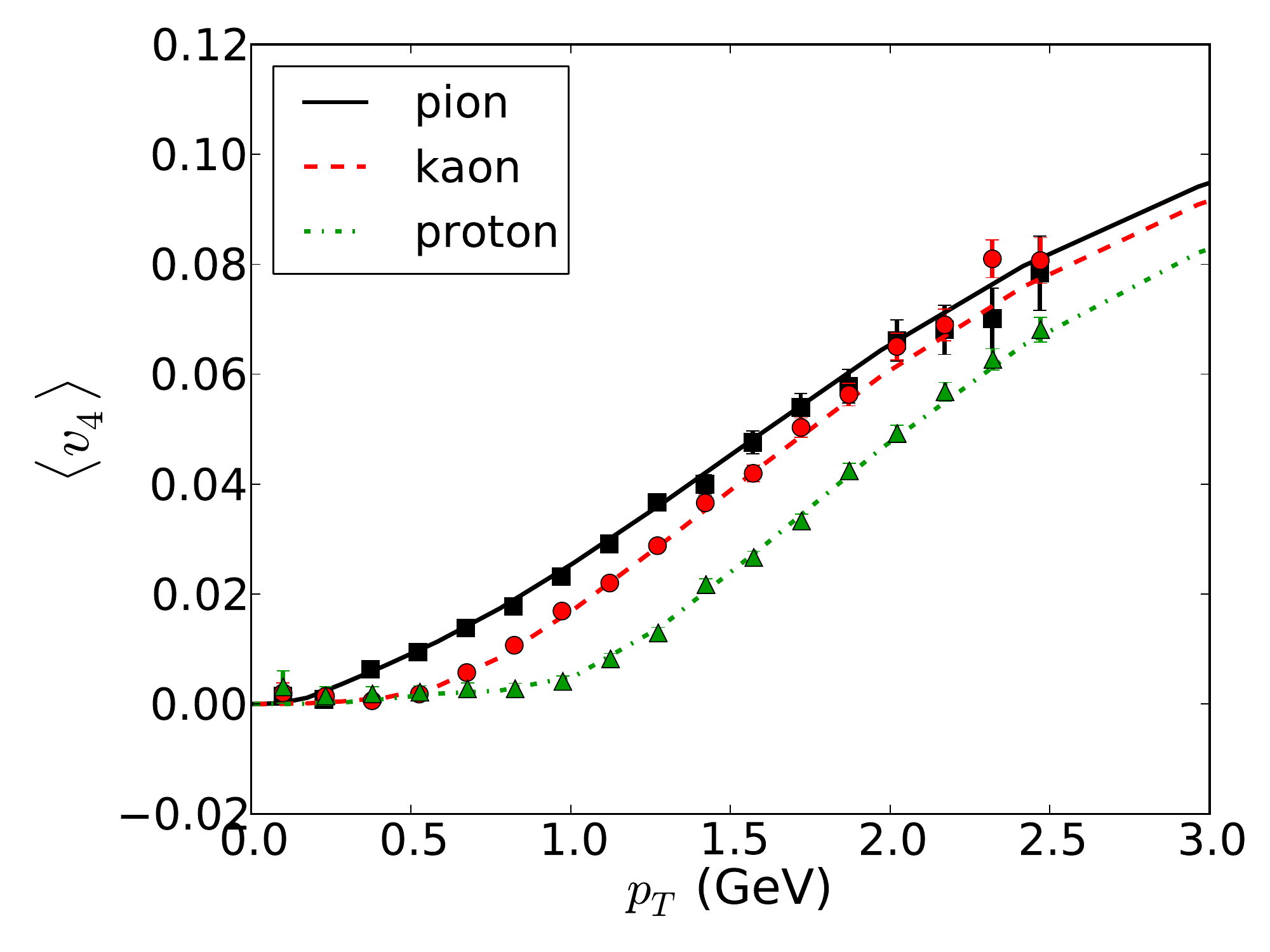}
  \end{tabular}
  \caption{Similar to Fig. \ref{iSS.fig2}, but for sampling method II of the purely numerical approach, which samples particle momenta first.}
  \label{iSS.fig6}
\end{figure*}
%=======================================

Finally, we shown in Figs. \ref{iSS.fig5} and \ref{iSS.fig6}, our results from the sampling method II within the purely numerical approach. In this method, since we first sample the momentum distributions of the particles, the sampled results reproduce the directly calculated particle momentum distributions from the Cooper-Frye formula very well. The spatial distributions of the particles, on the other hand, exhibit some noticeable deviations from the Cooper-Frye results, due to the removal of negative contributions.  

%%%%%%%%%%%%%%%%%%%%%%%%%%%%%%%%%%%%%%%%%%%%%
\section{OSCAR to URQMD: {\tt osc2u}}
\label{sec6}
%%%%%%%%%%%%%%%%%%%%%%%%%%%%%%%%%%%%%%%%%%%%%

The {\tt ISS} module produces an ensemble of hadrons on the particalization hyper surface -- each hadron being characterized by its production time and location, i.e. its position on the hyper surface, as well as its momentum, flavor-type, mass and quantum numbers. This ensemble is written out as a particle list in the standard {\tt OSCAR1997A} format.\footnote{\url{https://karman.physics.purdue.edu/OSCAR-old/docs/file/cascade_output_format/node8.html}} When running the code package in hybrid ({\tt VISHNU}) mode, this list needs to be converted into an initial condition file for the hadronic rescattering model {\tt URQMD}, which is the function of the {\tt Oscar} to {\tt UrQMD} converter {\tt osc2u}. 

Apart from reformatting the {\tt ISS} output to the {\tt UrQMD} initial condition format, {\tt osc2u} fulfills another important function, namely to synchronize all particles in the ensemble to the {\tt UrQMD} computational frame: in the {\tt ISS} output every particle is given with the time of its creation on the hyper surface. {\tt UrQMD}, however, requires all particles to be in the same computational frame in order to solve the Boltzmann collision integral. To achieve this, {\tt osc2u} propagates all hadrons {\em backwards} in time from their production time $\tau_f$ to $t=0$ and assigns each hadron $\tau_f$ as formation time. The {\tt UrQMD} calculation thus starts with an initial condition at time $t=0$ in its computational frame, propagating {\em forward} in time. During their formation time, hadrons are assigned zero cross section and do not interact. Therefore all hadrons in the initial condition will first start interacting at the location and time of their actual creation on the hyper surface.

%%%%%%%%%%%%%%%%%%%%%%%%%%%%%%%%%%%%%%%%%%%%%
\section{Hadronic Rescattering: {\tt UrQMD}}
\label{sec6b}
%%%%%%%%%%%%%%%%%%%%%%%%%%%%%%%%%%%%%%%%%%%%%

Ultra-relativistic Quantum Molecular Dynamics (UrQMD) is a microscopic transport model ideally suited for the description of the dynamics of a system of hadrons, both in and out of equilibrium
\cite{Bass:1998ca,Bleicher:1999xi}. The UrQMD approach is in spirit closely related to hadronic 
cascade \cite{Yariv:1979zz}, Vlasov--Uehling--Uhlenbeck \cite{Kruse:1985pg}
and (R)QMD transport models \cite{Aichelin:1988me,Sorge:1989dy} and has been extensively used to model
the evolution of hadronic systems in a variety of settings, from a wide range of heavy-ion collisions
to cosmic ray showers.

We shall describe here only the part of the model that is important
for the application at hand, namely the evolution of an expanding hadron gas
initially in local equilibrium (but subsequently not anymore constrained by any equilibrium assumptions), starting at the particalization or switching temperature $T_{sw}$.
The treatment of high-energy hadron-hadron scatterings, as it occurs in
the initial stage of ultrarelativistic collisions, is not discussed here.
A complete description of the model and detailed comparisons to experimental
data can be found in~\cite{Bass:1998ca,Bleicher:1999xi}.

The {\tt UrQMD} modeling package 
solves a Boltzmann equation for the distribution function of all hadrons in the 
system by evolving the system of hadrons through a sequence of binary collisions or $2-N$-body decays.
 
Binary collisions are performed in a point-particle sense:
Two particles collide if their minimum distance $d$, 
i.e.\ the minimum relative 
distance of the centroids of the Gaussians during their motion, 
in their CM frame fulfills the requirement: 
\begin{equation}
 d \le d_0 = \sqrt{ \frac { \sigma_{\rm tot} } {\pi}  }  , \qquad
 \sigma_{\rm tot} = \sigma(\sqrt{s},\hbox{ type} ).
\end{equation}
The cross section is assumed to be the free cross section of the
regarded collision type ($N-N$, $N-\Delta$, $\pi-N$, $\pi-\pi$, \ldots).

The {\tt UrQMD} collision term contains 53 different baryon species
(including nucleon, delta and hyperon resonances with masses up to 2 GeV) 
and 24 different meson species (including strange meson resonances), which
are supplemented by their corresponding anti-particle 
and all isospin-projected states.
The baryons and baryon-resonances which can be populated in {\tt UrQMD} are listed
in table~\ref{bartab}, the respective mesons in table~\ref{mestab} -- 
full baryon/antibaryon symmetry is included (not shown in the table), both,
with respect to the included hadronic states, as well as with respect to
the reaction cross sections.
All hadronic states can be produced in string decays, s-channel
collisions or resonance decays (string excitations and decays do not play a significant
role for the hadron gas evolution discussed here). 

\begin{table}
\centering
\begin{tabular}{cccccc}
\hline \hline
nucleon&delta&lambda&sigma&xi&omega\\  \hline \hline
$N_{938} $&$\Delta_{1232}$&$\Lambda_{1116}$&$\Sigma_{1192}$
&$\Xi_{1317}$&$\Omega_{1672}$\\
$N_{1440}$&$\Delta_{1600}$&$\Lambda_{1405}$&$\Sigma_{1385}$&$\Xi_{1530}$&\\
$N_{1520}$&$\Delta_{1620}$&$\Lambda_{1520}$&$\Sigma_{1660}$&$\Xi_{1690}$&\\
$N_{1535}$&$\Delta_{1700}$&$\Lambda_{1600}$&$\Sigma_{1670}$&$\Xi_{1820}$&\\
$N_{1650}$&$\Delta_{1900}$&$\Lambda_{1670}$&$\Sigma_{1775}$&$\Xi_{1950}$&\\
$N_{1675}$&$\Delta_{1905}$&$\Lambda_{1690}$&$\Sigma_{1790}$&$$&\\
$N_{1680}$&$\Delta_{1910}$&$\Lambda_{1800}$&$\Sigma_{1915}$&$$&\\
$N_{1700}$&$\Delta_{1920}$&$\Lambda_{1810}$&$\Sigma_{1940}$&$$&\\
$N_{1710}$&$\Delta_{1930}$&$\Lambda_{1820}$&$\Sigma_{2030}$&&\\
$N_{1720}$&$\Delta_{1950}$&$\Lambda_{1830}$&$$&&\\
$N_{1900}$&&$\Lambda_{2100}$&$$&&\\
$N_{1990}$&&$\Lambda_{2110}$&$$&& \\
$N_{2080}$ &&&&&\\
$N_{2190}$ &&&&&\\
$N_{2200}$ &&&&&\\
$N_{2250}$ &&&&&\\
\hline \hline
\end{tabular}
\caption{\label{bartab} Baryons and baryon-resonances treated in
{\tt UrQMD}. The corresponding antibaryon states are included as well.}
\end{table}

\begin{table}
\centering
\begin{tabular}{cccccc}
\hline \hline 
$0^-$ & $1^-$ &$ 0^+$ &$ 1^+$ &$ 2^+$ & $(1^-)^*$\\ \hline \hline
 $\pi$ & $ \rho$ & $ a_0$ & $ a_1$ & $ a_2$ & $ \rho(1450)$ \\
 $K  $ &$   K^*$ & $ K_0^*$ & $ K_1^*$ & $ K_2^*$ &$ \rho(1700)$ \\
 $\eta$&$  \omega$& $ f_0 $&  $f_1$ & $ f_2 $ & $ \omega(1420)$ \\
 $\eta'$&  $\phi $&  $f_0^*$ & $ f_1'$& $ f_2'$ & $ \omega(1600)$ \\
\hline\hline
\end{tabular}
\caption{\label{mestab} Mesons and meson-resonances, sorted with
respect to spin and parity, treated in {\tt UrQMD}.}
\end{table}

Tabulated and parameterized experimental 
cross sections are used when available. Resonance absorption, decays 
and scattering are handled via the principle of detailed balance. 
If no experimental information is
available, the cross section is either  calculated via
an One-Boson-Exchange (OBE) model or via a modified additive quark model
which takes basic phase space properties into account.

In the baryon-baryon sector, the total and elastic proton-proton and 
proton-neutron cross sections are well known \cite{Barnett:1996hr}. 
Since their functional dependence on $\sqrt{s_\mathrm{NN}}$ shows
a complicated shape at low energies, {\tt UrQMD} uses a table-lookup for those
cross sections. However, many cross sections involving strange baryons and/or 
resonances are not well known or even experimentally accessible -- for these
cross sections the additive quark model is widely used.

At RHIC and LHC energies, the most important reaction channels 
are meson-meson and
meson-baryon elastic scattering and resonance formation. 
For example, the total meson-baryon cross section for
non-strange particles is  given by
\begin{eqnarray}
\label{mbbreitwig}
\sigma^{MB}_{tot}(\sqrt{s}) &=& \sum\limits_{R=\Delta,N^*}
       \langle j_B, m_B, j_{M}, m_{M} \| J_R, M_R \rangle \nonumber \\
&&	\times        \frac{2 S_R +1}{(2 S_B +1) (2 S_{M} +1 )} 
\times        \frac{\pi}{p^2_{CMS}}\, 
        \frac{\Gamma_{R \rightarrow MB} \Gamma_{tot}}
             {(M_R - \sqrt{s})^2 + \frac{\Gamma_{tot}^2}{4}}
\end{eqnarray}
with the total and partial $\sqrt{s}$-dependent decay widths $\Gamma_{tot}$ and
$\Gamma_{R \rightarrow MB}$. 
The full decay width $\Gamma_{tot}(M)$ of a resonance is 
defined as the sum of all partial decay widths and depends on the
mass of the excited resonance:
\begin{equation}
\Gamma_{tot}(M)  
\label{gammatot}
       \,=\, \sum  \limits_{br= \{i,j\}}^{N_{br}} \Gamma_{i,j}(M) \quad.
\end{equation}
The partial decay widths $\Gamma_{i,j}(M)$ for the decay into the 
final state with particles $i$ and $j$ is given by
\begin{equation}
\label{gammapart}
\Gamma_{i,j}(M)
        \,=\,
       \Gamma^{i,j}_{R} \frac{M_{R}}{M}
        \left( \frac{\langle p_{i,j}(M) \rangle}
                    {\langle p_{i,j}(M_{R}) \rangle} \right)^{2l+1}
\times         \frac{1.2}{1+ 0.2 
        \left( \frac{\langle p_{i,j}(M) \rangle}
                    {\langle p_{i,j}(M_{R}) \rangle} \right)^{2l} } \quad,
\end{equation}
here $M_R$ denotes the pole mass of the resonance, $\Gamma^{i,j}_{R}$
its partial decay width into the channel $i$ and $j$ at the pole and
$l$ the decay angular momentum of the final state.
All pole masses and partial decay widths at the pole are taken from the Review
of Particle Properties \cite{Barnett:1996hr}. 
$\Gamma_{i,j}(M)$ is constructed in such a way that 
$\Gamma_{i,j}(M_R)=\Gamma^{i,j}_R$ is fulfilled at the pole.
In many cases only crude estimates for $\Gamma^{i,j}_R$ are given
in \cite{Barnett:1996hr} -- the partial decay widths must then be fixed by
studying exclusive particle production in elementary proton-proton
and pion-proton reactions. 
Therefore, e.g., the total pion-nucleon cross section depends on the
pole masses, widths and branching ratios of all $N^*$ and $\Delta^*$
resonances listed in table~\ref{bartab}. 
Resonant meson-meson scattering
(e.g. $\pi + \pi \to \rho$ or $\pi + K \to K^*$)
is treated in the same formalism.

In order to correctly treat equilibrated matter~\cite{Belkacem:1998gy}
(we repeat that the hadronic
matter with which {\tt UrQMD} is being initialized in our approach is in local
chemical and thermal equilibrium), the principle of detailed balance is
of great importance.
Detailed balance is based on time-reversal invariance 
 of the matrix element of the reaction. It is most commonly found
in textbooks in the form:
\begin{equation}
\label{dbgl3}
\sigma_{f \rightarrow i } \,=\, \frac{\vec{p}_i^2}{\vec{p}_f^2} \frac{g_i}{g_f}
\sigma_{i \rightarrow f} \quad ,
\end{equation}
with $g$ denoting the spin-isospin degeneracy factors.
{\tt UrQMD} applies the general principle of detailed balance to the 
following two process classes:
\begin{enumerate}
\item 
Resonant meson-meson and meson-baryon interactions: Each resonance created
via a meson-baryon or a meson-meson annihilation may again decay into
the two hadron species which originally formed it. This symmetry is only
violated in the case of three- or four-body decays and string fragmentations, 
since N-body collisions with (N$>2$) are not implemented in {\tt UrQMD}. 
\item
Resonance-nucleon or resonance-resonance interactions: the excitation
of baryon-resonances in {\tt UrQMD} is handled via parameterized cross sections
which have been fitted to data. The reverse reactions usually have not
been measured - here the principle of detailed balance is applied.
Inelastic baryon-resonance de-excitation is the only method in {\tt UrQMD}
to absorb mesons (which are {\em bound} in the resonance). Therefore
the application of the detailed balance principle is of crucial
importance for heavy nucleus-nucleus collisions.
\end{enumerate}

Equation~(\ref{dbgl3}), however, is only valid in the case of stable
particles with well-defined masses. Since in {\tt UrQMD} detailed balance
is applied to reactions involving resonances with finite lifetimes
and broad mass distributions, equation~(\ref{dbgl3}) has to be 
modified accordingly. For the case of one incoming resonance the
respective modified detailed balance relation has been derived
in \cite{Danielewicz:1991dh}. Here, we generalize this expression for
up to two resonances in both, the incoming and the outgoing channels.

The differential cross section for the reaction 
$(1\,,\,2) \rightarrow (3\,,\,4)$ is given by:
\begin{equation}
\label{diffcx1}
{\rm d} \sigma_{12}^{34} \,=\,
    \frac{| {\cal M} |^2}{64 \pi^2 s} \, \frac{p_{34}}{p_{12}} 
        \,{\rm d \Omega}\,
     \prod_{i=3}^4    \delta(p_i^2 -M_i^2) {\rm d}p_i^2  \quad,
\end{equation}
here the $p_i$ in the $\delta$-function denote four-momenta.
The $\delta$-function ensures that the particles are on mass-shell,
i.e. their masses are well-defined. If the particle, however, has 
a broad mass distribution, then the $\delta$-function
must be substituted by the respective mass distribution (including
an integration over the mass):
\begin{equation}
\label{diffcx2}
{\rm d} \sigma_{12}^{34} \,=\,
    \frac{| {\cal M} |^2}{64 \pi^2 s} \, \frac{1}{p_{12}} 
        \,{\rm d \Omega}\,
     \prod_{i=3}^4  p_{34} \cdot
   \frac{ \Gamma}{\left(m-M_i\right)^2+ \Gamma^2/4} \frac{{\rm d} m}{2\pi}
\, .
\end{equation}
Incorporating these modifications into equation~(\ref{dbgl3}) and
neglecting a possible mass-dependence of the matrix element we
obtain:
\begin{equation}
\label{uqmddetbal}
     \frac{ {\rm d} \sigma_{34}^{12} }{{\rm d} \Omega }   
      \, =\, \frac{\langle p_{12 }^2 \rangle   }
       {\langle p_{34 }^2 \rangle  } \,
       \frac{(2 S_1 + 1) (2 S_1 + 1)}
            {(2 S_3 + 1) (2 S_4 + 1)}
\times      \sum_{J=J_-}^{J_+} 
      \langle j_1 m_1 j_2 m_2 \| J M \rangle \, 
        \frac{ {\rm d} \sigma_{12}^{34} }{{\rm d} \Omega } \, .
\end{equation}
Here, $S_i$ indicates the spin of particle $i$ and 
the summation of the Clebsch-Gordan-coefficients is over the isospin of the
outgoing channel only. For the incoming channel, isospin is 
treated explicitly. The summation limits are given by:
\begin{eqnarray}
  J_- &=& \max \left( |j_1-j_2|,  |j_3-j_4| \right) \\  
  J_+ &=& \min \left( j_1+j_2,  j_3+j_4     \right)  \quad.
\end{eqnarray}
The integration over the mass distributions of the resonances  
in equation~(\ref{uqmddetbal}) has been denoted by the brackets 
$\langle\rangle$,
e.g.
\begin{equation}
p_{3,4}^2 \Rightarrow \langle p_{3,4}^2 \rangle \,
 = \,
 \int  \int 
  p_{CMS}^2(\sqrt{s},m_3,m_4)\, A_3(m_3) \, A_4(m_4) \, 
  {\rm d} m_3\; {\rm d} m_4 
\end{equation}
with the  
mass distribution $A_r(m)$ given by a free Breit-Wigner distribution
with a mass-dependent width according to equation~(\ref{gammatot}):
\begin{equation}
\label{bwnorm}
A_r(m) \, = \, \frac{1}{N} 
        \frac{\Gamma(m)}{(m_r - m)^2 + \Gamma(m)^2/4}
\end{equation}
with
\begin{displaymath}
        \lim_{\Gamma \rightarrow 0} A_r(m) = \delta(m_r - m) \, ,
\end{displaymath}
and the normalization constant
\begin{equation}
\label{bwnorm_norm}
N \,=\, \int\limits_{-\infty}^{\infty} 
\frac{\Gamma(m)}{(m_r - m)^2 + \Gamma(m)^2/4} \, {\rm d}m \, .
\end{equation}
Alternatively one can also choose a Breit-Wigner distribution with a fixed
width, the normalization constant then has the value $N=2\pi$.

The most frequent applications of equation~(\ref{uqmddetbal}) in {\tt UrQMD}
are the processes $\Delta_{1232}\, N \to N\, N$ and
$\Delta_{1232}\, \Delta_{1232} \to N\, N$.

%%%%%%%%%%%%%%%%%%%%%%%%%%%%%%%%%%%%%%%%%%%%%
\section{Interface for thermal photon emission}
\label{sec9}
%%%%%%%%%%%%%%%%%%%%%%%%%%%%%%%%%%%%%%%%%%%%%

\begin{figure}[h!]
  \centering
  \includegraphics[width=0.9\linewidth]{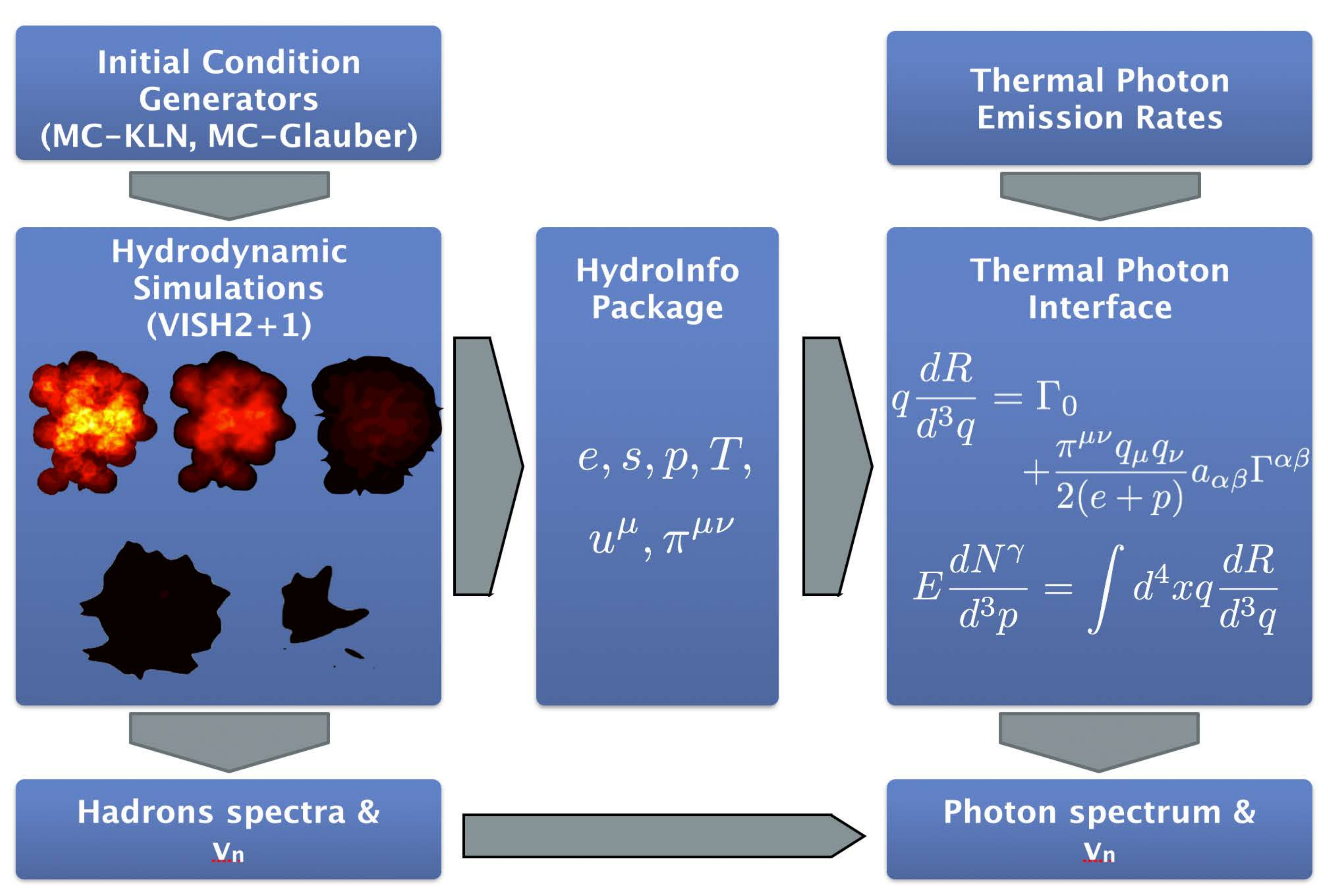}
  \caption{Work flow for event-by-event hydrodynamic simulation with photon emission.}
  \label{photonWorkflow}
\end{figure}

In the {\tt iEBE} package, we provide a separate branch to allow users to perform calculations for electromagnetic probes from relativistic heavy-ion collisions. In Fig. \ref{photonWorkflow}, we illustrate the work flow of such integrated calculations. 
%=======================================
%=======================================

In order to compute thermal photon emission from an evolving viscous hydrodynamic medium, we need to output the evolution history of the local temperature,  flow velocity and shear stress tensor and fold them with thermal photon emission rates to compute electromagnetic observables. Since the hydrodynamic evolution information is very demanding in terms of storage space, we output it in {\tt HDF5} binary format to minimize the storage requirement and increase the I/O efficiency. {\tt HDF5} format (Hierarchical Data Format) is a data model\footnote{See \url{http://www.hdfgroup.org/HDF5/} for details.}, library, and file format for storing and managing data. It supports an unlimited variety of datatypes, and is designed for flexible and efficient I/O and for high volume and complex data. The {\tt HDF5} library also provides multi-language support, which enables us to build  our interface in both {\tt Fortran} and {\tt C++} for future support. 

The momentum spectrum of thermal photons emitted from the expanding fireball can be written as 
\begin{equation}
E \frac{dN^\gamma}{d^3 p} = \int d^4 x \left(\Gamma_0 + \frac{\pi^{\mu\nu}}{2(e+P)} \Gamma_{\mu\nu} \right),
\label{photon.dN_1}
\end{equation}
where the integral goes over the space-time volume occupied by the radiating hot medium, $\Gamma_0$ is the thermal equilibrium emission rate, and the second term $\sim \pi_{\mu\nu}$ describes the shear viscous correction to the thermal emission rate. We can decompose $\Gamma_{\mu\nu}$ in a complete tensor basis and use the properties of the shear stress tensor, $\pi^{\mu\nu} = \Delta^{\mu\nu}_{\alpha \beta} \pi^{\alpha \beta}$, to write Eq. (\ref{photon.dN_1}) in the form
\begin{equation}
E \frac{dN^\gamma}{d^3 p} = \int d^4 x \left(\Gamma_0(u \cdot q, T) + \frac{\pi^{\mu\nu} \hat{q}_\mu \hat{q}_\nu}{2(e+P)} a_{\alpha \beta} \Gamma^{\alpha \beta}(u \cdot q, T) \right),
\label{photon.dN}
\end{equation}
where $\hat{q}^\mu = q^\mu/(u\cdot q)$, $a_{\alpha \beta}$ is a projection operator: 
\begin{equation}
a_{\alpha\beta} = \frac{3}{2( u \cdot q)^2} q_\alpha q_\beta + u_\alpha u_\beta + g_{\alpha\beta} - \frac{3}{2 (u \cdot q)}(q_\alpha u_\beta + q_\beta u_\alpha).
\end{equation}
The use of tensor decomposition in Eq. (\ref{photon.dN}) is particularly efficient numerically, because the anisotropic correction factors into a product of Lorentz scalars of which the first, $\pi^{\mu\nu} \hat{q}_\mu \hat{q}_\nu$, is most easily evaluated in the laboratory frame (where we know $\pi^{\mu\nu}$ from the solution of the hydrodynamic equations) while the second, $\Gamma_1 \equiv a_{\alpha \beta} \Gamma^{\alpha \beta}$, is best worked out in the local rest frame of the fluid cell (where $u \cdot q$ reduces to the local rest frame energy of the photon). This helps to avoid performing extensive Lorentz boosts and 3-D rotations of $\pi^{\mu\nu}$ for each fluid cell when coupled to hydrodynamic simulations. Besides speeding-up the calculation, it allows us to tabulate the viscous corrections into one convenient table that can easily be used for phenomenological studies.

Please note that the work flow in Fig. \ref{photonWorkflow} is generic for the calculation of all rare probes coupled to the evolving bulk medium that probe its temperature and flow velocity, such as jet energy loss and heavy quark diffusion. Similar modules for medium-induced jet quenching and jet shape modification as under construction.

%%%%%%%%%%%%%%%%%%%%%%%%%%%%%%%%%%%%%%%%%%%%%
\section{Conclusions}
\label{sec7}
%%%%%%%%%%%%%%%%%%%%%%%%%%%%%%%%%%%%%%%%%%%%%

In this work, we presented in detail the implementation of the {\tt iEBE-VISHNU} package for event-by-event numerical simulations of relativistic heavy-ion collisions.  We added multiplicity fluctuation in the MC-Glauber model based on the phenomenological KNO scaling observed in p-p collisions. This model can correctly reproduce the measured multiplicity distribution for p+Pb collisions at $\sqrt{s_\mathrm{NN}} = 5.02$ TeV. The multiplicity fluctuations also change the initial eccentricity distribution, $\{\varepsilon_n\}$. In {\tt VISHNew}, we improved the numerical stability of the (2+1)-d viscous hydrodynamic code when handling fluctuating initial conditions. We studied and documented the sensitivities of the final flow observables on the choice of the regulation strength parameters.  {\tt VISHNew} was tested against semi-analytical solutions derived based on Gubser's flow. The code {\tt iSS} converts the fluid cells into individual particles according to the Cooper-Frye formula. We proposed three different ways of dealing with the negative probability issues related to generating Monte-Carlo samples for different simulation purposes and pointed out ways to minimize these problems in various settings. This document includes performance tests and numerical checks for all three methods of sampling thermally emitted particles. 

In the present document we did not discuss the comparison of numerical results obtained with the iEBE-VISHNU code package for hadron and photon spectra and their anisotropy coefficients with experimental data. Such comparisons have been \cite{Qiu:2011hf, Heinz:2013bua, Shen:2013vja, Shen:2013cca, Heinz:2014uga, Shen:2014cga, Shen:2014lpa, Shen:2015qta, Plumberg:2015eia, Shen:2015qba, Bernhard:2015hxa, Goldschmidt:2015qya, Goldschmidt:2015kpa} and will continue to be published elsewhere. Recent improvements of the code package include a module for pre-equilibrium evolution Landau-matched to viscous hydrodynamics \cite{Liu:2015nwa}, and the inclusion of bulk viscous effects as well as a module for the computation of Hanbury-Brown Twiss (HBT) two-particle correlations from the iEBE-VISHNU output are in progress. These and additional future improvements of the code package will be made available at \url{https://u.osu.edu/vishnu/} as soon as code testing of the new components is completed.

%%%%%%%%%%%%%%%%%%%%%%%%%%%%%%%%%%%%%%%%%%%%%
\section*{Acknowledgments}
This work was supported in part by the U.S. Department of Energy, Office of Science, Office of Nuclear Physics, under Awards No. DE-SC0004286, DE-FG02-05ER41367, and (within the framework of the JET Collaboration) DE-SC0004104, and in part by the Natural Sciences and Engineering Research Council of Canada. We gratefully acknowledge important contributions by Scott Moreland during the early stages of the development of the {\tt superMC} module.
%%%%%%%%%%%%%%%%%%%%%%%%%%%%%%%%%%%%%%%%%%%%%



\bibliographystyle{elsarticle-num} 
\bibliography{iebe}


%\begin{thebibliography}{00}


%\bibitem{}

%\end{thebibliography}
\end{document}